\documentclass[iop]{emulateapj-rtx4} 
\shortauthors{Sekanina \& Kracht}
\shorttitle{Comets X/1106 C1, 1138, and 363 in History of Kreutz Sungrazer System}
\slugcomment{Version \today }
\newcommand{\Rsun}{$R_{\mbox{\scriptsize \boldmath $\odot$}}$}

\begin{document}
\title{THE GREAT COMET OF 1106, A CHINESE COMET OF 1138, AND DAYLIGHT COMETS IN LATE
 363\\AS KEY OBJECTS IN COMPUTER SIMULATED HISTORY OF KREUTZ SUNGRAZER SYSTEM}
\author{Zdenek Sekanina$^1$ and Rainer Kracht$^2$}
\affil{$^1$Jet Propulsion Laboratory, California Institute of Technology,
  4800 Oak Grove Drive, Pasadena, CA 91109, U.S.A.\\
  $^2$Ostlandring 53, D-25335 Elmshorn, Schleswig-Holstein, Germany}
\email{Zdenek.Sekanina@jpl.nasa.gov\\
{\hspace*{2.59cm}}R.Kracht@t-online.de{\vspace{-0.2cm}}}

\begin{abstract} 
We present the results of our orbital computations in support of the recently   
proposed contact-binary model for the Kreutz sungrazer system (Sekanina 2021, 2022).
We demonstrate that comet Ikeya-Seki (C/1965~S1) previously passed perihelion
decades after the Great Comet of 1106 (X/1106~C1) and argue that, like the Great
September Comet of 1882 (C/1882~R1), it evidently was a fragment~of~the comet
recorded by the Chinese in September 1138.  The 1106 sungrazer appears instead to
have been the previous appearance of the Great March Comet of 1843 (C/1843~D1).
With no momentum exchange involved, fragments of a Kreutz sungrazer breaking up
tidally near perihelion are shown to end up in orbits with markedly different
periods because their centers of mass are radially shifted by a few kilometers
relative to the parent.  The daylight comets of AD~363, recorded by a Roman
historian, are accommodated in our computations as the first appearance of
the Kreutz sungrazers after their bilobed progenitor's breakup.  We link the
\mbox{1843--1106--363} (Lobe~I) and \mbox{1882--1138--363} (Lobe~II) returns
to perihelion by single nongravitational orbits and gravitationally with minor
center-of-mass shifts acquired in fragmentation events.  We also successfully
model the motion of Aristotle's comet as the rotating progenitor that at aphelion
split (at a few m~s$^{-1}$) into the two lobes, the precursors of, respectively,
the 1843 and 1882 sungrazers; and provide a 1963--1041--363 link for comet
Pereyra (C/1963~R1).  Material fatigue could contribute to sungrazers' fragmentation
throughout the orbit, including aphelion. --- Continuing problems with the
nongravitational law in orbit software are noted.
\end{abstract} 
\keywords{comets general: Kreutz sungrazers; comets individual: 372 BC, 363, 1041,
 X/1106 C1, 1138, C/1843 D1, C/1882 R1, C/1963 R1, C/1965 S1; methods: data
 analysis{\vspace{-0.1cm}}}
\section{Introduction}  
Among the Kreutz sungrazers seen in the past 200 years, two objects have stood out as by
far the intrinsically brightest and presumably the most massive:\ the Great March Comet
of 1843 (C/1843~D1) and the Great September Comet of 1882 (C/1882~R1).  Moving about
the Sun in somewhat different orbits (though sharing a common line of apsides), they
are the quintessential members of two fundamental subgroups or populations, I and II,
respectively.  Their possible history was described more than 50~years ago by Marsden
(1967).  Although the number of known populations has gradually been growing (e.g.,
Marsden 1989), Populations I and II have remained.  Bright Kreutz sungrazers have
repeatedly been observed to fragment or disintegrate at perihelion, but the existence
of the distinct populations cannot be explained by perihelion breakups, unless
unrealistically long periods of time are invoked.

While the populations are in principle understood as products of cascading fragmentation
at all heliocentric distances, with the primary breakup in the general proximity of {\it
aphelion\/} (Sekanina 2002; Sekanina \& Chodas 2004, 2007), a meaningful modeling of the
Kreutz system's evolution requires at least limited knowledge of the history of its members.
Given that the typical orbital periods of the Kreutz comets exceed 700~yr, the parent
objects of the 19th to 21th century sungrazers are necessarily historical comets, for
which little, if any, information is available.

\section{Past Work Related to the Comet of 1106} 
The Great Comet of 1106 (X/1106 C1), long suspected from historical records to have been
a sungrazer (e.g., Pingr\'e 1783, Hall 1883, Kreutz 1888, 1901), should constitute an
integral part of any evolutionary investigation of the Kreutz system.  Although limited
data are available on the comet's tail, no orbit could be computed.  Granted it indeed
was a Kreutz comet, the issue of whether it was a previous appearance of the Great March
Comet of 1843 (of Population~I) or the Great September Comet of 1882 (of Population~II)
has long been a point of contention.{\vspace{0.05cm}}

A pivotal role in efforts to settle this issue has in fact been played by another sungrazer,
comet Ikeya-Seki (C/1965~S1), whose orbit in many respects resembles that of the 1882 comet
so closely that it has served as an indispensable proxy because of its much higher accuracy.
Perihelion fragmentation and the resulting post-perihelion nucleus multiplicity complicate
matters in either case, but comet Ikeya-Seki was observed after perihelion to possess two
persisting fragments, while the 1882 sungrazer displayed up to six.  Integrating the orbit
of the primary nucleus of Ikeya-Seki back in time, Marsden (1967) established that it would
have previously been at perihelion in March 1115 or September 1116 depending on whether
the relativistic effect was or was not included.  His integration of Kreutz's
(1891) nonrelativistic orbit for the primary nucleus (No.\,2 in Kreutz's notation) of the
1882 comet gave the previous perihelion in April 1138.  Marsden concluded that the Great
Comet of 1106 ``seems by far the most promising candidate for the previous appearance
of comet [Ikeya-Seki]'' and that it was ``virtually proven that [the 1882 comet and
Ikeya-Seki] were indeed one at their previous approach to the Sun.''

While evidence for the second conclusion is strong, the first is questionable because of
as yet unresolved issues in the orbital motions of the 1882 comet and Ikeya-Seki.  While
for the first comet we have nothing to add to Kreutz's (1891) results, we develop a novel
technique to apply to the orbit of comet Ikeya-Seki in order to settle the question of
whether or not the 1106 comet was this sungrazer's previous appearance.   

The ``best-determined'' orbit of each nucleus fragment of the Great September Comet of
1882 or comet Ikeya-Seki was computed (by Kreutz or by Marsden) by linking the comet's
pre-fragmentation astrometric positions before perihelion with that fragment's positions
after perihelion.  On the one hand, such an orbital solution is --- to say the least ---
problematic because it smoothes the motion prior to and after the point of dynamical
discontinuity associated with the splitting --- the very effect of interest.  On the
other hand, the options of overcoming the discontinuity problem are in practice very
limited because separate preperihelion and post-perihelion orbital arcs are inadequate
to serve as a basis for accurate orbital-period determination.

Although subject to occasional criticism, Marsden's (1967, 1989) preference for the
1106~comet as the parent to the 1882 sungrazer and Ikeya-Seki's comet (rather than
to the 1843 comet) was incorporated into the two-superfragment model of the Kreutz
system, proposed by Sekanina (2002) and implemented by Sekanina \& Chodas (2004).
The subsequently developed alternative model (Sekanina \& Chodas 2007) adopted the
1106~comet as the parent to the 1843~comet.  Following the appearance of comet Lovejoy
(C/2011~W3) and other relevant new developments in the past 10 years, these models
were upgraded by the recent introduction of a novel conceptual model (Sekanina 2021;
referred to hereafter as Paper~1), in which the birth of the Kreutz system was linked
to a breakup of the {\it contact-binary\/} progenitor near aphelion about two millennia
ago.  The 1106 comet was presented in this model as the most massive fragment of
Lobe~I and the precursor of the 1843~comet.  The perihelion time of the 1106 comet was
then found to have fitted a uniform time sequence of three generations of Population~I
fragments, derived from Aristotle's comet of 372~BC as the progenitor.  However, temporal
effects triggered by the indirect planetary perturbations, by the outgassing-driven
nongravitational forces, and/or by other possible mechanisms in the course of recurrent
near-perihelion fragmentation events were all ignored, and this is being corrected in
the present paper.

\section{New Evidence from the Orbital Motion of\\Comet Ikeya-Seki} 
The 119 observations used by Marsden (1967) to compute the orbit of the primary
nucleus, A, of Ikeya-Seki are all included in the list presented in the Minor Planet
Center's database\footnote{See {\tt https://minorplanetcenter.net/db\_search}.}
The list contains a total of 129~astrometric observations, of which 78 are
preperihelion data points from a period of 1965 September~21 through October~16;
nine~data points from the early post-perihelion period of October~28 through
November~1, before discovery of the companion nucleus, B; and 42~positions of
nucleus~A from November~5 on.  The actual number of reported astrometric observations
was much higher, but none of the grossly inaccurate ones made it into the list.
For example, missing are 23 of 27~positions originally published by the Central
Bureau for Astronomical Telegrams on 1965~December~1 and all of those reported on
December~17 and 30 (Gingerich 1965, Marsden 1965).

The key astrometric observations were made by Z.\ M.\ Pereyra and J.\ J.\ Rodr\'{\i}guez
with the 33-cm f/10.5 astrograph of the C\'ordoba Observatory (Pereyra 1971):\ the
first measured images were exposed by Pereyra on 1965 September~21.38~UT, less than
three days after discovery; the last preperihelion and the first post-perihelion
observations were made by Rodr\'{\i}guez on October~16.36 and 28.35~UT, respectively.
The length of the preperihelion arc covered by the observations was merely 25~days.
Further positional data in the early post-perihelion period, until November~1
(11~days after perihelion), were secured by Roemer (Roemer \& Lloyd 1966) and by
Lourens (1966).  The secondary nucleus B was detected first on November~4.53~UT
(Pohn 1965), but no astrometry was available from the period before November~12.
The last reduced images of either fragment were acquired by Tammann (1966) with
the Palomar Observatory's 122-cm Schmidt telescope on 1966 January~14.33~UT, 85~days
after perihelion;\footnote{Porter (1967) noted that the comet was seen on films
exposed with the Baker-Nunn cameras at two stations of the Smithsonian Astrophysical
Observatory in late January and, possibly, mid-February, but none of these images
was ever measured.} they were measured by Marsden (1967).  Overall, the comet's
orbital arc covered by the astrometric observations extends over 115~days; however,
there is a data gap from 5~days before to 7~days after perihelion.{\hspace{0.1cm}}

\subsection{Approximating the Motion of the Pre-Split Nucleus} 

Our first objective was an in-depth review of the orbit of comet Ikeya-Seki based on the
available set of preperihelion observations to confirm that they were utterly inadequate for
determining the comet's past motion.  We tested the veracity of this conclusion by deriving
several orbital solutions.  Our computer code accounted for the perturbations by the planets,
Pluto, and the three most massive asteroids, and for the relativistic effect.  A general
solution using all 78 preperihelion data points left residuals from three observations
that exceeded 4$^{\prime\prime}$ in one coordinate, and these were subsequently removed
from the set.  Eventually we obtained the orbital period from solutions based on 78, 75,
74, 72, and 70~data points.  Summarized in Table~1, they show that the orbital period came
out to be indeterminate, with enormous mean errors, confirming that this was not the way
to proceed.{\hspace{-0.05cm}}
\begin{table}[t]
\vspace{-4.2cm}
\hspace{5.2cm}
\centerline{
\scalebox{1}{
\includegraphics{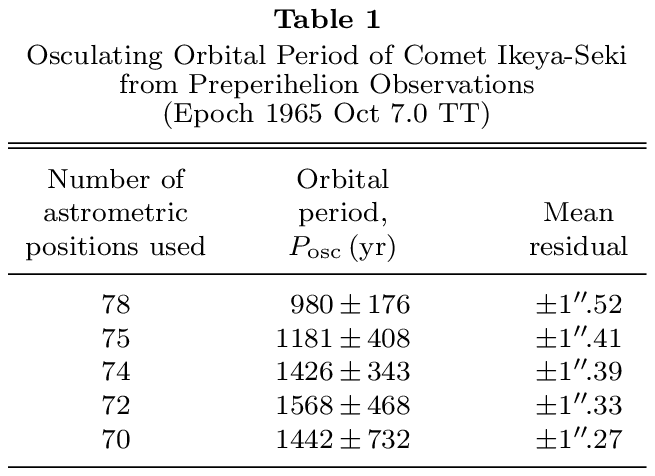}}}
\vspace{-20.3cm}
\end{table}

Instead, we followed a rather different path.  We~began by assuming, for the sake of
argument, that~the 1106~comet {\it was\/} the previous appearance~of~\mbox{Ikeya-Seki},
as suggested by Marsden (1967).  We subtracted the Julian date of the perihelion time
of the 1106~comet, nominally 1106~January~26 according to Hasegawa \& Nakano (2001), from
the Julian date of the 1965 perihelion passage of comet Ikeya-Seki to yield a pre-split
barycentric orbital period of \mbox{$P_{\rm bar} = 859.68$ years}, equivalent to a
barycentric reciprocal semimajor axis of \mbox{$(1/a)_{\rm bar} = +0.011055$ AU$^{-1}$}.
With Everhart \& Ragha\-van's (1971) barycentric correction for the epoch of 1965~Oct~7.0~TT
equaling \mbox{$u_{\rm b} = -0.000255$ AU$^{-1}$}, this exercise led to an osculating
reciprocal semimajor axis of \mbox{$(1/a)_{\rm osc} = +0.010800$ AU$^{-1}$} and an orbital
period of \mbox{$P_{\rm osc} = 891$ years}.

We next employed this value of $(1/a)_{\rm osc}$ in {\it forced\/} orbital solutions to
examine the degree of compatibility of the comet's preperihelion astrometric observations
with the premise on its previous perihelion passage in 1106.  We found that {\it all 78
fitted this premise\/} to within $\pm$4$^{\prime\prime}$ and 74 to within $\pm$3$^{\prime
\prime\!}.5$, a greater degree of conformity than in the formally optimized solutions in
Table~1.  This implied that the orbital periods much longer than 900~years were unrealistic.
%
%

We accepted that the comet's fragmentation had taken place at a time of less than one
hour from perihelion (Sekanina 1982, Sekanina \& Chodas 2007) and pursued an objective
of deriving an orbital period that was representative of the comet's true motion before
splitting, thus eliminating or mitigating the problems that plagued the computations
leading to the unsatisfactory results in Table~1.  We based our strategy on the fact
that orbital positions of the most massive fragment of a split comet get affected by
the fragmentation event only gradually, a number of days, if not weeks, later.  This
is documented by an often considerable lapse of time between the breakup (as determined
by an appropriate method) and{\vspace{-0.02cm}} the first detection of nucleus duplicity
(or{\vspace{-0.02cm}} multiplicity).\footnote{For comet Ikeya-Seki this time amounted
to two weeks.}  For example,{\vspace{-0.03cm}} a typical nuclear fragment's velocity
of 1~m~s$^{-1}$ relative to the comet's center of mass would not generate a potentially
measurable position shift exceeding, say, 1000~km (equivalent in an extreme case to a
projected position shift of $\sim$3$^{\prime\prime}$ at a geocentric distance of 0.5~AU)
until approximately 12~days after the breakup, which for Ikeya-Seki would be November~2.

This kind of argument led us to a conclusion that the comet's credible pre-split
orbital period could successfully be approximated by extrapolating a sequence of
orbital runs, each of which links the set of preperihelion astrometric observations
with a different set of {\it early\/} post-perihelion observations.  We obtained
four such sets of orbital solutions based on the 72~preperihelion data linked with,
successively, two post-perihelion observations made on October~28 (the earliest
post-perihelion data); four observations from October~28--29; six observations from
October~28--31; and nine observations from October~28--November~1.  If these early
post-perihelion observations are accurate, the runs should offer a set of barycentric
orbital periods, $P_{\rm bar}$, and times of the previous perihelion, $T_{\rm prev}$,
that display a systematic trend when plotted as a function of the termination date.
Extrapolation of these values of $P_{\rm bar}$ and $T_{\rm prev}$ back to the termination
date equaling the time of fragmentation should provide the desired credible estimates
of the pre-split values.  The values of $P_{\rm bar}$ and $T_{\rm prev}$ derived from
runs whose termination dates are located beyond this early post-perihelion period are
expected to be noisy, slowly converging to Marsden's (1967) values for nuclear fragment~A,
and of lesser interest to this numerical exercise.

The results are presented in Table 2.  The individual columns list:\ the time of
termination observation, $t_{\rm fin}$, and its reference to the perihelion time,
\mbox{$t_{\rm fin} \!-\! T$} (the first observation, $t_{\rm first}$, having
always been September~21.38~UT and \mbox{$t_{\rm first}\!-\!T = -29.80$ days});
the osculation value of the reciprocal semimajor axis, $(1/a)_{\rm osc}$, for the
epoch of 1965~October~7.0~TT and its mean error; the corresponding past barycentric
value, $(1/a)_{\rm bar}$; the past barycentric orbital period, $P_{\rm bar}$, and its
mean error; the derived nominal time of the previous perihelion passage, $T_{\rm prev}$;
the mean residual of the orbital solution; the number of observations used, separately
before and after perihelion; and the computed offset of the secondary nucleus B from
the primary A at $t_{\rm fin}$.  It is a measure of the offset of A from the center
of mass of the pre-split comet, which is expected to be its very small,
approximately proportionate fraction.
\begin{table*}[th]
\vspace{-4.2cm}
\hspace{0.5cm}
\centerline{
\scalebox{1}{
\includegraphics{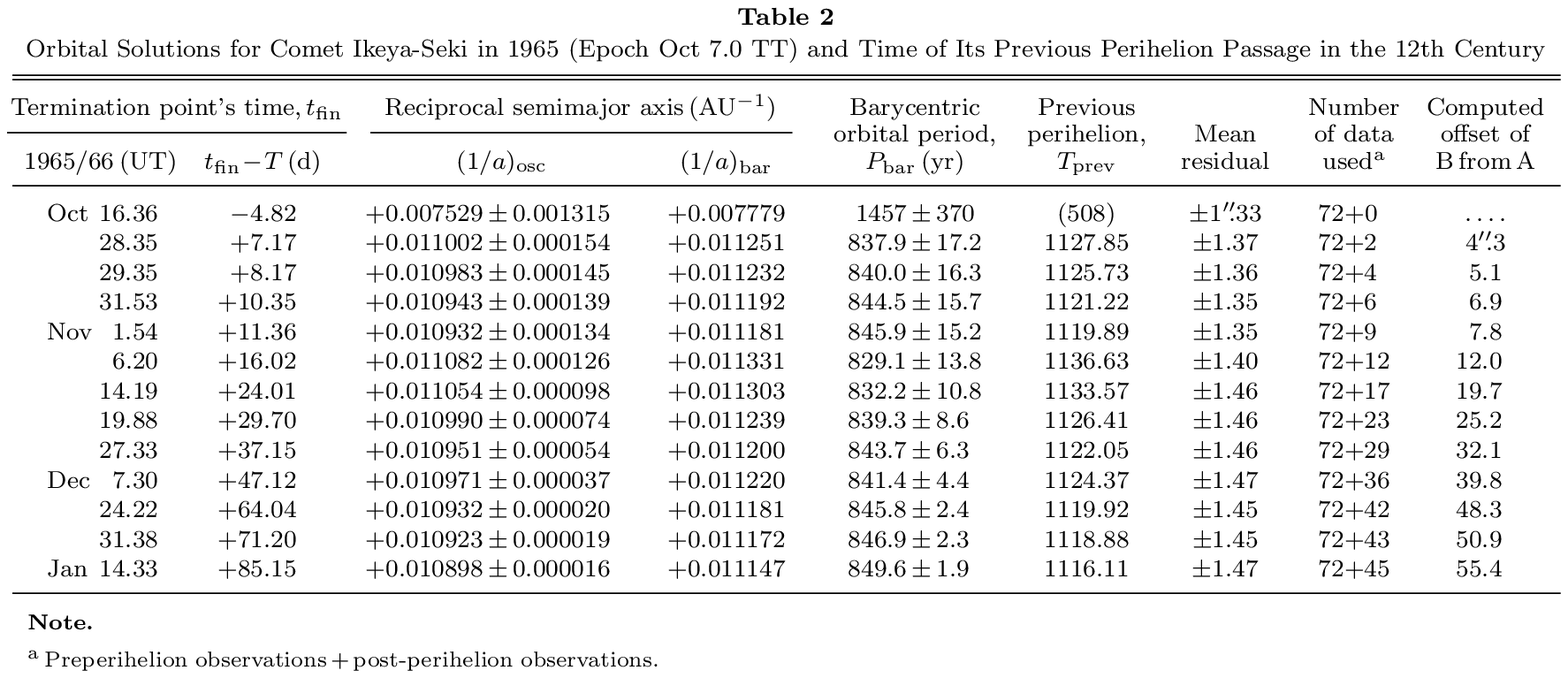}}}
\vspace{-17.15cm}
\end{table*} 

For the sake of comparison, the first row of the table shows yet another useless solution
based solely on the preperihelion observations.  The numbers in the subsequent rows show
that adding a few early post-perihelion observations to the 72~preperihelion data points
reduces the barycentric orbital period by some 300~yr and its formal error by a factor of
more than 20.  And regardless of these errors, the orbital solutions in rows 2 to 5 of the
table, whose termination observations were between Oct 28 and Nov 1, consistently display,
as expected, a modest systematic trend in the predicted perihelion time and have the mean
residuals measurably lower than the subsequent solutions based on longer arcs and including
the later post-perihelion observations.  When extrapolated from the early post-perihelion
termination dates to the time of fragmentation, Oct~21.20~UT, the orbits nominally imply
November~1139 for the time of comet Ikeya-Seki's previous perihelion $T_{\rm prev}$ and
825.9~yr for its barycentric orbital period $P_{\rm bar}$, with an error of $\pm$2.0~yr
introduced by the extrapolation.  Including the formal errors from the orbital solutions,
the derived perihelion time still differs by 2$\sigma$ from the time of appearance of the
1106~comet.

Starting with row 6, Table~2 provides information on the orbital solutions that include
the termination dates of Nov~6, two days after the discovery of the secondary nucleus,
or later.  In terms of the time $T_{\rm prev}$ these solutions are incompatible
with the four solutions in rows 2 to 5, and are more noisy, as seen from their mean
residuals.  The disparity in $T_{\rm prev}$ is clearly apparent from Figure~1.  Yet,
{\it all\/} values of $T_{\rm prev}$ in the plot deviate by one to at least three
decades from the time of appearance of the 1106 comet in the same direction.

The formal error of the orbital period in Table~2 depends strongly on the
length of the orbital arc covered by the observations.  The errors from the solutions
with the termination points after Nov~1 are lower than those in rows 2 to 5, even though
the mean residuals vary  in the opposite direction.  One may question the meaning of
these errors and suggest that the orbital periods and the implied perihelion times from
the solutions with the termination dates of October~28 through November~1 could in fact
be more accurate than they appear to be.

This argument is supported by the distribution of individual residuals listed for the
orbits in rows 2--7 of Table~2 in Table~3.  It is discouraging to see most of the
highly consistent observations from Oct~28--Nov~1 to leave systematic residuals of
2$^{\prime\prime}$ to 3$^{\prime\prime}$ in right ascension from the solutions
whose termination dates were after Nov~1.  In fact, this trend continued to the
final orbit of fragment A and was only slightly reduced by incorporating
nongravitational terms into the equations of motion.

Comparison of Marsden's (1967) relativistic set of the orbital elements for nucleus
fragment~A with our best gravitational and nongravitational orbits is offered in
Table~4.  The nongravitational solutions were computed using Style~II formalism of
Marsden et al.\ (1973).  We were rather surprised that it was possible to derive the
radial nongravitational parameter $A_1$ with a formal error as small as 20~percent.
However, the value of $A_1$ appears to be about an order of magnitude higher than
expected and it may be a product of the computing methodology, namely, the linkage
of the preperihelion observations of the pre-split nucleus with the full set of
the post-perihelion observations of fragment~A.

We were even more surprised that we were able to derive both the radial parameter
$A_1$ and the transverse parameter $A_2$, even though the former was now
determined with a higher error.  The most unexpected result of this two-parameter
nongravitational run was the predicted time of the previous perihelion passage,
nearly 50~years after the appearance of the Great Comet of 1106.  While this very late
time is not necessarily very significant, it accentuates the difficulties with the
1106 comet as Ikeya-Seki's parent.  
%
%

\subsection{Effects of Ikeya-Seki's Fragmentation on\\the Orbital Period}
%
Intuitively, the sets of orbital elements for the two fragments into which a comet has
split are plausible limits for confining the set of pre-split elements, as the fragments
presumably acquired differential momenta in the opposite directions.  Indeed, in
implementing their two-superfragment model, Sekanina \& Chodas (2004) argued that the
osculating orbital period of the pre-split nucleus of comet Ikeya-Seki should have been
between 880 and 1055~years (i.e., the orbital periods of the two fragments, as computed
by Marsden 1967), but much closer to the first because the primary fragment should have
been substantially more massive than the secondary and subjected to a much smaller
effect.  It so happened that by choosing the orbital period slightly longer than the
period of the primary fragment, it was possible to readily link Ikeya-Seki with the
1106~comet, with an implied primary-to-secondary mass ratio of 15:1.

A view from another angle suggests that the orbital period of the pre-split comet does
not have to lie between the orbital periods of the two fragments and can in fact be
the shortest of the three.  Let $\Delta P$ be the difference between the osculating
orbital periods of the primary, A, and secondary, B, fragments of the comet at the
adopted epoch of 1965~October~7.0~TT, as derived from Marsden's (1967) standard orbital
solutions:
\begin{equation}
\Delta P = P_{\rm B} \!-\! P_{\rm A} = +175 \!\pm\! 5 \; {\rm yr}. 
\end{equation}
This result offers a total effect but says nothing about the nature of the forces that
were responsible for the different motions of the two fragments.  The hypothesis that
it was due to an outgassing-driven differential acceleration comes from a study of the
fragments' relative motion; in the orbital computations the differential acceleration
is known to masquerade as an effect in the orbital period (or the eccentricity).

\begin{figure*}
\vspace{-8.05cm}
\hspace{-0.15cm}
\centerline{
\scalebox{0.86}{
\includegraphics{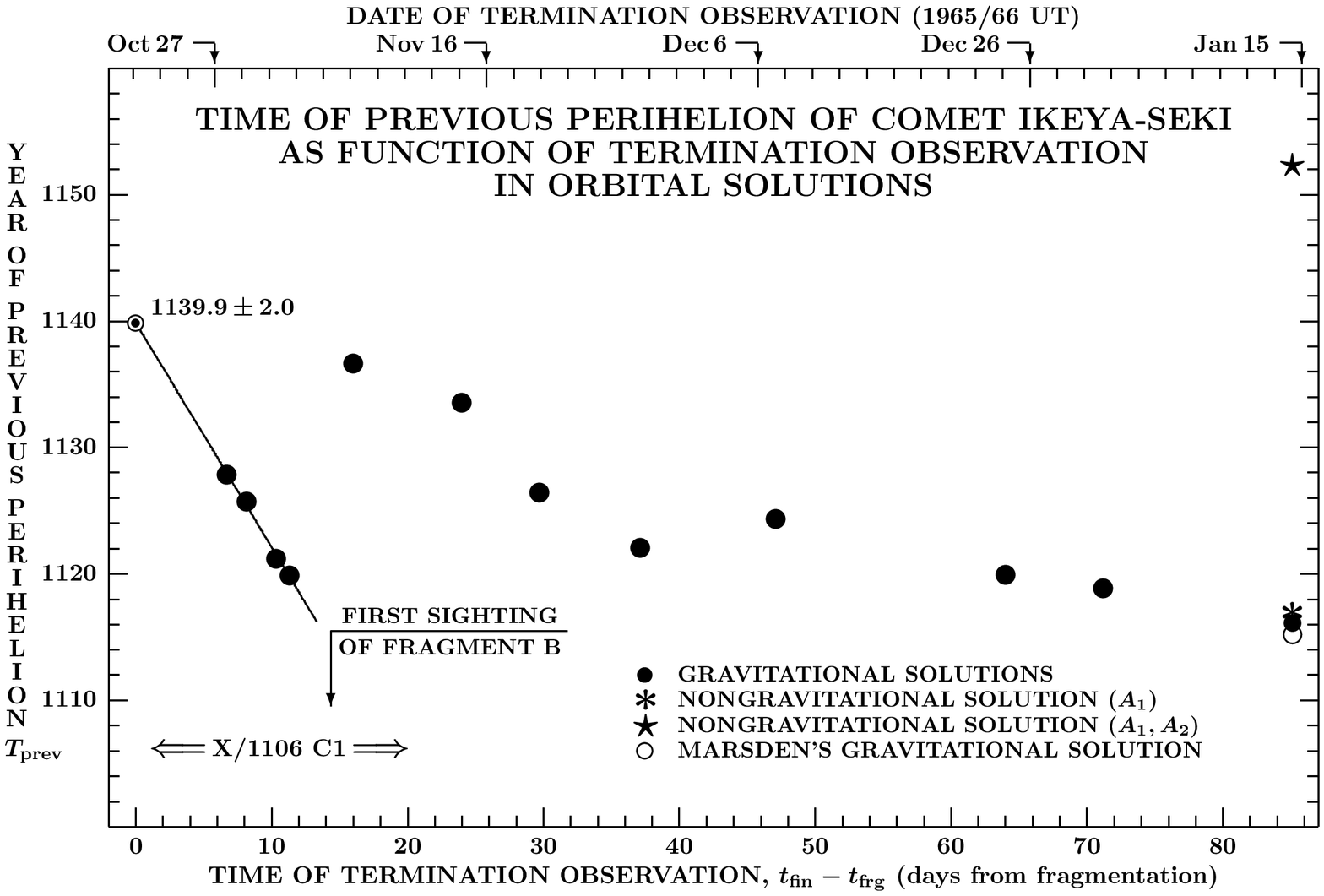}}}
\vspace{-5.34cm}
\caption{The time of the previous perihelion passage of comet Ikeya-Seki in the 12th
century, implied by the orbital solutions with different termination points in the
orbit following fragmentation.  Only the early post-perihelion observations, not later
than November 1 (or 12~days after perihelion), could be used as termination points to
orbital solutions that offer consistent results, with the previous perihelion nominally
in late AD~1139.  Fragmentation appears to have taken place only a fraction of an hour
after the 1965 perihelion.  Note that the ordinate of not a single data point in the
plot is near the time of appearance of the 1106 comet (X/1106~C1).{\vspace{0.7cm}}}
\end{figure*}

The motion of fragment~B relative to fragment~A of comet Ikeya-Seki was investigated
by Sekanina (1982), using his model for the split comets.  The nucleus was found to
have broken up 23\,$\pm$\,7~minutes after perihelion and{\vspace{-0.03cm}} the
(radial) nongravitational acceleration was 0.67\,$\pm$\,0.03~units of 10$^{-5}$\,the
Sun's gravitational acceleration.  The magnitude of the force, assumed to vary inversely
as the square of heliocentric distance, $r$, was at 1~AU from the Sun equivalent
to\\[-0.4cm]

\begin{equation}
\gamma = 0.20 \times\!10^{-8} \, {\rm AU} \; {\rm day}^{-2}. 
\end{equation}
The acceleration's contribution to the orbital velocity, $V$, of B relative to A,
integrated over the post-fragmentation orbital arc equaled in parabolic-approximation
\begin{equation} 
\Delta V_{\rm subl} = \int_{t_{\rm frg}}^{\infty} \gamma \left( \! \frac{r_0}{r}
 \! \right)^{\!2} \sin {\textstyle \frac{1}{2}} u \,dt = \frac{\gamma}{k}
 \sqrt{\frac{2}{r_{\rm frg}}} \, r_0^2,\\[-0.1cm] 
\end{equation}
where $t_{\rm frg}$ and $r_{\rm frg}$ are, respectively, the fragmentation time and
heliocentric{\vspace{-0.05cm}} distance; $k$ the Gaussian gravitational constant,
\mbox{$k = 0.0172021$ AU$^{\frac{3}{2}}$\,day$^{-1}$}; $r_0$ the unit heliocentric
distance, \mbox{$r_0 = 1$ AU}; and $\sin {\textstyle \frac{1}{2}} u$ the acceleration's
contribution in the direction~of~the~orbital-velocity {\nopagebreak}vector, $u$ being
the true anomaly at time $t$.
\begin{table*}
\vspace{-4.2cm}
\hspace{0.55cm}
\centerline{
\scalebox{1}{
\includegraphics{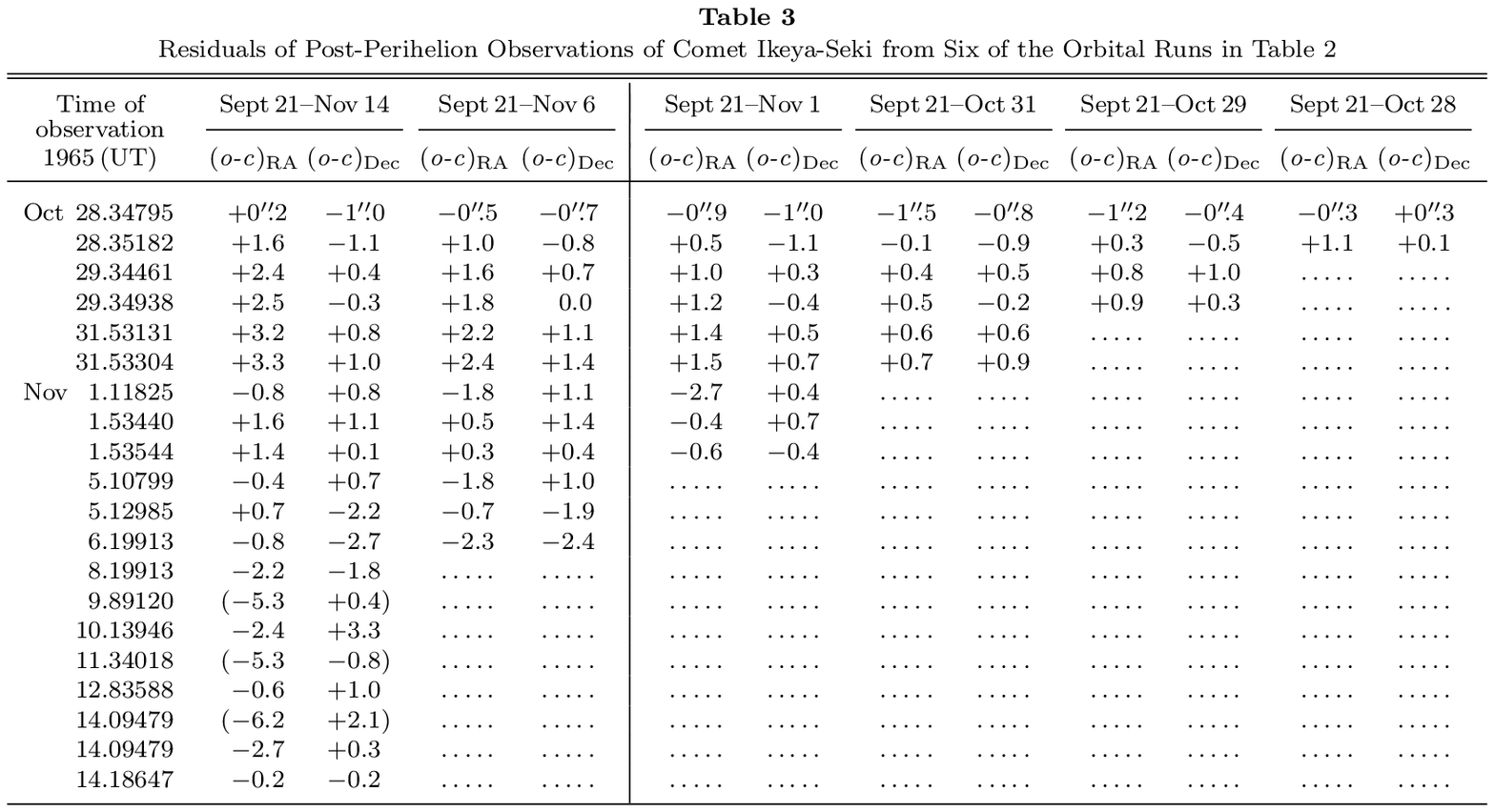}}}
\vspace{-15.4cm}
\end{table*}

Because the magnitude of $\Delta V_{\rm subl}$ is being gradually built up by
the incremental contributions from the nongravitational acceleration in the
course of the apparition, only a fraction of its total is included in the value
of $\Delta P$ in (1), which implies a finite upper limit of the integral in
(3).  This fraction is estimated to equal
\begin{equation}
\Delta V_{\rm subl}^\ast = \Delta V_{\rm subl} \left( \! 1 - \sqrt{\frac{r_{\rm
  frg}}{r_{\rm eff}}} \! \right) \!, \\[-0.05cm] 
\end{equation}
where $r_{\rm eff}$ is an effective heliocentric distance that describes the degree
of incompleteness at which the effect is accounted for.  If approximated by an
average between the companion's first and last{\vspace{-0.06cm}} post-perihelion
observations, used in the orbit determination, in terms of $r^{-\frac{1}{2}}$, it
amounts to
\begin{equation}
r_{\rm eff} = 4 \! \left( \! r_{\rm first}^{-\frac{1}{2}} \!+ r_{\rm last}^{-\frac{1}{2}}
 \! \right)^{\!\!-2} \!\!.  
\end{equation}
With the first astrometric observation of fragment B on 1965 Nov 12.8 and its last
observation on 1966 Jan 14.3~UT, we have \mbox{$r_{\rm first} = 0.87$ AU},
\mbox{$r_{\rm last} = 2.12$ AU},{\nopagebreak} \mbox{$r_{\rm eff} = 1.29$ AU}, and
\begin{equation}
\Delta V_{\rm subl}^\ast = 0.92 \, \Delta V_{\rm subl}. 
\end{equation}

In near-parabolic approximation, a change of $\Delta V$ in the orbital velocity implies a
change of $\Delta P$ in the orbital period that equals generally
\begin{equation}
\Delta P = \kappa P^{\frac{5}{3}} r_{\rm frg}^{-\frac{1}{2}} \Delta V, 
\end{equation}
where
\begin{equation}
\kappa = 3 (\sqrt{2} k \pi^2)^{-\frac{1}{3}}. 
\end{equation}
Inserting (3), (4), and (8) into (7), we obtain for $\Delta P_{\rm subl}$, the part of the
difference between the orbital periods of the two fragments caused by the sublimation,
\begin{equation}
\Delta P_{\rm subl} = 3 \left( \! \frac{\sqrt{2}}{\pi k^2} \! \right)^{\!\frac{2}{3}}
 \!\!\! P^{\frac{5}{3}} \,\frac{\gamma r_0^2}{r_{\rm frg}} \left( \! 1 \!-\!\sqrt{\frac{r_{\rm
 frg}}{r_{\rm eff}}} \right). 
\end{equation}
Inserting next the value of $\gamma$ from (2), \mbox{$P = 880$ yr}, and \mbox{$r_{\rm frg}
= 0.00807$ AU}, one gets
\begin{equation}
\Delta P_{\rm subl} = 373 \; {\rm yr,} 
\end{equation}
a result that is very different from $\Delta P$ in (1).  This means that the
difference between the fragments' orbital periods was {\it not\/} caused by the
outgassing effect {\it alone\/}.

\begin{table*}
\vspace{-4.05cm}
\hspace{0.26cm}
\centerline{
\scalebox{0.97}{
\includegraphics{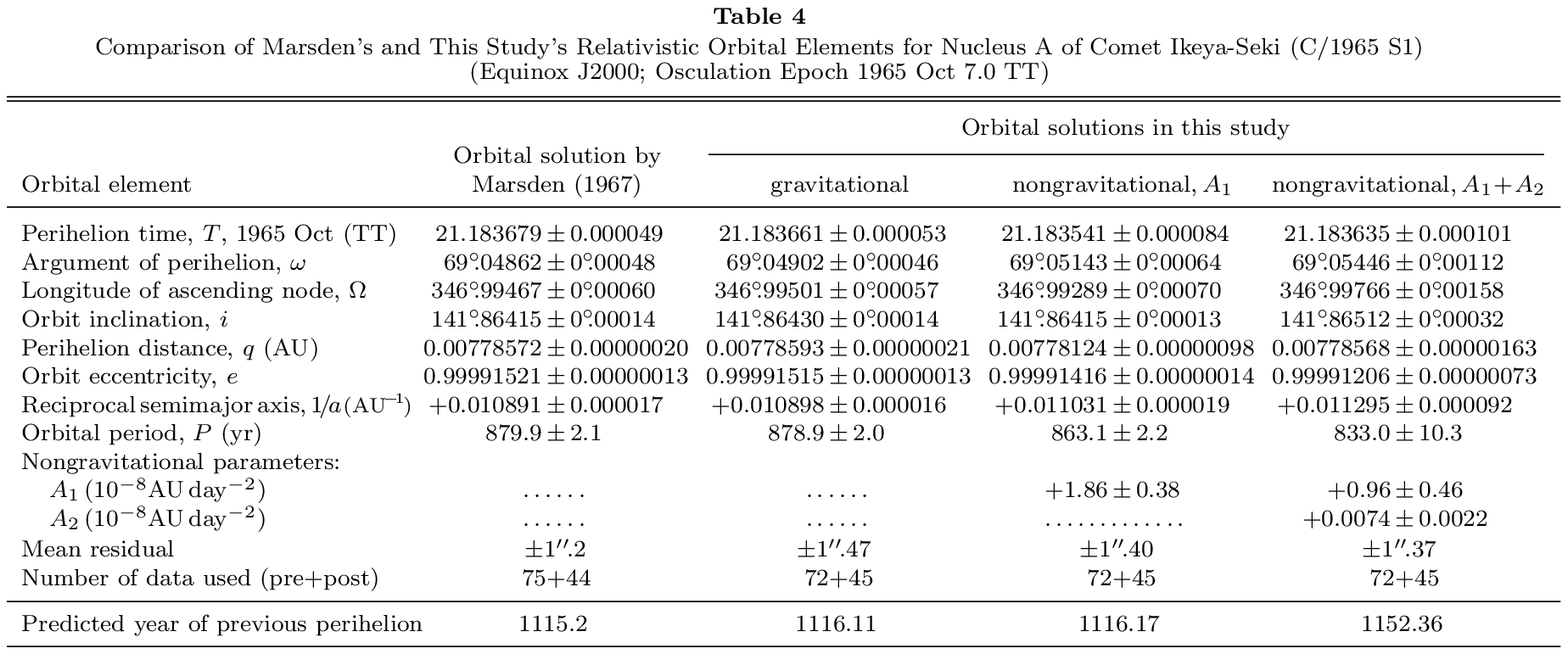}}}
\vspace{-16.5cm}
\end{table*}

The other effect is schematically shown in Figure~2, which depicts an
irregularly shaped nucleus very shortly before and after it split tidally in
two along a section approximately perpendicular to the radius vector.  The
breakup is most likely to occur when the longest dimension of the nucleus is
aligned with the direction to the Sun, as the tidal effect reaches then a maximum.
The nucleus is broken up in one of two possible ways:\ the primary, more
massive fragment A ends up on the sunward side, or on the antisunward side.
At the moment of the breakup, both fragments have the same orbital velocity, but
their centers of mass are at uneven heliocentric distances, whose difference
equals the distance between their centers of mass.  As a result, the two
fragments begin to move in different orbits, with absolutely {\it no momentum
exchange\/} involved.  Differentiating the expression for the orbital velocity
$V$, we have at fragmentation:
\begin{equation}
2 V dV = k^2 \! \left( a^{-2} da - 2 r_{\rm frg}^{-2} \, dr \! \right), 
\end{equation}
where $a$ is the semimajor axis of the pre-split comet, $da$ is the difference
between the semimajor axes of the separated fragments, and $dr$ is the
difference between the heliocentric distances of the fragments' centers of
mass.  Differentiating the relationship between the semimajor axis and the
orbital period, we have
\begin{equation}
da = {\textstyle \frac{2}{3}} (k/2\pi)^{\frac{2}{3}} P^{-\frac{1}{3}} \, dP. 
\end{equation}
Next, we note that $dr$ in (11) equals $\ell$, the difference in the distance from
the Sun between the centers of mass at the time of the breakup, in the sense
fragment~B minus fragment~A and, similarly, $dP$ in (12) equals $\Delta P_{\rm sep}$,
the difference in the fragments' orbital periods triggered by this finite separation
distance $\ell$, also in the sense fragment~B minus fragment~A.  Since \mbox{$dV = 0$}
in (11), we get for $\ell$ by inserting from (12) into (11)
\begin{equation}
\ell = {\textstyle \frac{1}{3}} (2 \pi/k)^{\frac{2}{3}} P^{-\frac{5}{3}} \,
  r_{\rm frg}^2 \, \Delta P_{\rm sep}. 
\end{equation}
This relation means that when the secondary fragment is farther from the Sun at
breakup (\mbox{$\ell > 0$}), it is injected, by this virtue alone, into an orbit
of longer period than the primary fragment as shown in Scenario I in Figure~2,
and vice versa.  

We now propose that the two effects, one driven by the outgassing, $\Delta P_{\rm
subl}$, and given by (9), {\it plus\/} the other, triggered by the radial separation
of the fragments' centers of mass at breakup, $\Delta P_{\rm sep}$, and given by
(13), should, when added together, result in the difference in the orbital period,
$\Delta P$, ascertained from the standard orbital solutions for the fragments and
given by (1):
\begin{equation}
\Delta P_{\rm subl} + \Delta P_{\rm sep} = \Delta P. 
\end{equation}
This condition gives for Ikeya-Seki
\begin{equation}
\Delta P_{\rm sep} = 175 - 373 = -198 \; {\rm yr} 
\end{equation}
and implies that, from (13), the separation between the two fragments'
centers of mass equaled
\begin{equation}
\ell = -8.0 \; {\rm km.} 
\end{equation} 
Being negative, the distance $\ell$ fits Scenario~II in Figure~2:\ the
secondary fragment, B, was at the time of breakup at the sunward end of the parent
nucleus.  It is further apparent from Figure~2 that the center of mass of the
pre-split comet was also sunward of the center of mass of the primary fragment and
that therefore the pre-split comet moved in an orbit of {\it shorter period than the
primary nucleus\/}.  In addition, of course, it is very likely that the primary
fragment was subjected to a higher nongravitational acceleration than the parent
nucleus, which should increase the difference in their orbital periods further.
The purpose of this exercise was to call attention to an effect involving the center
of mass of fragments in events of tidal disruption and allowing one to offer limited
information on the fragments' sizes, which we comment on in Section~3.3. 
%

Using the differentiation was very convenient, but made the results approximate
especially because the differences were relatively large.  For example, an expression
for $\ell$ more accurate than (13) is
\begin{equation}
\ell = {\textstyle \frac{1}{2}} \!\left( \! \frac{2 \pi}{k} \! \right)^{\! \frac{2}{3}}
 \!\!r_{\rm frg}^2 \!\left( \! P_{\rm A}^{-\frac{2}{3}} \!-\! P_{\rm B}^{-\frac{2}{3}}
 \right). 
\end{equation}

\begin{figure}
\vspace{-0.52cm}
\hspace{2.45cm}
\centerline{
\scalebox{0.865}{
\includegraphics{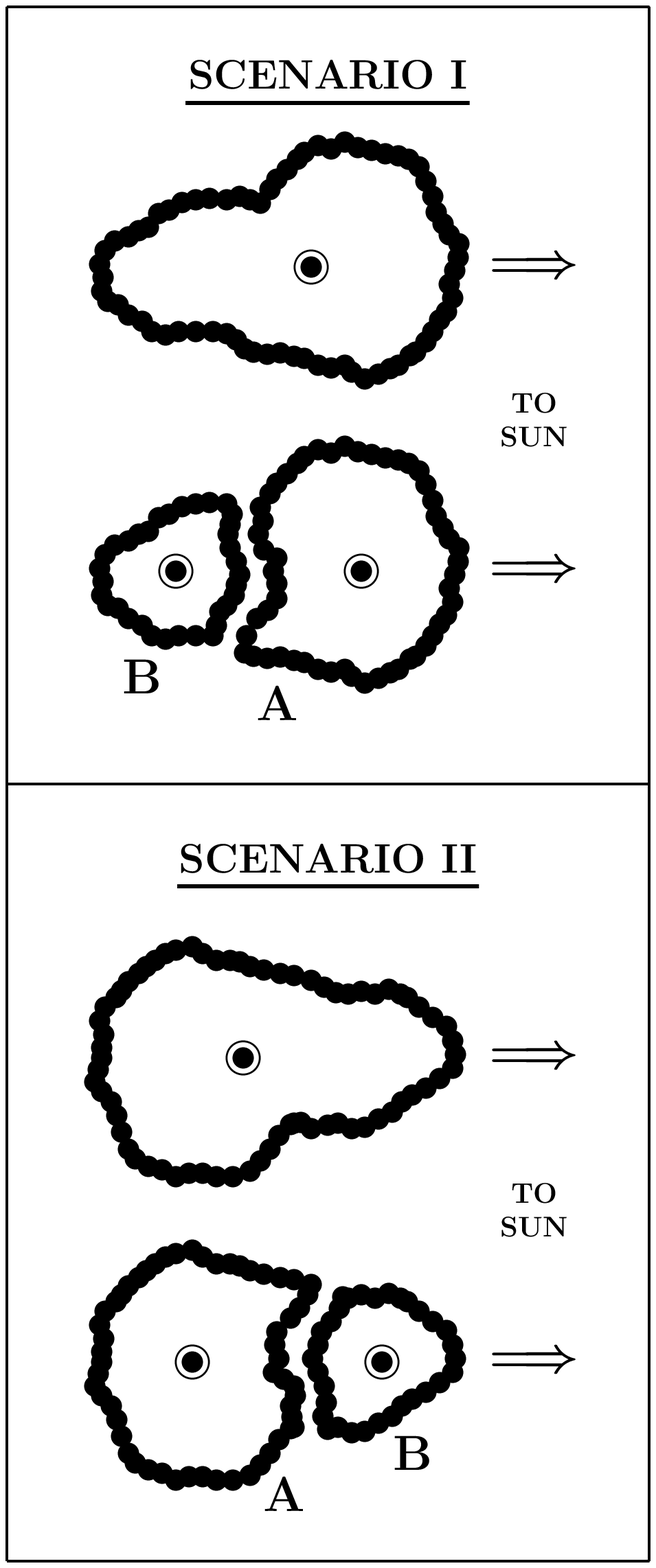}}}
\vspace{-5.33cm}
\caption{Elongated cometary nucleus of a sungrazing comet shortly before and after
breaking up tidally into two uneven fragments at perihelion.  Turned to the Sun (to
the right) at the time of breakup is the larger end of the nucleus, to become the
primary fragment A, in Scenario~I, but the smaller end, to become the secondary
fragment B, in Scenario~II.  The large circled dots are the positions of the centers
of mass of the parent (or pre-split) nucleus, at the top of either panel, and of the
fragments A and B at the bottom.  Accordingly, the orientation relative to the Sun
alone assures that fragment~A ends up in an orbit of a shorter period and fragment~B
in an orbit of a longer period than was the parent in Scenario~I, whereas the opposite
is true in Scenario~II.{\vspace{0.7cm}}}
\end{figure}
%
%

As a word of caution, we note that reports of additional, temporary fragments
may suggest other possible explanations for the short orbital period of the
pre-split comet Ikeya-Seki, not necessarily negating a momentum-exchange effect.
Hirayama \& Moriyama (1965) reported that only minutes after perihelion they
observed coronagraphically the comets's head splitting into three components, one of
them being much brighter than the other two.  We believe that this event was not the
one giving birth to fragment~B.  Independently, Pohn (1965) remarked on a potential
third nucleus on November~4.5~UT, while Andrews (1965) noted that on November~6.1~UT
the secondary nucleus was possibly a triple complex.  It is beyond the scope of this
paper to judge the potential influence of these transient phenomena on the orbit of
the pre-split nucleus, and we limit ourselves to merely mentioning them for the
benefit of the reader.

\subsection{The Ramifications} 
In Section 3.1 we argued that before splitting at its 1965 perihelion, comet Ikeya-Seki
was moving in an orbit of a period that was shorter than that of either fragment.  The
comet's previous return to perihelion was thus found to have occurred decades after the
appearance of the Great Comet of 1106, most probably near the year 1139, and offering
strong evidence for ruling out the 1106 comet as Ikeya-Seki's parent and previous
appearance.  And because there is no doubt that the Great September Comet of 1882 and
Ikeya-Seki were on their approach to their previous perihelion a single object, our
result rules out an association between the comets of 1106 and 1882 as well.

In Section 3.2 we showed that it is dynamically feasible for a sungrazer, prior to its
breaking up tidally in close proximity of perihelion, to move in an orbit with a shorter
period than that of either of its two nuclear fragments, thereby corroborating the results
of Section~3.1.  This problem deserves further attention in the future.

The probable time of the previous perihelion return of Ikeya-Seki around the year 1139 opens
two questions:\ one is the role of the Great Comet of 1106 in the history of the Kreutz
system, the other is the whereabouts of the parent to Ikeya-Seki and the 1882 comet.  The
first question was addressed in part by Sekanina \& Chodas (2008), when they concluded that
the orbit for the Great March Comet of 1843 derived from the best available astrometric
observations did not rule out its previous return to perihelion as the Great Comet of 1106.
 
The two giant sungrazers of the 19th century, the Great March Comet of 1843 and the Great
September Comet of 1882, moving in orbits similar but not nearly identical and representing
two major populations of the Kreutz system, implied the existence of two major sungrazers
in the first half of the 12th century, given that their barycentric orbital periods were
close to 750~years.\footnote{Unfortunate was the lingering belief
that the orbital period of the 1843 sungrazer was much shorter, near 500~yr, predicated
on the earlier results by Kreutz (1901) and by Hubbard (1852) before him, while still
others (e.g., Nicolai 1843) noted that the choice of the orbital period made almost no
difference in the representation of observations.  The modern revision, with a period
longer than 700~yr deemed feasible, had been unavailable until fairly recently (Sekanina
\& Chodas 2008).} Yet, no candidate other than the 1106~comet has ever been seriously
proposed.  The solution to this problem was usually circumvented by pointing out that
a sungrazer arriving at perihelion between late May and early August approaches from
behind the Sun and recedes in that same direction, thus being missed unless it is
bright enough to be seen in broad daylight.  Closer inspection of relevant circumstances
shows that this is not necessarily so, but it is true that such a sungrazer could at
best be detected as an unimpressive naked-eye object whose record in historical annals
would be far less notable than that of the 1106~comet.

We did not undertake a concerted effort in search for the ``missing'' sungrazer, but
merely conducted a cursory perusal of Ho's (1962) catalogue of the ancient and medieval
observations of comets in the Far East.  Under the entry 403 we came across brief
descriptions of two obviously different objects from late August and September 1138,
one Japanese, the other Chinese.  The Japanese object was of no interest, but the text
on the Chinese comet caught our attention.  The primary source was the 29th volume of
{\it Sung Shih\/}, a historical account of the Sung Dynasty compiled from older sources
in 1345, and the secondary source was the 14th volume of {\it Hs\"{u} Thung Chien Kang
Mu\/}.  The relevant record for 1138 says: \\[0.23cm]
{\hspace*{0.8cm}}{\parbox{7cm}{\large \it On September~3 a broom{\vspace{-0.08cm}} star
 (hui) was observed in the east.  It{\vspace{-0.08cm}} went out of sight on
 September~29.\/}}\\[0.23cm]
The used terminology (hui) suggests that the comet had a tail (even though its length
was not given) and, while not explicitly stated, the comet must have been seen in
the morning sky, before sunrise.

This Chinese comet of 1138 is neither among the Kreutz candidates proposed by Hasegawa
\& Nakano (2001) nor on England's (2002) list of possible early Kreutz sungrazers.
Hasegawa's (1980) catalogue contains this object under the entry number 636 and notes
that it also was included in the compilations by Pingr\'e (1783) and by Kanda (1935).
The circumstances at the time of appearance of this possible major sungrazer are
examined in some detail in Section~4.

We call attention to a case of remarkable coincidence, as Marsden's (1967) integration
of Kreutz's (1888, 1891) nonrelativistic orbit of the brightest fragment B (No.\,2
in Kreutz's notation) of the Great Comet of 1882 back in time gave the previous
perihelion passage in April 1138, only months before arrival of the proposed Kreutz
candidate.\footnote{The relativistic orbit of fragment B of the 1882 comet was
computed by Hufnagel (1919), who derived a period shorter than the period of Kreutz's
nonrelativistic orbit by more than 10~yr.  This large difference makes Hufnagel's result,
which according to Marsden (1967) implies November 1149 for the time of the previous
perihelion passage of the 1882 comet, rather suspect.}  The center of mass of fragment~B
was apparently slightly sunward of the center of mass of the 1882 parent comet at
the time of its perihelion fragmentation, which together with a contribution from
the differential outgassing-driven nongravitational acceleration could provide a
near-zero effect in the orbital period between the parent and fragment~B.  If so,
the times of the previous perihelion for nucleus fragment~B and the parent comet
should be about the same and the amazing agreement of Marsden's prediction with
the timing of the 1138 comet may not be fortuitous.

The data on the fragmentation of Ikeya-Seki offer limited information on the size
of the comet's nucleus.  In a first approximation, the distance of 8~km between the
centers of mass of the two fragments in (16) equals the semidiameter of the long
axis of the pre-split nucleus.  If the axial ratios are, for example, 1:1:2, the
effective diameter of the pre-split nucleus is 10~km.  If the comet of 1138 is
indeed the previous appearance of Ikeya-Seki, the difference between the orbital
periods of nuclear fragment~A (our gravitational or $A_1$ nongravitational solution)
and the pre-split comet is about 22~yr.  Interpreted in its entirety as an effect
of separation at fragmentation (rather than a sublimation effect), it offers a lower
limit to the fragment~A-to-fragment~B mass ratio equaling 8, equivalent to a lower
limit of 2 for the fragment~A-to-fragment~B size ratio.  The effective diameters
of fragments~A and B then come out to be about 9.6~km and 4.8~km, respectively.    

\section{The Chinese Comet of 1138}  
Even though the description of the Chinese comet of 1138 in Ho's (1962) catalogue
is brief, important insights into the issue of the object's association with the Kreutz
system can be gained.  We approximate the angular elements and perihelion distance
of the orbit of the 1138 comet by the values originally derived for the presumed
1106 apparition of the Great September Comet of 1882 (Sekanina \& Chodas 2002),
assume the orbit to be a parabola, and address the membership issue in some detail
in the following.

\subsection{Observing Site}  
To examine the observing conditions, we need to have an idea on the geographic
location of the probable observing site.  The (southern) Sung Dynasty's capital was
Jiangning (today a southern district of Nanjing, the capital of Jiangsu Province)
until 1138 and Lin'an (nowadays a district of Hangzhou,{\vspace{-0.025cm}} the capital
of Zhejiang Province) from 1138 on.\footnote{The northern territories of the Sung
Dynasty were lost to the Jin Dynasty in 1127.}  The distance between the two cities,
where information on the comet's sightings was presumably coming from, is only about
200~km and their locations can approximately be depicted by geographic longitude
120$^\circ$ east of Greenwich, latitude 31$^\circ$ north of the equator, and 20~meters
above sea level.  The observing window up to sunrise on September~3, between 4:00 and
5:40 local time, is equivalent to Sept~2.83--2.90~UT.

\subsection{Limiting Magnitude}  
Before addressing the issue of light curve of the 1138 comet, we examined the
observing conditions during the critical period of time in the morning of
September~3, the day of the first recorded sighting, and September~29, the
day of disappearance.  The conditions for detecting, with the unaided eye,
a stellar object of apparent visual magnitude $H_{\rm app}$ at a given location
of the sky from a given observing site are described by a limiting magnitude,
$H_{\rm lim}$, determined from an algorithm developed by Schaefer (1993, 1998)
as a function of (i)~the object's solar and lunar elongations; (ii)~the object's,
Sun's, and Moon's elevations above the local horizon, and the Moon's phase; and
(iii)~the atmospheric and other conditions at the observing site.  The algorithm
also allows for seasonal and long-term effects.  In a recent paper, \mbox{Sekanina}
(2022; referred to hereafter as Paper~2) employed the Schaefer algorithm, strictly
valid for stellar objects, to comets.  By introducing and applying a {\it visibility
index\/} $\Im$,
\begin{equation}
\Im = H_{\rm lim} - H_{\rm app},
\end{equation}
he found meaningful results when tested on daylight magnitude estimates of the
sungrazing comet Ikeya-Seki in October 1965.  The more positive the value of
$\Im$ is, the better prospect there is to detect, with the naked eye, a comet
of magnitude $H_{\rm app}$ at a location of the sky at which the limiting
magnitude is $H_{\rm lim}$.  A value of $\Im$ near zero, within a few tenths
of a magnitude, means a marginal chance of seeing the object or the point of
disappearance.

Moonlight interfered with observation of the 1138 comet on both September~3 and
September~29.  The new moon was on September~6.29~UT and October~5.96~UT.  We adopted
an obvious view that because disappearing 26~days after the first sighting the comet
was during September gradually fading and therefore receding from the Sun, having
passed perihelion before September.

An estimate of the perihelion time was a key parameter for investigating the comet
as a potential Kreutz sungrazer.  A crude guess was obtained by straightforwardly
comparing the 1138 comet to comet Pereyra (C/1963~R1), a definite member of the Kreutz
system discovered on 1963~September~14 and passing perihelion 21~days earlier.  By
simply shifting the dates, the perihelion passage of the 1138 comet was predicted to
have taken place on August~13.  In spite of the vastly different circumstances at
the two comets' arrival times more than 800~years apart, this guess eventually
turned out to be not too far off the mark, to the extent we can judge.

\begin{table*}[t]
\vspace{-4.2cm}
\hspace{0.55cm}
\centerline{
\scalebox{1}{
\includegraphics{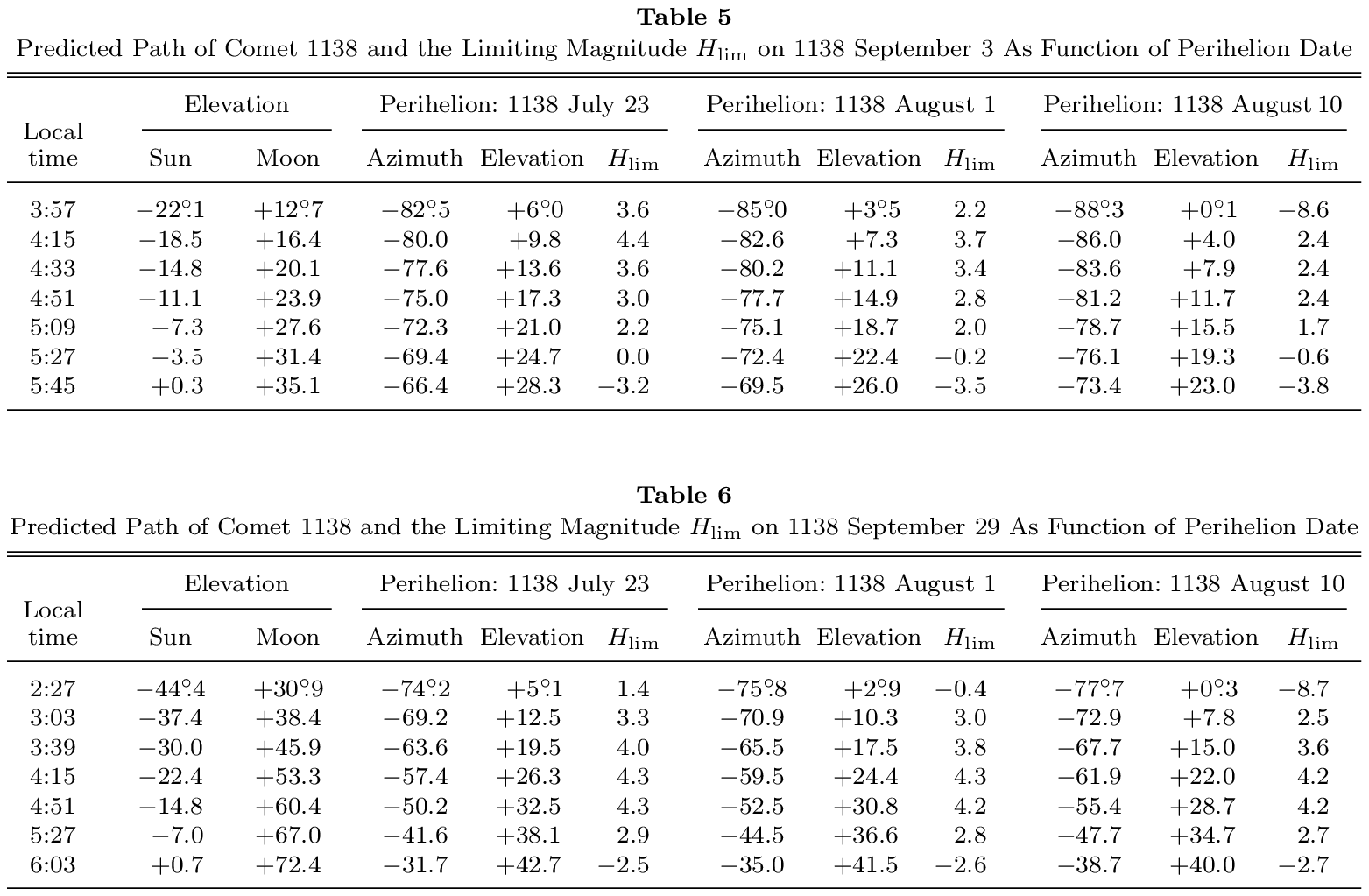}}}
\vspace{-14.2cm}
\end{table*}

Based on this guess, we selected three candidate dates --- July~23, August~1, and
August~10 --- for the 1138 perihelion time.  For the observing site defined in
Section~4.1 and the dates of 1138~September~3 and 29 we computed, for a number of
times (between the comet's rise above the eastern horizon and sunrise) at a step
of 18~minutes, the horizontal coordinates (the azimuth reckoned from the south
clockwise) of the Sun, the Moon, and the comet on each of the three perihelion-time
assumptions, and derived the limiting magnitudes as a function of time.  On September~3
the comet was found to have risen between 3:43 and 3:57 local time (19:43 to 19:57~UT
on September~2) as the perihelion time advanced from July~23 to August~10, while sunrise
occurred on 5:43 local time (21:43~UT on September~2).  This means that the comet could
be observed for quite a bit less than two hours in early September.  On September~29
the comet could be seen over three hours, so that the more restrictive viewing conditions
on September~3 offered tighter constraints on the perihelion time.

\begin{table}[b]
\vspace{-3.6cm}
\hspace{5.2cm}
\centerline{
\scalebox{1}{
\includegraphics{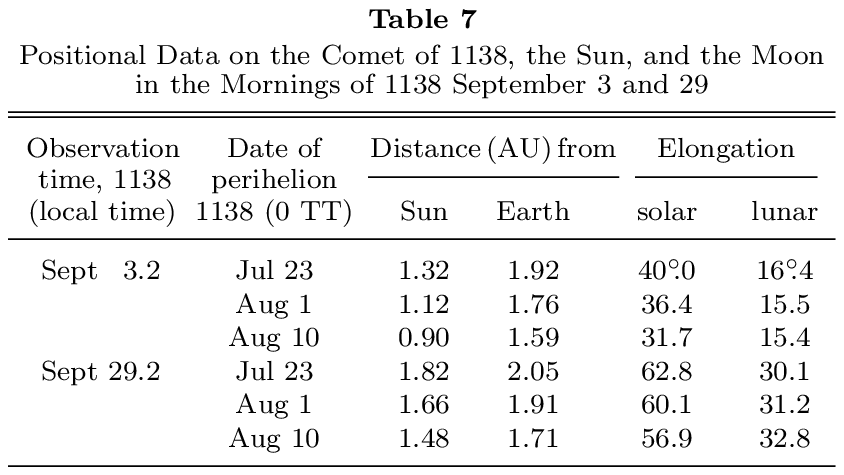}}}
\vspace{-20.83cm}
\end{table}

The quantities that varied during observation are presented in Table~5 for the morning
hours of September~3 and in Table~6 for September~29, while the other positional data on the
comet, the Sun, and the Moon on September~3 and 29 as a function of the comet's assumed
perihelion time are listed in Table~7.  The comet's azimuths in Table~5 show that the
condition of the comet appearing on September~3 in the east was closely satisfied; at
elevation 12$^\circ$, for example, to 9$^\circ$ if perihelion occurred on August~10, to
better than 12$^\circ$ if on August~1, and to better than 13$^\circ$ if on July~23.

The curves of the limiting magnitude in Tables~5 and 6 are U-shaped.  On September~3 the
bright ends reflect, respectively, the high atmospheric extinction as the comet rose above
the horizon before 4:00, and the dawn effect before sunrise a minute or so before 5:45.
Only before about 4:30 is the limiting magnitude on either date affected significantly
by moonlight.  The comet should have been the fainter --- but also the sooner above the
horizon --- the earlier it passed through perihelion.  The brightness effect favors a
later perihelion passage, the horizon-crossing effect an earlier one.  Perihelion on
August~10 would require the comet to have been of magnitude~1--2 or brighter on
September~3.  An important conclusion from Table~6 is that the comet should have been
close to magnitude 4.2--4.3 on September~29 regardless of the perihelion time, given
its disappearance on that day.  Combined with a brightness estimate on September~3,
based on the limiting magnitudes in Table~5, the assumption of August~10 perihelion
implies that the comet should have faded steeply with heliocentric distance $r$, more
steeply in fact than $r^{-4}$, otherwise it could be seen on September~3 for only
a short period of time at very low elevations.  These arguments suggest that the
comet's perihelion passage occurred with high probability {\it before\/} August~10.

\begin{table*}[t]
\vspace{-4.18cm}
\hspace{0.8cm}
\centerline{
\scalebox{1}{
\includegraphics{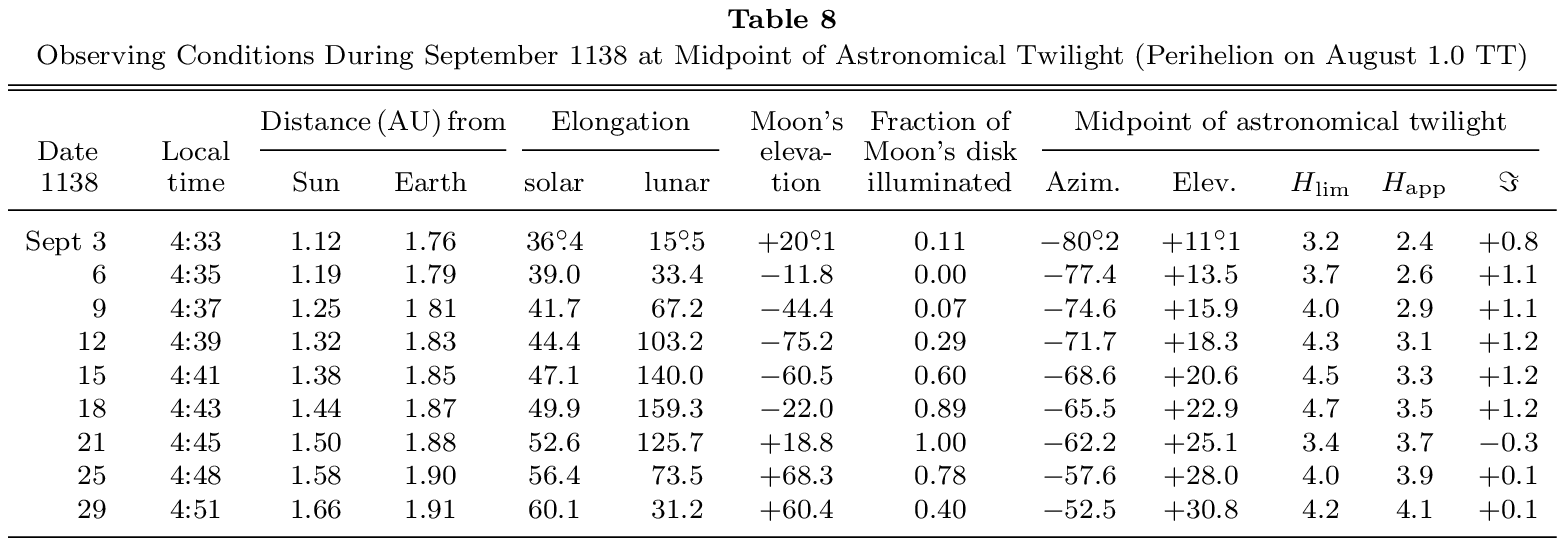}}}
\vspace{-19.3cm}
\end{table*}

We further note that a sungrazer of apparent magnitude 4.2--4.3 six weeks after
perihelion --- the scenario with perihelion on July 23 --- implies an intrinsically
very bright comet.  Such a comet should have been sighted sooner at a smaller
solar elongation than 40$^\circ$ (Table~7).  It is improbable that perihelion
took place as early as July~23.  While the exact date of perihelion remains
unknown, we find it most likely that the comet passed perihelion within several
days of August~1.  Below we adopt the Chinese comet of 1138 as the presumed
parent to the Great September Comet of 1882 and comet Ikeya-Seki, and August~1
as the date of its perihelion passage.

In Table 8 we present an ephemeris of the 1138 comet, assuming perihelion on
August~1.0~TT.  The ephemeris is short in terms of the number of entries, which
are the midpoints of astronomical twilight (with the Sun $\sim$15$^\circ$ below
the horizon), but it is extended in terms of the number of listed quantities.
The table allows one to acknowledge that, of the nine listed dates, moonlight was
disturbing on four.  The interference on September~21, just hours after the
full moon was so strong that the comet may have been lost only to be marginally
detected again before its ultimate disappearance, to which moonlight also
contributed its share.  The final two columns of Table~8 will be commented on
in the next section.

\subsection{Light Curve of the Comet of 1138}  
In the absence of any direct information, we turn for help to Paper~2, in which the
problem of sungrazers' light curves was discussed in broad terms.  A comet's
visual brightness was investigated using the usual power-law formula, which on a
magnitude scale is
\begin{equation}
H_{\rm app}(r,\Delta;H_0,n) = H_0 + 2.5 n \log r + 5 \log \Delta, 
\end{equation}
where $r = r(t)$ is the heliocentric distance and $\Delta = \Delta(t)$ the geocentric
distance (both in AU) of the comet at the time of observation, $t$; $H_0$ is the absolute
magnitude (normalized to \mbox{$r = \Delta = 1$ AU}); and $n$ is a photometric exponent
equal to the power of heliocentric distance with which the brightness, normalized to
a unit geocentric distance, inversely varies, $r^{-n}$.  Since the light curves of major
Kreutz sungrazers are known to be, in general, asymmetric relative to perihelion, the
preperihelion parameters, $H_0^-$, $n^-$, and the post-perihelion parameters, $H_0^+$,
$n^+$ were examined in Paper~2 separately.  The preperihelion photometric exponents of
all major Kreutz sungrazers were assumed to be constant\footnote{This approximate rule
does under no circumstances apply to the SOHO Kreutz sungrazers.} and equal to \mbox{$n^-
\! = 4$}, while the post-perihelion photometric exponents were deemed a function of
perihelion fragmentation.  For a sungrazer with the nucleus breaking up at perihelion
into $N_{\rm frg}$ fragments, the post-perihelion value of the exponent was adopted to
vary as
\begin{equation}
n^+ = 4.4 - 0.2 N_{\rm frg},  
\end{equation}
where \mbox{$N_{\rm frg} = 1$} when the comet does not fragment.  This equation
expresses the well-known fact that major sungrazers fade rather rapidly, unless
they break into persisting fragments at perihelion.  This empirical relationship
is based on an extensive amount of data on comet Ikeya-Seki, a modest set of data
on the Great September Comet of 1882, and fragmentary data on the Great March
Comet of 1843, collected in Paper~2.  The preperihelion absolute
magnitudes $H_0^-$ were estimated at 5.9 for Ikeya-Seki, 3.4 for the 1882 sungrazer,
and 3.5 for the 1843 sungrazer.  The post-perihelion absolute magnitude $H_0^+$ was
linked to the preperihelion value by requiring perihelion continuity,
\begin{equation}
H_0^- + 2.5 n^- \log q + 5 \log \Delta_q = H_0^+ + 2.5 n^+ \log q + 5 \log
 \Delta_q, 
\end{equation}
where $q$ is the perihelion distance and $\Delta_q$ the geocentric distance of the
comet at perihelion (both in AU).  With \mbox{$n^- \! = 4$} and $n^+$ from (20),
the post-perihelion absolute magnitude was found to equal
\begin{equation}
H_0^+ = H_0^- - (1 \!-\! {\textstyle \frac{1}{2}} N_{\rm frg}) \log q. 
\end{equation}
The post-perihelion photometric exponent amounted to 4.0 for comet Ikeya-Seki
(\mbox{$N_{\rm frg} \!= 2$}), 3.3 for the 1882 comet (assuming \mbox{$N_{\rm frg}
\! = 5$ to 6}), and 4.2 for the 1843 comet (\mbox{$N_{\rm frg} = 1$}).  From (22)
the post-perihelion absolute magnitude came out to be 5.9 for Ikeya-Seki, $-$0.2
for the 1882 comet, and 4.6 for the 1843 comet.  The parent sungrazers were
arbitrarily assigned in Paper~2 the preperihelion absolute magnitudes 0.6~mag
brighter than their primary fragments (the 1843 and 1882 comets).  We now recognize
the comet of 1138 as one of the two parents.

The number of fragments into which the 1138 comet broke up is of course unknown,
but it could not be smaller than two --- the 1882 comet and Ikeya-Seki.  Potentially,
the comet may also be a parent to Strom's (2002) sun-comet of 1792 as well as to the
probable sungrazer X/1702~D1 (Kreutz 1901; Marsden 1967); that would make the number
of the persistent fragments equal four.  We can only speculate whether another major
fragment might arrive in the coming decades.  If we accept that the nucleus of the
1138 comet split at perihelion into the four pieces, then its light-curve parameters
\mbox{$n^+ \! = 3.6$} and \mbox{$H_0^+ = 0.7$} (assuming that \mbox{$H_0^- = 2.8$}
or 0.6~mag{\vspace{-0.055cm}} brighter than $H_0^-$ for the 1882 comet).  The
post-perihelion light curve then follows the formula
\begin{equation}
H_{\rm app} = 0.7 + 9 \log r + 5 \log \Delta. 
\end{equation}
The ephemeris in Table 8 provides the apparent magnitudes from this formula in the
penultimate column.  The table's last column offers the values for the visibility
index of the comet at the midpoint of astronomical twilight.  One can see that
for more than two weeks after September~3 the computed magnitude was compatible with
the comet's visibility with the unaided eye.  The situation worsened suddenly with
the arrival of the full moon on September~21, when the predicted visual perception
of the comet apparently indicated it dropped below the detection threshold.  Toward
the end of the month the interference by moonlight subsided a little, but the comet
grew gradually fainter until eventually vanishing, as the limiting magnitude after
September~29 remained essentially constant.

The tabulated values of the visibility index $\Im$ were computed assuming constant
atmospheric conditions, an assumption that is unavoidable but in practice never
satisfied.  For example, the observing conditions deteriorate as air humidity
climbs; an increase from 50~percent to 80~percent can cause the visibility limit
can shoot up by 0.5~mag or more.  A few days with higher humidity beginning on
September~29 may have been all that was needed for this date to be recorded as the
point of disappearance.\footnote{Given that the full moon{\vspace{-0.02cm}} occurred
on Aug~22.5 and Sept~20.9 Chinese time, a scenario that is perfectly compatible with
the published record is rather heretic:\ if Ho's dates were off (late) by eight
days, the first sighting would have been on Aug 26, about four days after the full
moon, when the observing conditions greatly improved relative the previous days (as
Table~8 shows a month later) and the comet would have disappeared on Sept 21, just
hours after the next full moon.  Although Ho (1962) provides no information on
his methods of conversion of the Chinese calendar to the Julian/Gregorian calendar,
we are certain that, barring an inexplicable blunder, the chance of a converted
date in the 12th century being off by eight days is nil.}   

Our last item on the light curve proposed for the 1138 comet is a plot of the apparent
magnitude against the solar elongation when the brightness variation with heliocentric
distance follows the law (23) after{\vspace{-0.06cm}} perihelion and the corresponding
law (\mbox{$H_0^- \!= 2.8$}, \mbox{$n^- \!= 4.0$}) before perihelion.  Displayed in
Figure~3, the suggested light curve is compared with the naked-eye limiting magnitude
in both broad daylight and twilight.  The twilight curve is constructed for the most
favorable case of the comet and the Sun sharing the same azimuth, i.e., when the
comet's solar elongation equals the difference between the elevations of the comet
and the Sun.

\begin{figure}[t]
\vspace{-5.44cm}
\hspace{2.04cm}
\centerline{
\scalebox{0.66}{
\includegraphics{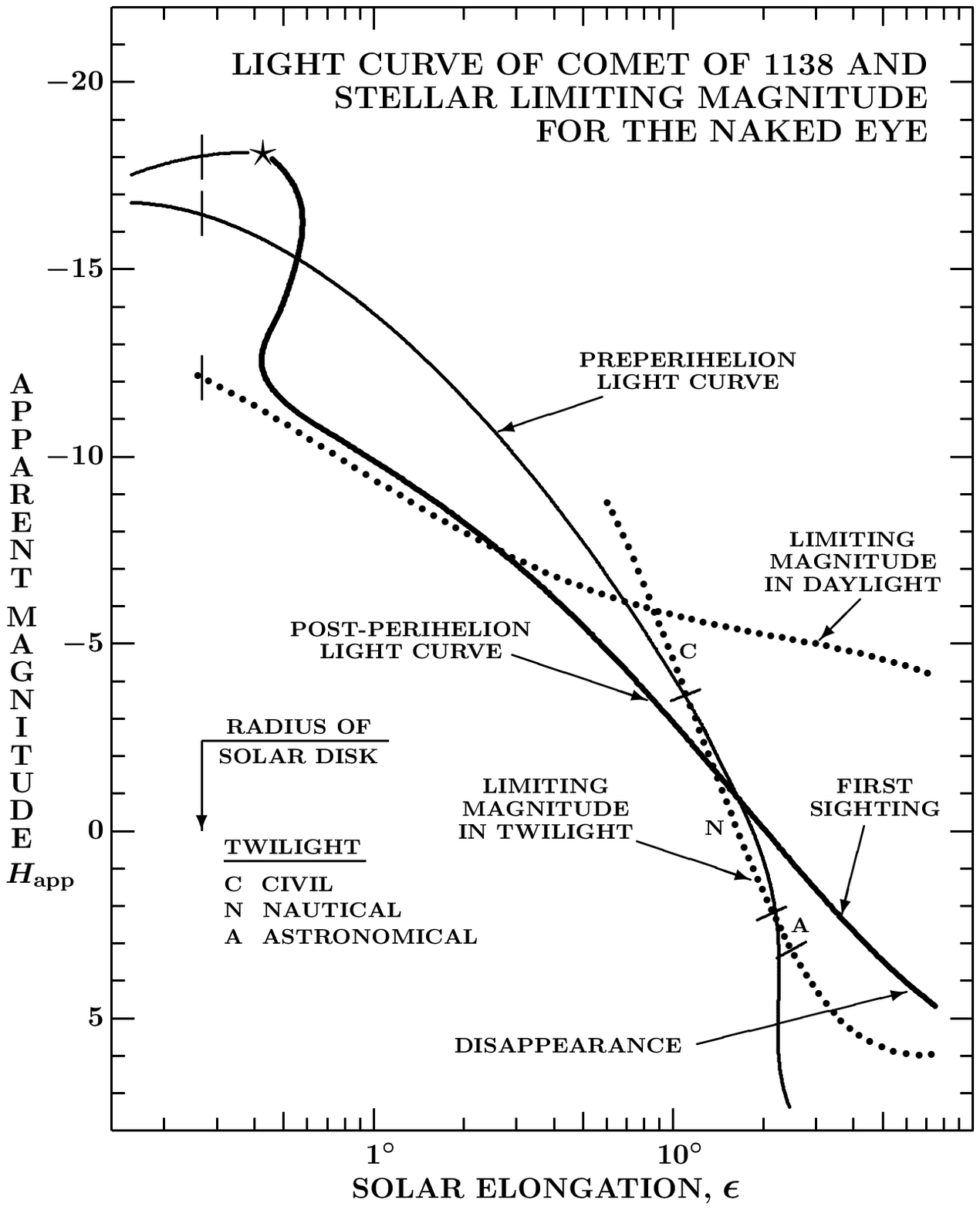}}}
\vspace{-3.75cm}
\caption{The proposed light curve for the Chinese comet of 1138, plotted as a function
of solar elongation.  The thin line refers to the preperihelion branch, the thick line
to the post-perihelion branch.  Perihelion is marked by a star.  The points of first
sighting on September~3 and the disappearance on September~29 are also indicated.  The
extent of the solar disk is depicted by short vertical lines.  Just before perihelion
the comet was passing almost centrally across the Sun.  The comet's light curve is
compared with the limiting magnitude of a stellar object for the unaided eye according
to Schaefer's (1993, 1998) algorithm.  One limiting-magnitude curve applies to the
conditions in broad daylight, the other under twilight's most favorable conditions when the
comet and the Sun have the same azimuth (the solar elongation equaling the difference
between the elevations of the comet and the Sun).  Boundaries between astronomical,
nautical, and civil twilight are shown.  No moonlight effects are plotted.{\vspace{0.65cm}}}
\end{figure}

The plot of the light curve in the figure looks rather bizarre, in part because the comet
stayed extremely close to the Sun after perihelion for so long.  While 12~hours before
perihelion the comet was 3$^\circ\!$.5 from the Sun, the separation was less than 1$^\circ$
12~hours after perihelion!  At solar elongations greater than 20$^\circ$ the comet is
predicted to have been significantly fainter before perihelion and to have attained
apparent magnitude 6 only when 22$^\circ$ from the Sun.  It thus must have been entirely
out of reach of naked-eye detection.  At solar elongations smaller than 15$^\circ$, the
comet is predicted to have been brighter before perihelion and at elongations smaller
than 6$^\circ$ it should have been brighter than the naked-eye limiting magnitude.  At
that time the comet had just a little over 24~hours to get to perihelion.  In about six
or so hours after perihelion the comet's brightness dropped to a level close to the limit
and stayed near or below that level for at least two weeks, until the comet was some
20$^\circ$ from the Sun.  It took only two more weeks (with the Moon above the horizon
in the meantime) before the comet was discovered.
  
\subsection{Tail of the Comet of 1138}  
Although the historical record quoted by Ho (1962) referred to the comet as a {\it broom
star\/} --- a term used for objects with a tail --- no information was provided on the
tail's length.  This should not be surprising if the observed tail was unimpressive.  An
independent investigation of the tails of Kreutz sungrazers (Sekanina, in preparation)
shows that some five weeks after perihelion the tails of the 1882 comet and Ikeya-Seki
were dominated by microscopic dust subjected to solar radiation accelerations not
exceeding 0.6 the solar gravitational acceleration.  On September~3 the tail of the
1138 comet should on this condition have been less than 8$^\circ$ long.  Theoretically,
the length could have grown up to nearly 12$^\circ$ by September~29, but the data on
other sungrazers suggest that two months after perihelion only a small fraction of the
tail's computed length could be seen with the naked eye.  Because of the geometry, the
tail pointed away from the earth.  Its end on September~3 is calculated to have been
at a geocentric distance of 2.5~AU and it must have been rather faint.  The unfavorable
geometry (next to moonlight) also explains the short visibility period of the comet.

Kreutz sungrazers become spectacular objects over short periods of time thanks to their
early post-perihelion tails (from several days to two weeks after perihelion for comet
Ikeya-Seki, for example).  The tails are sometimes 10 to 100 times brighter than the
head.  To become a spectacle, the sungrazer does not have to be a giant, but needs to
arrive at the ``right'' time of the year.  The 1138 comet could serve as an example of
a sungrazer that arrived at the ``wrong'' time of the year.  Its path in the sky appears
to have stayed within one degree of the Sun in the first 12~hours after perihelion, when
its predicted light curve in Figure~3 is shaped like a ``spiral to obscurity,'' losing
8~magnitudes.  If the comet had arrived seven weeks later, it could have been comparable
in brightness to the Great September Comet of 1882, if 80~days later, it could have been
more prominent than Ikeya-Seki.  And if it arrived six months earlier or later, it could
successfully compete with the appearance of the Great Comet of 1106.

On the other hand, the record of the 1138 comet in historical sources and our sungrazer
diagnosis suggest that intrinsically bright Kreutz sungrazers reaching perihelion in
early August were not necessarily missed completely --- when undetected in broad daylight
--- by pre-telescopic northern-hemisphere observers, notwithstanding the numerous
instances of claim to the contrary in the literature.  However, like the comet of 1138,
they were likely to have appeared to the naked eye as unimpressive {\it broom stars\/}
in the morning sky and followed for a few weeks.  We suspect that the same argument
applies to other bright Kreutz sungrazers that reach perihelion between late May and
late July.

\section{Perihelion Asymmetry of Production Curve and Nongravitational Perturbations
 of A Comet's Orbital Motion} 
The outgassing-driven nongravitational perturbations of the sungrazers' orbital
motions, explored in Section~3, and the apparent asymmetry of their light curves,
examined in Section~4, are of critical importance to our further investigation
of the Kreutz system, because there appears to be a degree of correlation between
the two parametric functions.  This issue was brought to the forefront of attention
following the 1986 return of 1P/Halley, as Rickman (1986) and Rickman et al.\ (1987)
argued that the comet's water production curve showed a strong asymmetry with respect
to perihelion, implying that the contributions from the radial component of the
sublimation-driven nongravitational force (parameter $A_1$) integrated over the
revolution about the Sun did not cancel out, basically a variation of Bessel's (1836)
concept.  In line with this argument, Festou et al.\ (1990) and Rickman et al.\ (1992)
proposed a hypothesis according to which the water-production asymmetry essentially
determines the magnitude of the nongravitational perturbations impacting the orbital
period, relegating effects of sublimation lag, expressed by the transverse component
(parameter $A_2$), to the status of a minor factor.  While it is inadmissible to use
the production-rate asymmetry for determining a magnitude of the outgassing effect
on the orbital period (Sekanina 1992, 1993), the two quantities are statistically
correlated.  And even though Festou et al.'s dataset consisted exclusively of
short-period comets, the correlation appears to be valid generally.

In relation to the Kreutz system, this result implies that, as a matter of rule, the
orbital motions of the sungrazers that are brighter before perihelion should be slightly
accelerated, whereas the motions of the sungrazers brighter after perihelion should
be decelerated.  This dependence is further correlated with tidal fragmentation near
perihelion, as profusely fragmenting sungrazers have a tendency after perihelion to
fade less steeply than those subjected to insignificant or no disruption (cf.\ Paper~2).  

It is unfortunate that the symmetric sublimation law still is a universal tool in
Marsden et al.'s (1973) orbit-determination software package available to numerous
users at present, almost 50~years after it was experimentally incorporated into
the nongravitational model that at the time was still in the process of continuing
refinement.  While it is possible to vary the law's parameters (such as its steepness,
for example), it is not possible to use different functional relations before and after
perihelion.  Similarly, the law's peak cannot be moved away from perihelion.  There is
a version of the Style~II formalism with an asymmetric sublimation law in existence,
conceived (Sekanina 1988) and implemented (Yeomans \& Chodas 1989) to accommodate the
nongravitational motions of the short-period comets whose gas production (and activity
in general) peaks either profoundly before perihelion (such as 2P/Encke in the early
times, 3D/Biela) or after perihelion (such as 1P/Halley, 6P/d'Arrest).  However, this
type of asymmetric law is not appropriate for application to the sungrazers, whose
activity peaks always very near perihelion but the preperihelion and post-perihelion
slopes of production curves may be widely uneven.  To incorporate a broad range of
nongravitational laws in the orbital code should be straightforward, yet no such
software package is widely available to our knowledge.

In the broader context, we emphasize the urgent need for solving the long-overdue
problem with the sublimation law in orbit-determination software.  A wide range
of options must be available to the user, in line with the recognized enormous
behavior diversity of comets, increasingly perceived in their motions thanks to the
steadily improving quality of astrometric observations.  In the current versions of
orbital software the selection of sublimation laws is mostly limited to the choice
of the standard function's constants.  By further postponing a radical solution to
this problem, the cometary community runs the risk that continuing chronic difficulties
will prevent orbital investigations from keeping up with the rapid progress in cometary
physics, with which they are increasingly intertwined.    

%
%
\section{Long-Term Orbit Integration Into the Past:\\A Feasibility Study}
%
To the extent that the Chinese comet of 1138 indeed is the parent body of the Great
September Comet of 1882 and comet Ikeya-Seki, the problem of the missing second major
Kreutz sungrazer in the early 12th century has been settled.  Under these circumstances,
the Great Comet of 1106 is perceived, almost by default, as the previous appearance
of the Great March Comet of 1843 and a precursor of countless smaller fragments of
Population~I.

\subsection{The Great Comet of 1106} 

The tail of the comet of 1106 is discussed elsewhere (Sekanina, in preparation); here we
display in Figure~4 this spectacular comet's predicted light curve, as a function of
solar elongation, to be compared with the predicted light curve of the 1138 comet in
Figure~3.  In line with Paper~2, we are adopting for the comet of 1106 a set of
preperihelion{\vspace{-0.055cm}} parameters \mbox{$H_0^- \!= 2.9$}, \mbox{$n^- \!= 4.0$} and
post-perihelion parameters \mbox{$H_0^+ \!= 4.0$} and \mbox{$n^+ \!= 4.2$}.  The assumed
preperihelion parameters of the comets of 1106 and 1138 are nearly the same, but after
perihelion the 1106 comet is intrinsically four magnitudes {\it fainter\/} at 1~AU from
the Sun.  Although its light curve is steeper, it does not surpass the light curve for
the 1138 comet even at that comet's perihelion distance of 0.008~AU.\footnote{The peak
of the 1106 comet's light curve in Figure~4 exceeds the peak of the 1138 comet's light
curve in Figure~3 only because the perihelion distance of the former was about two
thirds the perihelion distance of the latter.}  Yet, the 1106 sungrazer was a spectacle
after perihelion, while the performance of the 1138 comet was lackluster. 

The perihelion time of the 1106 comet is a point of contention.  Hasegawa \& Nakano
(2001) provide nominally January~26, but there are two caveats:\ one is the large
error, $\pm$5~days, that the authors put on their value; the other is a report of the
comet observed on February~2 only one {\it cubit\/} (about 1$^\circ$) from the Sun
in broad daylight (e.g., Kronk 1999).  This condition requires the comet to have been
no more than several hours past perihelion.  On the curve in Figure~4 we accordingly
mark the first sighting of the comet for two perihelion times.  The observing
circumstances long before perihelion appear to have been very favorable, but the comet
was then at high southern declinations.  As seen from Figures~3 and 4, he 1106 and 1138
comets were both likely to have been several magnitudes brighter than the limit for
naked-eye detection in broad daylight just days before perihelion, yet neither was
detected.  Except for the early 1106 daylight sighting, the observing periods of the
two comets were rather similar, supporting the notion that the 1138 comet was the
missing sungrazer.

\begin{figure}[t]
\vspace{-4.97cm}
\hspace{1.47cm}
\centerline{
\scalebox{0.64}{
\includegraphics{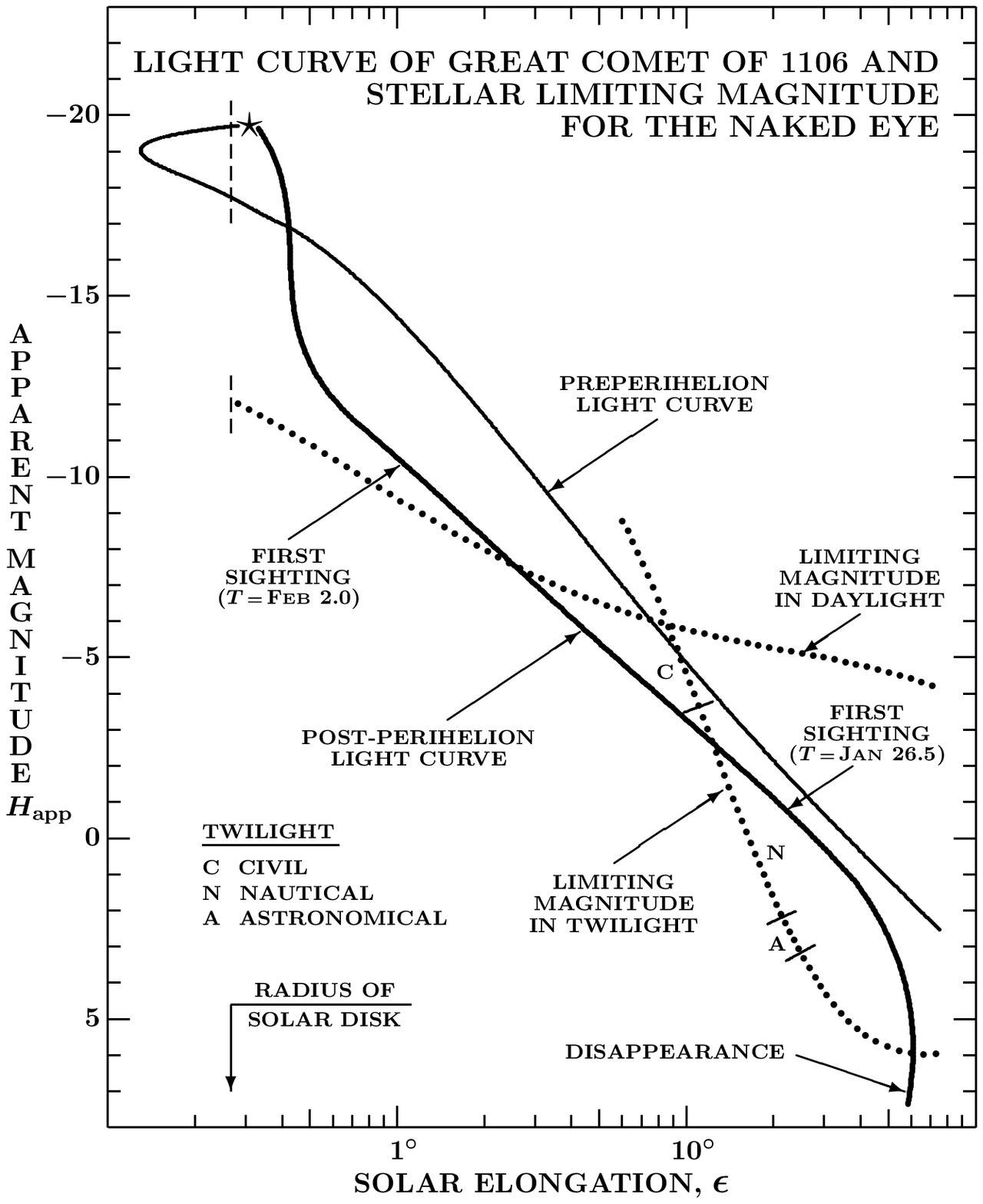}}}
\vspace{-3.69cm}
\caption{The proposed light curve for the Great Comet of 1106 plotted as a function
of solar elongation.  The point of first sighting is plotted for two perihelion times,
January 26.5 TT and February 2.0 TT.  It was argued in Paper~2 that under moderate
or better observing conditions historical sungrazers often did not disappear until the
head was near magnitude~7, because it was the tail, a few magnitudes brighter, that
was last seen.  For further comments on the plotted curves, see caption to
Figure~3.{\vspace{0.5cm}}}
\end{figure}

\subsection{Progenitor's Lobe I and Its Main Surviving Mass\\As the
 Great March Comet of 1843} 

The issues that have remained untouched as yet have been those of the early evolution of
the Kreutz system.  In the contact-binary model (Paper~1), a daytime swarm of brilliant
comets in late 363 recorded by Ammianus Marcellinus, a Roman historian, was proposed
to be the first perihelion appearance of the Kreutz sungrazers.  As separate bodies the
sungrazers were less than 500~years old, having originated as a product of fragmentation
of the massive progenitor in the general proximity of aphelion near the beginning of the
Christian Era.

The primary event was proposed~to~have~been~a~break\-up of the contact-binary progenitor
into essentially two lobes --- the early precursors of the main populations, I and II
--- depicted schematically in Figure~5.  As already pointed out, the 1843 and 1882
sungrazers were deemed the largest surviving masses of Lobe~I and Lobe~II, respectively.
In line with this scenario, we undertook orbital computations to test the feasibility
of the two fundamental steps in the proposed early evolutionary path of the Kreutz
system: (i)~Could the daylight comets of AD~363 be orbitally linked with the comets
of 1106 and 1843 on the one hand and with the comets of 1138 and 1882 on the other
hand; and (ii)~could Aristotle's comet of 372~BC be the fragmenting progenitor?

\begin{figure}[t]
\vspace{-8.15cm}
\hspace{1.73cm}
\centerline{
\scalebox{0.76}{
\includegraphics{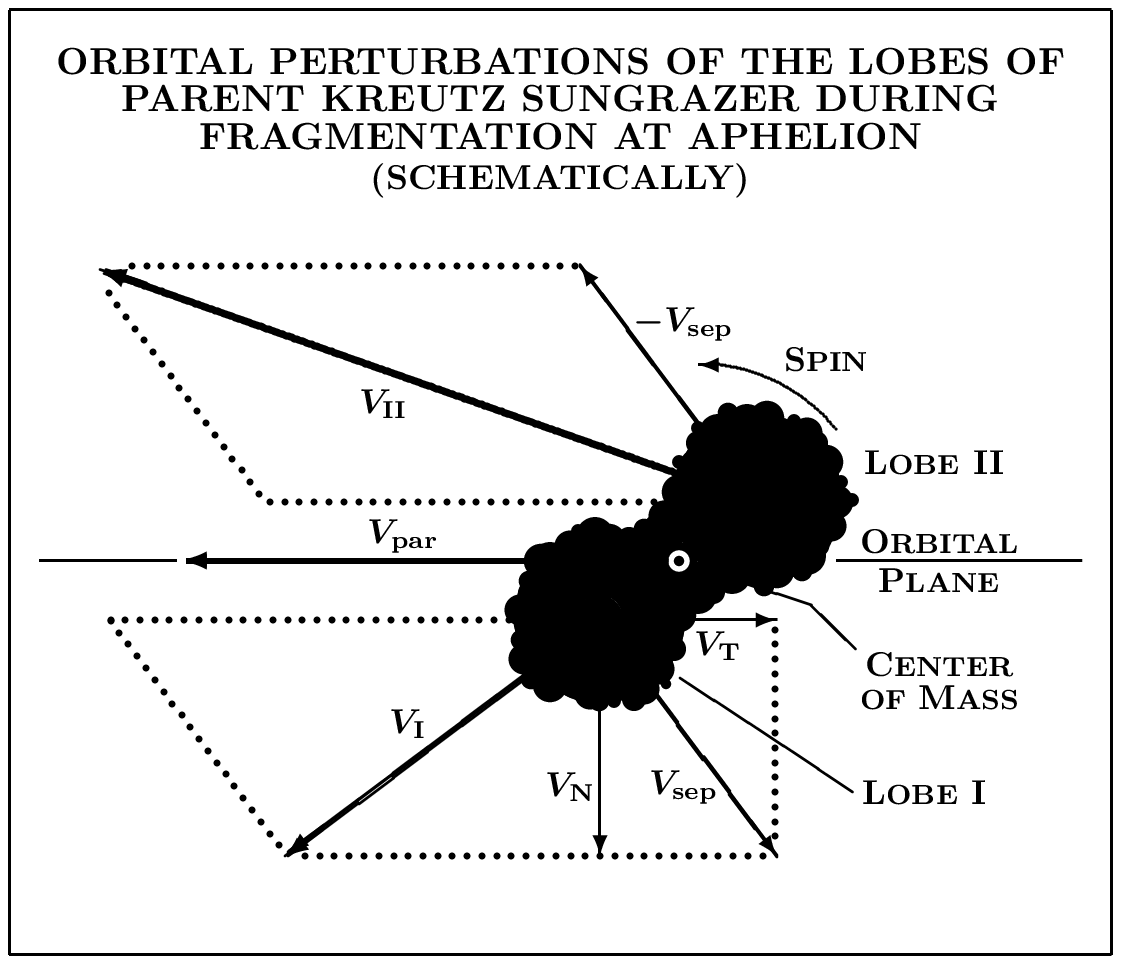}}}
\vspace{-7.24cm}
\caption{Schematic representation, at the time of breakup, of the Kreutz system's
parent (progenitor) nucleus, modeled as a rotating contact binary in Paper~1,
consisting of Lobe~I, Lobe~II, and the connecting neck.  The view is from the
direction of the Sun and the position of the original orbital plane is defined by
the parent's pre-breakup orbital velocity vector {\boldmath $V_{\bf par}$}.  The dot
in the middle of the neck is the center of mass of the parent, coinciding with the
projected spin axis, which at aphelion is assumed to point at the Sun.  As a result
of the breakup, the lobes are subjected to orbital perturbations.  At the time of
breakup the comet rotates counterclockwise, so that Lobe~I is released to the lower
right, moving relative to the center of mass in the direction of the separation
velocity vector {\boldmath $V_{\bf sep}$}, while Lobe~II is released to the upper
left, moving in the direction of the separation velocity vector {\boldmath $-V_{\bf
sep}$}.  The separation velocity consists of its transverse, {\boldmath $V_{\bf T}$},
and normal, {\boldmath $V_{\bf N}$}, components, the radial one is assumed to be
\mbox{\boldmath $V_{\bf R}$ = 0}.  Summed up with the parent's pre-breakup
orbital-velocity vector, the separation velocities insert Lobe~I into a new orbit
defined by the velocity vector {\boldmath $V_{\bf I}$} and Lobe~II into an orbit
defined by the velocity vector {\boldmath $V_{\bf II}$}.  The Great March Comet of
1843 is the largest surviving mass of Lobe~I, the Great September Comet of 1882 is
the largest surviving mass of Lobe~II.  For clarity, the orbital and separation
velocities are not drawn to scale; in the scenario in Paper~1 the ratio \mbox{{\boldmath
$|V_{\bf sep}|$}/{\boldmath $|V_{\bf par}|$} = 0.13}.{\vspace{0.7cm}}}
\end{figure}

In Section 3 we examined the motions of two observed sungrazer fragments, the products
of a presumed tidal breakup of their parent at or near perihelion.  Their orbital
periods differed dramatically on account of two effects.  One of them was triggered
by the differential outgassing-driven nongravitational acceleration and accounted
for by integration of the continuous contributions from outgassing along the orbit.
Long-term effects in the orbital period of this {\it sublimation scenario\/} are
handled by nongravitational orbital solutions.

Major changes in the orbital period could also be generated as a corollary of tidal
fragmentation in close proximity of perihelion.  When no momentum exchange is involved,
the nascent fragments have the same orbital velocity as the parent, but their centers
of mass are located at slightly different heliocentric distances, causing the fragments
to end up in orbits with different semimajor axes and periods.  This is a {\it
fragmentation scenario\/} whose long-term effects in the orbital period could only
be handled by modeling the main features of the fragmentation events, with help of
gravitational and/or nongravitational orbital solutions.

In the following we maintain that the evolution of the Kreutz system has been governed
by combined effects articulated by the sublimation scenario on the one hand and by the
fragmentation scenario on the other hand.  While the two categories of effects mix at
some unknown variable ratio to scatter sungrazers and their fragments around, we focus
below on finding out whether either category can explain --- in the entirety or to an
extent --- the observed properties of the Kreutz system on its own, obviously a more
difficult task to accomplish.  We have employed the same orbit-determination package
of software as in Section~3, fully accounting for both the perturbations by the planets,
Pluto, and the three most massive asteroids, as well as for the relativistic and
nongravitational effects, using the Style~II formalism by Marsden et al.\ (1973).

Numerical experimentation used in the extensive investigation of comet Ikeya-Seki
in Section~3 showed two perplexing features: (i)~considerable influence and total
unpredictability of the indirect planetary perturbations, which implied the absence
of any uniformity or pattern in the sequence of perihelion times; and (ii)~problems
related to the sublimation law symmetric relative to perihelion, the issue that
already was addressed in Section~5.  In the absence of an appropriate law, the
asymmetry was matched imperfectly by introducing the transverse component of the
nongravitational acceleration via the parameter $A_2$.

Turning first to the {\it sublimation scenario\/}, we attempted to link, in
a single orbital run, three consecutive perihelion returns of Lobe~I, ending
chronologically with the Great March Comet of 1843.  The starting set of elements
for integration back in time was this comet's {\it gravitational\/} orbit
referred to as Solution~II in Sekanina \& Chodas (2008).  It already was optimized
to satisfy the 1106 comet's perihelion time of January~26, proposed by Hasegawa
\& Nakano (2001), yet it provided an essentially equally good fit to the re-reduced
1843 astrometric observations as the nominal least-squares Solution~I.  When
integrated back to the 4th century, Solution~II provided a perihelion time on
392~September~19.  The difference of about 29~years needed to fit the adopted
perihelion time of 363~November~15 suggested that the comet had been accelerated
over the period of nearly 15~centuries (363--1843).

\begin{table*}[t]
\vspace{-4.2cm}
\hspace{0.5cm}
\centerline{
\scalebox{1}{
\includegraphics{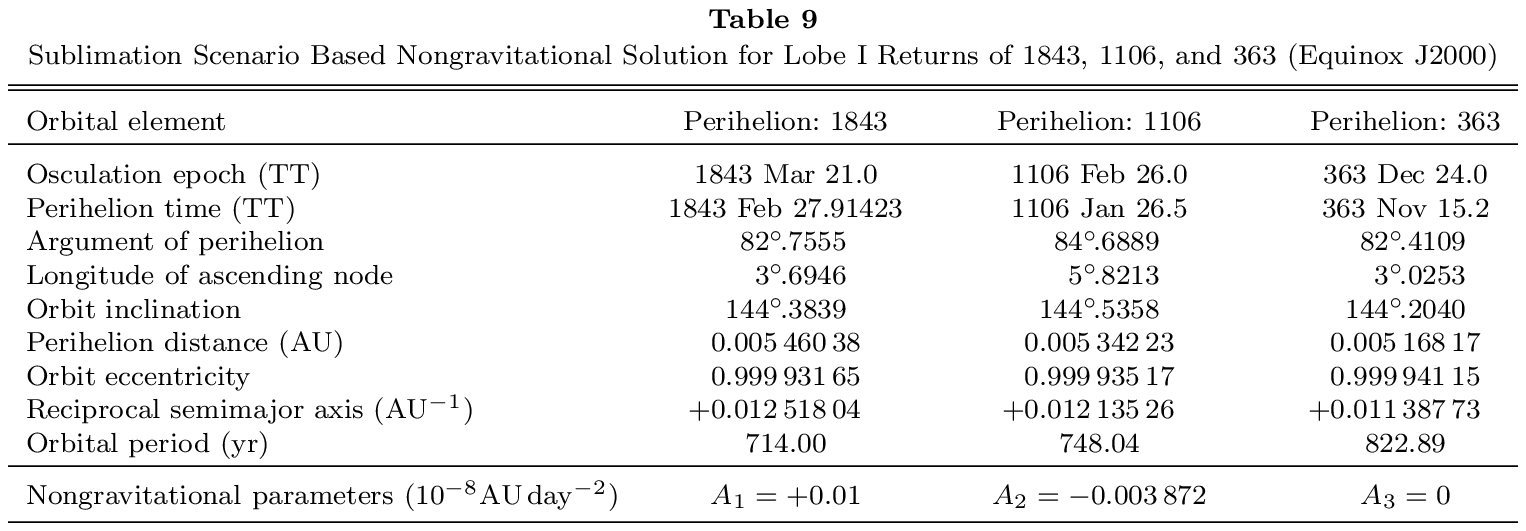}}}
\vspace{-19.3cm}
\end{table*}

To investigate the magnitude of the acceleration, we incorporated the
nongravitational terms into the equations of motion and searched, by trial and
error, for the optimum value of the parameter $A_2$, while simultaneously adjusting
slightly the eccentricity (by about 30~percent of the difference between Solutions~I
and II in the paper by Sekanina \& Chodas 2008).  The{\vspace{-0.04cm}} adopted
radial nongravitational parameter $A_1$ of 10$^{-10}$\,AU~day$^{-2}$ represents a
reasonable value for the nucleus estimated at some 50~km in diameter but, as noted
in Section 5, its value has nearly no effect on the results because of the symmetric
nongravitational law that the method uses.  Very similar orbital solutions could
thus be derived with other values of $A_1$.  The successful linkage of the
three~consecutive returns of Lobe~I, presented in Table~9, is manifested by the
parametric value of $A_2$, which was optimized by first re-fitting the perihelion
time in 1106 and then iterating the relationship between the eccentricity and
the parameter $A_2$ until the {\it previous\/} perihelion time proposed in
Paper~1, 363~Nov~15, was matched.  This value of $A_2$, shown in Table~9, is
smaller than $A_2$ for any long-period comet in the catalogue by Marsden \&
Williams (2008).  A successful linkage of the perihelion returns in 1843, 1106,
and 363 within the framework of the {\it sublimation scenario\/} alone is thus shown
to be feasible.  The parameter $A_3$ of the normal component of the nongravitational
acceleration was assumed to be zero, because this component could not affect the
perihelion time.

The linkage of the three consecutive perihelion returns implies that the orbital
effect of fragmentation on the Great March Comet of 1843 was insignificant enough
that ignoring it did not prevent the formulation of a straightforward nongravitational
solution, which over a period of 15~centuries provides supporting evidence for the
evolutionary model proposed in Paper~1.
%
%

We next turned to the {\it fragmentation scenario\/} for Lobe~I.  The modus operandi was
predicated on the fact that one corollary of a breakup is the instantaneous redistribution
of a single mass into fragments, accompanied by repositioning of the original center
of mass into the centers of mass of the fragments.  The small value of $A_2$ in Table 9
encouraged us to neglect the nongravitational effect altogether.  Accordingly, we started
from the same gravitational Solution~II for the 1843 sungrazer (Sekanina \& Chodas 2008).
As in the sublimation scenario, we ran the orbit back to 1106 to the point of birth of
the 1843 comet, which became the dominant fragment of the Great Comet of 1106, its parent
and itself a fragment of the original Lobe~I.

The focus of our interest at this point was the nature of the fragmentation event,
specifically the separation between the centers of mass of the parent and the fragment
along the radius vector (Figure~2).  The time of osculation coincided with the time
of fragmentation, $t_{\rm frg}$, assumed to have taken place at perihelion. when in
the ecliptical coordinate system the position vector of the center of mass of the
nascent 1843 comet was \mbox{{\boldmath $U_{\bf frg}$} = $(X_{\rm frg}, Y_{\rm frg},
Z_{\rm frg})$}.  The unknown position vector of the center of mass of the parent comet
in the ecliptical coordinate system was \mbox{{\boldmath $U_{\bf par}$} = $(X_{\rm par},
Y_{\rm par}, Z_{\rm par})$}.  The{\vspace{-0.07cm}} orbital-velocity vector of the
fragment at time $t_{\rm frg}$, \mbox{{\boldmath $V_{\bf frg}$} = $(\dot{X}_{\rm frg},
\dot{Y}_{\rm frg}, \dot{Z}_{\rm frg})$}, together with the position vector determined
the fragment's orbit.  From Figure~2, the{\vspace{-0.08cm}} orbital-velocity vector
of the parent at $t_{\rm frg}$, \mbox{{\boldmath $V_{\bf par}$} = $(\dot{X}_{\rm par},
\dot{Y}_{\rm par}, \dot{Z}_{\rm par})$} equaled the fragment's velocity vector.
The difference between the position vectors of the 1106 and 1843 comets at $t_{\rm
frg}$ imply the existence of one or more additional fragments, which we comment on
briefly later.

At a given heliocentric distance of the point of fragmentation, $r_{\rm frg}$, the
position vector of the parent can be derived from the position vector of the fragment
and vice versa.  The vectorial difference \mbox{{\boldmath $\Delta U$}$_{{\bf
f}\rightarrow{\bf p}}$ = {\boldmath $U_{\bf par}$} $\!-\!$ {\boldmath $U_{\bf frg}$}}
in 1106 depends according to Equation~(17) on the orbital period of the parent comet,
$P_{\rm par}$, between the perihelion passages in 1106 and 363, and on the hypothetical
orbital period of the fragment, $P_{\rm frg}$, between 1106 and its past projected
perihelion in 392, established by the computations made for this particular case.
The derived periods are \mbox{$P_{\rm par} = 742.186$ yr} and \mbox{$P_{\rm frg} =
713.339$ yr}.

\begin{table*}
\vspace{-4.2cm}
\hspace{0.5cm}
\centerline{
\scalebox{1}{
\includegraphics{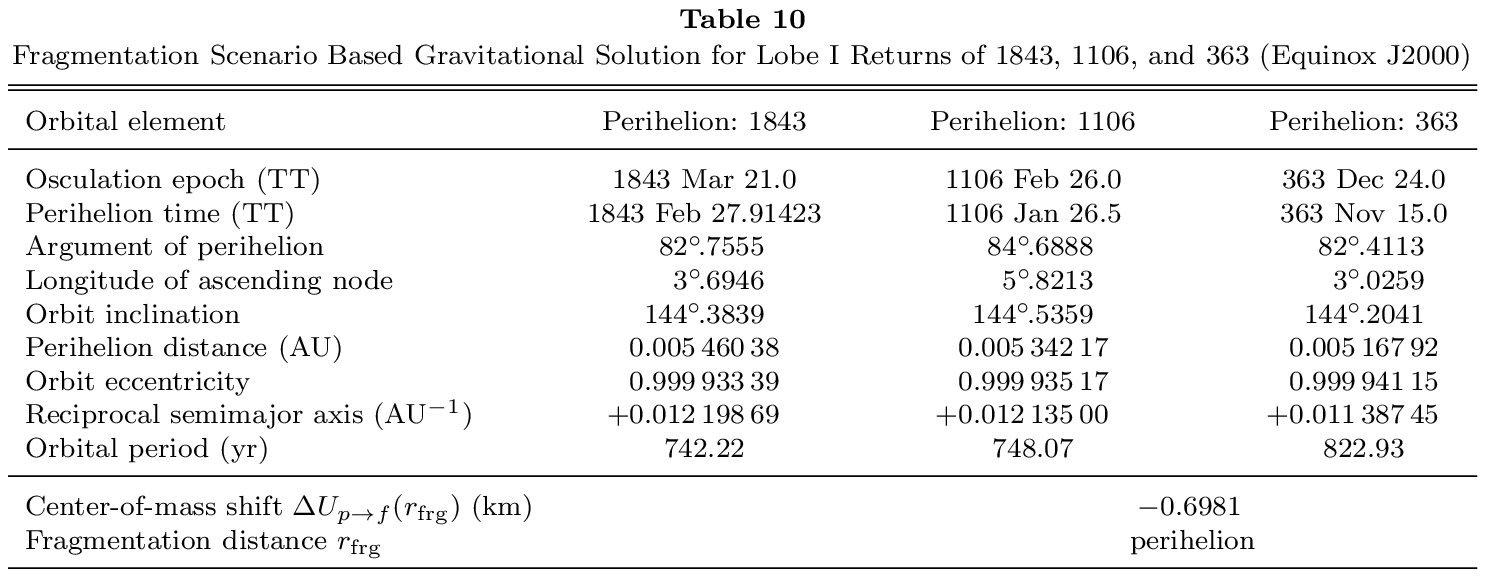}}}
\vspace{-18.9cm}
\end{table*}

In the arguments leading to Equation (17) we concluded that, because of the nature of
the tidal disruption, the vector {\boldmath $\Delta U_{{\bf f}\rightarrow{\bf p}}
(r_{\bf frg})$} points essentially either in the direction of the radius vector
{\boldmath $r_{\bf frg}$} (when positive) or in the opposite direction (when negative).
We replace {\boldmath $\Delta U_{{\bf f}\rightarrow{\bf p}}(r_{\bf frg})$} with
$\Delta U_{f\rightarrow p}(r_{\rm frg})$ and simplify the formula for its magnitude to
\begin{equation}
\Delta U_{f \rightarrow p}(r_{\rm frg}) = {\textstyle \frac{1}{2}} r_{\rm frg}^2 \!
 \left( \! P_{\rm frg}^{-\frac{2}{3}} \!-\! P_{\rm par}^{-\frac{2}{3}} \! \right),
\end{equation}
where $P_k^{-\frac{2}{3}}$ (\mbox{$k$ = frg, par}) expressed in years are numerically
equal{\vspace{-0.03cm}} to the reciprocal semimajor axes, $1/a_k$, expressed in
AU$^{-1}$, so that $\Delta U_{f \rightarrow p}(r_{\rm frg})$ is in AU.  Conversely,
$\Delta U_{f\rightarrow p}(r_{\rm frg})$ determines the relationship between the
orbital periods $P_{\rm par}$ and $P_{\rm frg}$:
\begin{equation}
P_{\rm par} = P_{\rm frg} \! \left( \! 1 - \frac{2 \Delta U_{f \rightarrow p}(r_{\rm
frg})}{r_{\rm frg}^2} P_{\rm frg}^{\frac{2}{3}} \! \right)^{\!\!-\frac{3}{2}} \!\!\!.
\end{equation}

Since we assumed that the 1106 sungrazer fragmented exactly at perihelion, from
Table~9 we insert into (24) \mbox{$r_{\rm frg} = q = 0.005\,342$ AU}, which with
the above values for $P_{\rm frg}$ and $P_{\rm par}$ yields
\begin{equation}
\Delta U_{f \rightarrow p}(q) = +4.662 \times \! 10^{-9}\; {\rm AU} = +0.6974
 \; {\rm km}. 
\end{equation}

Once $\Delta U_{f\rightarrow p}(r_{\rm frg})$ is known, we proceed to the
computation of the orbital elements of the 1106~comet.  In a general case,
when the point of fragmentation is at a heliocentric distance $r_{\rm frg}$,
the ecliptical coordinates of the parent's position vector at time $t_{\rm frg}$
are
\begin{equation}
\left( \!\! \begin{array}{ccc}
X_{\rm par} \\
Y_{\rm par} \\
Z_{\rm par}
\end{array}
\!\! \right) = \left( \!\! \begin{array}{ccc}
X_{\rm frg} \\
Y_{\rm frg} \\
Z_{\rm frg}
\end{array}
\!\! \right) + \Delta U_{f \rightarrow p}(r_{\rm frg})\!\left(\!\!\begin{array}{ccc}
P_x & \!Q_x \\
P_y & \!Q_y \\
P_z & \!Q_z
\end{array}
\!\! \right) \!\times\! \left( \!\! \begin{array}{cc}
\cos u_{\rm frg} \\
\sin u_{\rm frg}
\end{array}
\!\! \right)\!,   
\end{equation}
where $P_x$, \ldots, $Q_z$ are the standard direction cosines and $u_{\rm frg}$
is the true anomaly at fragmentation.  In the case of fragmentation at perihelion
the transformation formulas get simplified accordingly.  Since the indirect
planetary perturbations may change slightly the orbit, one may not obtain the
prescribed perihelion time of 363~November~15 exactly, and one or more iterations
may be necessary.  Indeed, the value of $\Delta U_{f \rightarrow p}$ had to be
changed from 0.6974~km to 0.6981~km to fit the adopted perihelion time in 363
exactly.

The sets of gravitational orbital elements of Lobe~I at the three returns to
perihelion between 363 and 1843, based on the fragmentation scenario, are listed
in Table~10.  Comparison with Table~9 shows that the differences in all the
elements between the two categories of orbits are trivial, with the possible
exception of the orbital period.  An obvious conclusion is that applications
of both the sublimation scenario and the fragmentation scenario, either one on
its own, satisfy the constraints of the evolutionary model of the Kreutz system
over the 15~centuries.  Orbit integration prior to the year 363 and its
ramifications will be addressed in Section~7.   
%
%

\subsection{Possible Nuclear Sizes of Comets 1106 and 1843\\and Population
 of Minor Kreutz Comets} 

The shift $\Delta U_{f \rightarrow p}$ between the parent and a fragment is an
important quantity that provides information on the fragment's approximate
dimensions and their relationship to the parent's dimensions.  In general terms,
the smaller the shift, the larger the fragment's size.  Although no reliable data
are available, the nucleus of the 1882 sungrazer was crudely estimated at 50~km in
diameter (Sekanina 2002), and the 1843 and 1882 sungrazers are probably of comparable
sizes.  From this vantage point, a shift in the position of the center of mass by mere
0.7~km is strong evidence that during the perihelion breakup of the 1106 comet most
mass remained concentrated in one fragment --- the 1843 sungrazer.  Since the shift
is sunward --- measured from the parent to the fragment --- minor fragments were
at the antisunward end of the parent, while its sunward half did not fragment at
all; in Figure~2 this fits Scenario~I.

On certain assumptions about the shape of the parent object, one can provide even some
quantitative estimates.  Let the parent be approximated by a prolate spheroid of an
arbitrary axial ratio, whose long axis pointed at the time of fragmentation toward the
Sun, and let the dominant fragment, in this case the 1843 sungrazer, be represented by
a truncated prolate spheroid, which extends from one end of the spheroid to a plane normal
to the long axis at a certain distance, $d$, from that end.  Assigning the parent's
spheroid a unit volume, we measure, in these units, the volume $\Re$ of the truncated
spheroid by a dimensionless parameter $\zeta$, which equals the ratio of the distance
$d$ to the extent of the long axis:
\begin{equation}
\Re(\zeta) = 3 \zeta^2 \! \left(1 - {\textstyle \frac{2}{3}} \zeta \right). 
\end{equation}
The parent's center of mass is given{\vspace{-0.05cm}} by \mbox{$\zeta = \frac{1}{2}$},
whereas the fragment's center of mass has obviously \mbox{$\zeta < \frac{1}{2}$}.  The
volume of the truncated spheroid needs to be expressed in terms of $\Delta U_{f
\rightarrow p}$.  In order to do that, we introduce a dimensionless quantity, $\psi$,
by
\begin{equation}
\Delta U_{f \rightarrow p} = \psi {\cal R}, 
\end{equation}
where $\cal R$ is the radius of the spheroid's long axis. The normalized volume
of the truncated spheroid is then
\begin{equation}
\Re(\psi) = 1 - {\textstyle \frac{3}{2}} \psi \! \left(1 \!-\! {\textstyle \frac{1}{3}}
 \psi^2 \right). 
\end{equation}
Inserting for $\Re$ from (30) into (28), we find the length $d$ of the truncated
spheroid by solving for $\zeta$:
\begin{equation}
d = 2 {\cal R} \zeta = \frac{{\cal R} \sqrt{\Re(\psi)}}{\cos \! \left[ {\textstyle
 \frac{1}{3}} \arccos \!\left( \!-\sqrt{\Re(\psi)} \right) \right]} 
\end{equation}
Approximating the size of the prolate spheroid for the 1106 comet's nucleus by
\mbox{${\cal R} = 28$ km}, we have \mbox{$\psi = 0.025$}, from (30) \mbox{$\Re =
0.9625$}, and from (31) \mbox{$d = 49.5$ km},~or~near the estimated nuclear
diameter of the 1843 sungrazer.

The remaining part of the spheroid, modeled as a spheroidal cap, is in this case
6.5~km long and has a volume of 3.75~percent of the parent.  Its center of mass is
23.5~km from the center of mass of the parent and its calculated orbital period
exceeds 20,000~years.  This mass is essentially lost to the system and it may
never be discovered.

Up to now we only considered tidal breakups taking place exactly at perihelion and
separation of fragments from the long end of a spheroidal nucleus.  In reality,
tidal fragmentation events may have happened (or continued happening) shortly
before or after perihelion, in which case the shifts $\Delta U_{f \rightarrow
p}(r_{\rm frg})$ increased in magnitude and the orbital periods of large numbers
of fragments were confined to a narrower range, many exceeding the orbital period
of the parent by less than a factor of two.  In addition, the parent's nucleus
was unquestionably an irregular body with variable tensile strength, so that tidal
fragmentation proceeded in parts other than one end of its longest dimension as
well.  Finally, especially minor fragments could have been shed from the body of
the parent/main fragment by forces other than tidal; in the extremely hostile
environment of the Sun's inner corona such mass shedding should be part of the
bulk fragmentation process.  As long as the shedding is accompanied by no momentum
exchange, the fragments end up essentially in the orbits governed by $\Delta U_{f
\rightarrow p}(r_{\rm frg})$.

In this context we note that the separation times for the individual nuclei of the
Great September Comet of 1882, derived from their observed separations by Sekanina
\& Chodas (2007), averaged 1.8~hours after perihelion, at which time the comet's
heliocentric distance was twice the perihelion distance, or 3.3\,{\Rsun}.  For the
1843 sungrazer the corresponding fragmentation distance would be 2.3\,{\Rsun}.  At
such a distance, a large number of Kreutz sungrazers and potential Kreutz sungrazers
of Population~I could be fitted as fragments of the 1106/1843 comet with moderate
$\Delta U$ shifts.  A shift of about $-$20~km could explain the concentration (or at
least a major contribution to the concentration) of potential Kreutz sungrazers in
the middle and the second half of the 17th century, including the candidate comets
of 1663, 1666, 1668, 1673, and 1695 on Hasegawa \& Nakano's (2001) list.  Positive
values of $\Delta U$ could apply to a number of known Kreutz sungrazers:\ for
example,{\vspace{-0.03cm}} $\sim$3.5~km fits C/1880~C1 and C/1882~K1, the eclipse
sungrazer;\footnote{However, the 1880 sungrazer may have separated from the 1106 comet
at large heliocentric distance; C/1887~B1 very probably separated subsequently from the
1880 comet (Paper~1).} $\sim$11.5~km fits the sungrazers picked up by the coronagraphs
on board the Solwind (P78-1) and Solar Maximum Mission satellites; and 12 to 14~km fits
the SOHO and STEREO sungrazers from the 1990s through 2020s, whose orbital periods
should be close to 900~years.  An exception is comet Pereyra (C/1963~R1), which
appears to be a fragment of another subcategory of Population~I and its evolution
is briefly described in Section~8.

\subsection{Orbital Relationship Between the Great September Comet of 1882 and
 Comet Ikeya-Seki}  
Before we examine the sublimation and fragmentation scenarios of the comet
of 1138, we return to the problem of the pre-fragmentation orbit of comet
Ikeya-Seki and its relationship to the orbit of the Great September Comet of
1882.  A major point of Marsden's (1967) paper was his virtual proof that on
approach to the previous perihelion in the 12th century the two comets were one.

Our surprising conclusion that Ikeya-Seki was previously at perihelion as the
Chinese comet of 1138 more than 30~years after the Great Comet of 1106 was not
accompanied by the full set of elements of the pre-fragmentation orbit.  In
the meantime, we were pursuing several avenues in an effort to determine such
an orbital set, but were repeatedly encountering a variety of problems.  Our
ultimate choice was based on an inference from the perturbation theory that
a fragmentation event at perihelion, not involving momentum exchange,
measurably affects the orbital period (or, equivalently, the
semimajor axis or eccentricity) but not the other five elements.  Their
post-perihelion variations have roots mostly in the nongravitational forces,
so that --- again with the exception of the eccentricity and equivalent elements
--- the orbit of the most massive fragment should provide the closest approximation
to the orbit of the pre-fragmentation comet.  For Ikeya-Seki, the most massive
fragment was unequivocally nucleus A.

The same argument suggested that among the fragments of the Great September
Comet of 1882 it was its nucleus~B (or No.\,2 in Kreutz's notation) that should
provide the best approximation to the comet's pre-fragmentation orbit.  The
orbit of this nucleus, presented in the Appendix, was our first choice for the
purpose of comparison with the orbit of comet Ikeya-Seki, in spite of the fact
that it is nonrelativistic.\footnote{The theory of general relativity, including
its orbital ramifications, was introduced by A.\,Einstein in 1915, just about a
quarter-century after Kreutz's publication of the definitive orbits for the nuclear
fragments of the 1882 sungrazer in the second paper of the series.}

Our first concern was the time of the previous perihelion of the comet's nucleus~B,
for which we obtained 1136~January~23; formal inclusion of the relativistic effect
moved the perihelion forward by merely 37~days.  As the mean error of the orbital
period is shown in the Appendix to amount to $\pm$2.8~yr, the estimated perihelion
time of the 1138 comet remained within 1$\sigma$ of the predicted time of the 1882
comet's previous return to perihelion.  The very fact that the integration of this
nucleus' orbital motion offered the ``correct'' time for the previous return to
perihelion of the pre-fragmentation nucleus of the 1882~sungrazer is astonishing
and its interpretations range from sheer coincidence to evidence that fragment B
was indeed the primary nucleus, orders of magnitude more massive than any other
among the five or six major fragments reported.

\begin{table*}[t]
\vspace{-4.15cm}
\hspace{0.29cm}
\centerline{
\scalebox{0.97}{
\includegraphics{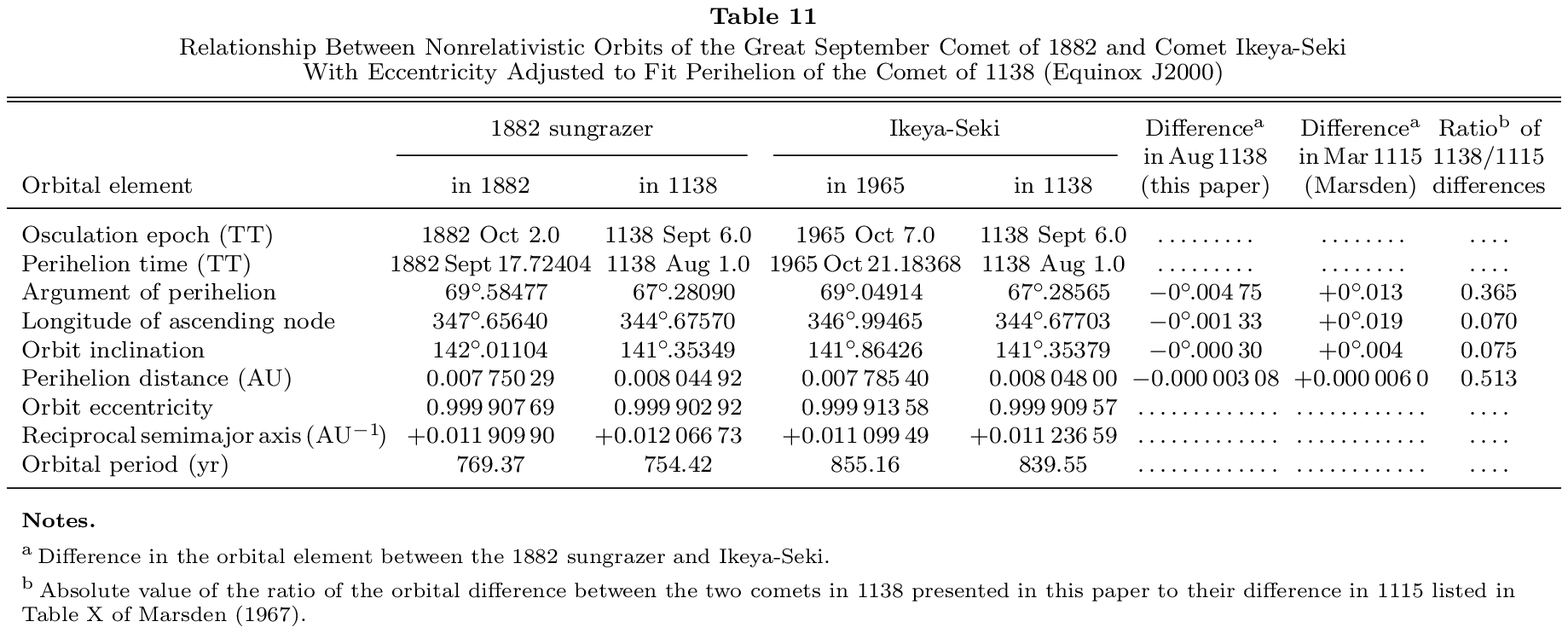}}}
\vspace{-17cm}
\end{table*}

Of primary interest to us were the sets of orbital elements of the 1882~sungrazer
and comet Ikeya-Seki in August 1138.  To make it meaningful, we computed the orbit
of Ikeya-Seki {\it without\/} the relativistic effect to match the same quality
of the orbital set available for the 1882 sungrazer.  We integrated the orbits
of both comets back in time, adjusting in either case the eccentricity to force
the previous perihelion on 1138~August~1.  In essence, we applied the same test
as Marsden (1967) did in his paper:\ aiming at comparison of the orbits of the
1882~comet's nucleus B and Ikeya-Seki's nucleus A at the time of their presumed
separation.  The difference was that Marsden assumed the separation in September
1115, whereas our target date was some 23~years later.  The outcome is presented
in Table~11; the elements for an osculation epoch in the start-up years of 1882
and 1965 are, respectively, in columns 2 and 4; the 1138 elements in columns 3
and 5.  The degree of similarity between the two comets in the four relevant
elements in 1138 is apparent from column~6; the degree of similarity that Marsden
(1967) found in 1115 is copied in column~7.  Besides the fact that all 1138
differences were of opposite sign than the 1115 differences, their comparison,
in terms of an absolute value of the 1138-to-1115 ratio, offers a stunning
result:\ the {\it orbital elements of the two comets were much more alike in
1138\/}, the nodal longitude and the inclination by more than one order of
magnitude(!), the argument of perihelion by a factor of nearly 3, and the
perihelion distance by a factor of about 2.  The near-coincidence of the orbits
of the two comets in 1138 is equally impressive when measured by the mean errors
of Kreutz's elements in the Appendix:\ the differences in the nodal longitude and
inclination are smaller than 0.5$\sigma$(!), in the argument of perihelion about
2$\sigma$, and in the perihelion distance, the worst case, a little over 3$\sigma$.
The tiny differences in the nodal longitude and inclination show that orbital planes
of the two comets deviated from each other in 1138 less than was the uncertainty of
the 1882 comet's orbital-plane determination.  The obvious conclusion is that the
year 1138 was a much better choice for separating Ikeya-Seki from the 1882~sungrazer
than the year 1115, not to mention 1106.  This is yet another piece of evidence
that contradicts the hypothesis of the 1106~comet being the parent to the 1882/1965
pair.   

%
%

%
\begin{table*}
\vspace{-4.2cm}
\hspace{0.55cm}
\centerline{
\scalebox{1}{
\includegraphics{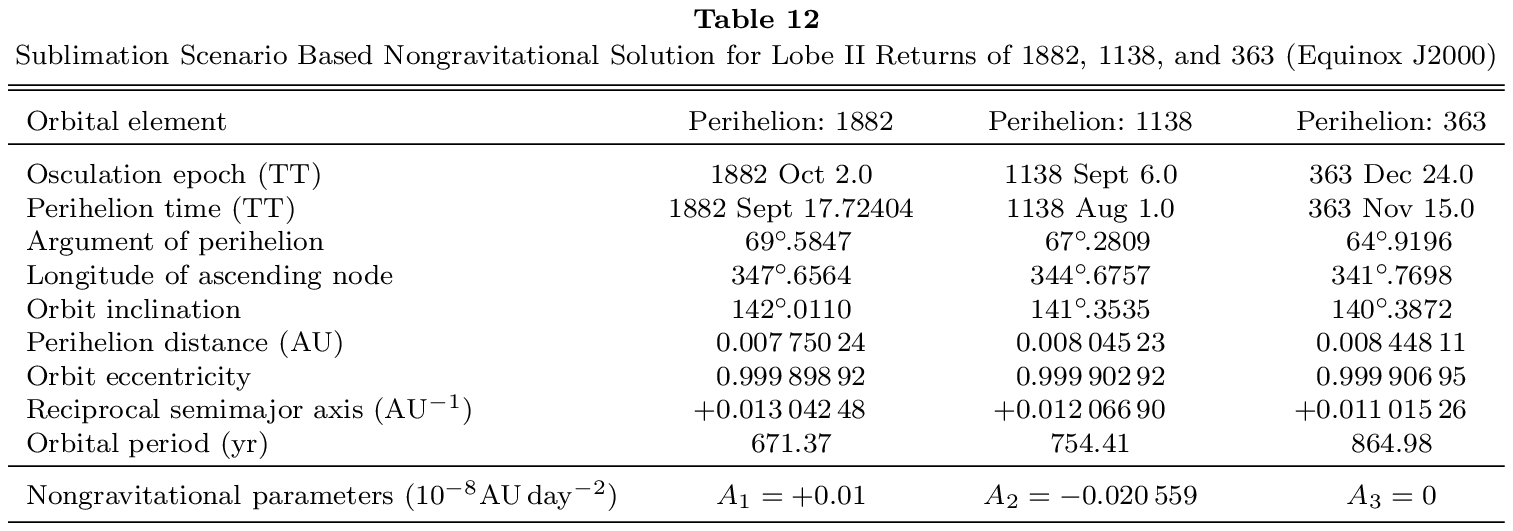}}}
\vspace{-19.3cm}
\end{table*}

\subsection{Progenitor's Lobe II and Its Main Surviving Mass\\As
 the Great September Comet of 1882} 
The resemblance of the orbital elements (except for the period and equivalent
quantities) between the 1882 sungrazer and Ikeya-Seki in the year 1138 is so
profound that a tight relationship between the two objects cannot be doubted.
In the context of the contact-binary model, their parent --- and later the
1882~sungrazer --- was the largest surviving mass of Lobe~II of the Kreutz
system's progenitor (Figure~5).  Long-term orbit integration of the motion
of Lobe~II was carried out in the same manner as that of Lobe~I in Sections~6.2
and 6.3, engaging, respectively, the {\it sublimation\/} and {\it fragmentation\/}
scenarios.

To test the feasibility of a single orbital run linking three consecutive returns
of Lobe~II, we again employed Kreutz's (1891) gravitational (nonrelativistic) orbit
of nucleus B (see the Appendix) but used, from now on, orbit-determination software
with the relativistic effect built in.  As expected, a fairly minor adjustment in
the eccentricity was needed to fit the adopted perihelion time of the 1138 comet.
The previous perihelion was computed to have occurred on 459~Jan~29, nearly a
century later than the expected year 363, implying that Lobe~II was subjected
to a much higher nongravitational acceleration than Lobe~I.  The results of
the search for a nongravitational solution consistent with the sublimation
scenario confirmed the suspicion.  As shown in Table~12, the required magnitude
of the transverse acceleration was now more than twice the adopted magnitude of
the radial acceleration and more than five times the transverse acceleration for
the 1843 comet (Table~9).  In case of the 1882 sungrazer the effects that
we tried to account for by the nongravitational parameter $A_2$ appear to have been
governed by fragmentation rather than by sublimation.

Application of the fragmentation scenario based procedures in the case of Lobe~II
was much more involved than it had been for Lobe~I in Section~6.2.  Before we began
to investigate Lobe~I, we did not know whether it fragmented profusely, modestly,
or not at all.  On the other hand, unquestionable evidence on fragmentation of
Lobe~II was offered by the breakup of the 1882~sungrazer and Ikeya-Seki, which
had to be accommodated by the contact-binary model.  Referring to them as fragments
$f_1$ (the 1882 sungrazer) and $f_2$ (Ikeya-Seki), respectively, the constraint
provided by their orbital periods (\mbox{$P_{f_1} = 744.084$ yr} and \mbox{$P_{f_2}
= 827.172$ yr}) was
\begin{equation}
\Delta U_{f_1 \rightarrow f_2}(r_{\rm frg}) = +4.02 \!\left(\!\frac{r_{\rm frg}}{q}
 \!\right)^{\!\!2}  {\rm km}\!. 
\end{equation}
Here $\Delta U$ has the meaning of the radial distance between the centers of mass
of the two fragments.  If the 1882 sungrazer and Ikeya-Seki had irregular nuclei
some 50~km and 10~km, respectively, in diameter, their centers of mass
could hardly be separated by less than $\sim$20~km, which means that the
fragmentation event of this kind could not have taken place closer to the Sun than
$\sim$2.2~times the perihelion distance, or about 0.018~AU (or nearly 4~solar radii)
from the Sun, at least 2.2~hr before or after perihelion.

The second constraint for Lobe II was provided by the difference between the projected
motion of the 1882 sungrazer integrated back in time from its 1138 perihelion to the
previous hypothetical perihelion passage, on 459~January~29 --- implying an orbital
period of \mbox{$P_{\rm frg} = 679.491$ yr} --- and the corrected motion dictated by
the expected perihelion of the parent on 363~November~15, implying \mbox{$P_{\rm par}
= 774.696$ yr}.  The computations showed that the required shift of the center of
mass at the 1138 perihelion equaled
\begin{equation}
\Delta U_{f \rightarrow p}(q) = +5.2499 \; {\rm km}, 
\end{equation}
an effect that was a factor of 7.5 greater than in the case of Lobe~I, shown in (26).

The motion of Lobe~II between the returns of 363 and 1138 is determined by the
condition (33); the sets of representative gravitational orbital elements in the
returns of 1882, 1138, and 363, based on the fragmentation scenario, are listed
in Table~13.  To accommodate the conditions (32) and (33) during the 1138 perihelion
passage, we assumed again that the parent comet, approaching perihelion was a prolate
spheroid of an arbitrary axial ratio and an unknown long-axis diameter $\cal D$,
which was to be determined from the constraint (33).  As in the case of Lobe~I,
we were addressing the problem in terms of dimensionless quantities, in units
of $\cal D$.  We further assumed that the parent split first into two at
perihelion, satisfying (33), followed by the separation of the major fragment
into the 1882~sungrazer and Ikeya-Seki hours later.

Let us adopt for the 1882 sungrazer a diameter of 50~km and for Ikeya-Seki a
diameter of 10~km in the direction of the spheroid's long axis, so that
\mbox{${\cal D} > 60$ km}.  In Figure~6 we draw the points on the spheroid's
axis that describe the fragmentation events.  With the Sun to the right, the
points refer to the first breakup in the upper panel and to the second breakup
in the lower panel.  In either panel $A$ is the sunward end of both the surviving
mass of Lobe~II and the 1882~sungrazer, $O$ the center of mass of Lobe~II, $C$
the center of mass of the 1882/1965 pair's parent, $B$ its antisolar end, and
$W$ the boundary between the 1882~sungrazer and Ikeya-Seki inside Lobe~II.  In
addition, in the upper panel, $Z$ is the antisolar end of Lobe~II, so that $B$
and $Z$ are the boundaries of the minor fragment that separated at perihelion,
and $G$ is its center of mass.  In the lower panel, $E$ and $F$ are the centers of
mass of, respectively,{\vspace{-0.075cm}} the 1882~sungrazer and Ikeya-Seki.  Since
\mbox{$\widehat{AZ} = 1$} and \mbox{$\widehat{AO} = \widehat{OZ} = \frac{1}{2}$},
the{\vspace{-0.105cm}} normalized size of the 1882~sungrazer{\vspace{-0.085cm}} is
\mbox{$\alpha = \widehat{AW} = 50/{\cal D}$} and the normalized size of Ikeya-Seki
\mbox{$\beta = \widehat{WB} = 10/{\cal D}$}.  The{\vspace{-0.085cm}} size of their
parent is of course \mbox{$\gamma = \widehat{AB} = \alpha \!+\! \beta = 60/{\cal
D}$} and the size of the minor{\vspace{-0.065cm}} fragment separating at perihelion
is \mbox{$\epsilon = \widehat{BZ} = 1 \!-\! \gamma = ({\cal D} \!-\! 60)/{\cal D}$}.
We further denote the dimensionless distances from $A$ of the centers of mass of,
respectively, the 1882~sungrazer,{\vspace{-0.065cm}} Ikeya-Seki, their parent, and
the minor{\vspace{-0.065cm}} fragment by \mbox{$\alpha^\prime = \widehat{AE}$},
\mbox{$\beta^\prime = \widehat{WF}$}, \mbox{$\gamma^\prime = \widehat{AC}$}, and
\mbox{$\epsilon^\prime = \widehat{GZ}$}.  We count the distances positive in the
direction away from the Sun, and vice versa.

\begin{table*}[t]
\vspace{-4.2cm}
\hspace{0.5cm}
\centerline{
\scalebox{1}{
\includegraphics{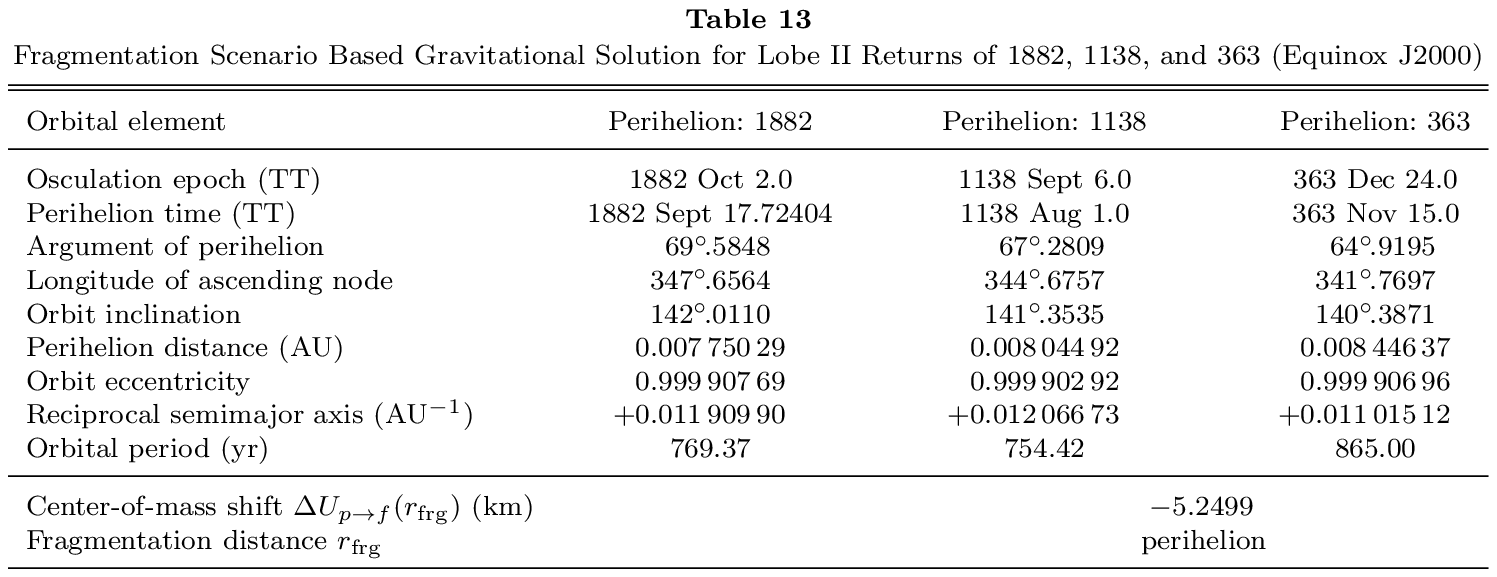}}}
\vspace{-18.9cm}
\end{table*}

Given that the normalized volume of a truncated spheroid, $\Re$, varies with
the distance from the apex as indicated by (28), we have for the parent of the
1882/1965 pair, separating from the minor fragment in the course of the perihelion
event, \mbox{$\Re(\gamma) = 3\gamma^2(1 \!-\! \frac{2}{3}\gamma)$},{\vspace{-0.05cm}}
as is schematically shown in the top panel of Figure~6.  Following (31),
\begin{equation}
\gamma = \frac{60}{\cal D} = \frac{\sqrt{\Re(\gamma)}}{2 \cos \left\{ {\textstyle
 \frac{1}{3}} \arccos \left[ -\sqrt{\Re(\gamma)} \right] \right\}}. 
\end{equation}
Since the coordinate of the center of mass of the truncated spheroid has to halve
its volume, we equate
\begin{equation}
\Re(\gamma^\prime) = {\textstyle \frac{1}{2}} \Re(\gamma). 
\end{equation}
On the other hand, from (33) and Figure~6 it follows that
\begin{equation}
\gamma^\prime = \frac{1}{2} - \frac{\Delta U_{f \rightarrow p}(q)}{\cal D}
 = \frac{1}{2} - \frac{5.244}{\cal D}. 
\end{equation}
The diameter $\cal D$ in (34) and (36) is given in km.  Combining (34) through
(36), we find that the ratio
\begin{equation}
K = \frac{{\textstyle \frac{1}{2}} \!-\! \gamma^\prime}{\gamma} = 0.0875,
\end{equation}
which allows one to determine the normalized volume of the 1882/1965 parent
shaped as a truncated spheroid, \mbox{$\Re(\gamma) = \Re$}, given that
\begin{eqnarray}
K & = & \Re^{-\frac{1}{2}} \cos \left[ {\textstyle \frac{1}{3}} \arccos (-\sqrt{\Re})
 \right] \nonumber \\[-0.1cm]
 & & \times \! \left\{ 1 - \frac{\sqrt{2 \Re}}{2 \cos \left[
 {\textstyle \frac{1}{3}} \arccos \left( - {\textstyle \frac{1}{2}} \sqrt{2 \Re}
 \right) \right]} \right\}. 
\end{eqnarray}
The normalized length $\gamma$ follows from (34) and so does $\cal D$.  Numerically,
\begin{eqnarray}
\Re(\gamma)   & = & 0.81142, \nonumber \\
\gamma        & = & 0.72225, \nonumber \\
\gamma^\prime & = & 0.43680, \nonumber \\
{\cal D}      & = & 83.07 \; {\rm km.} 
\end{eqnarray}
In addition, the diameter of the minor fragment separating at perihelion is determined
to equal about 23~km and the distance of its center of mass from the center of Lobe~II
25.8~km.  Following the separation, this fragment ended up in an orbit with a period of
about 1800~yr.

\begin{figure}[b]
\vspace{-8.4cm}
\hspace{2.87cm}
\centerline{
\scalebox{0.78}{
\includegraphics{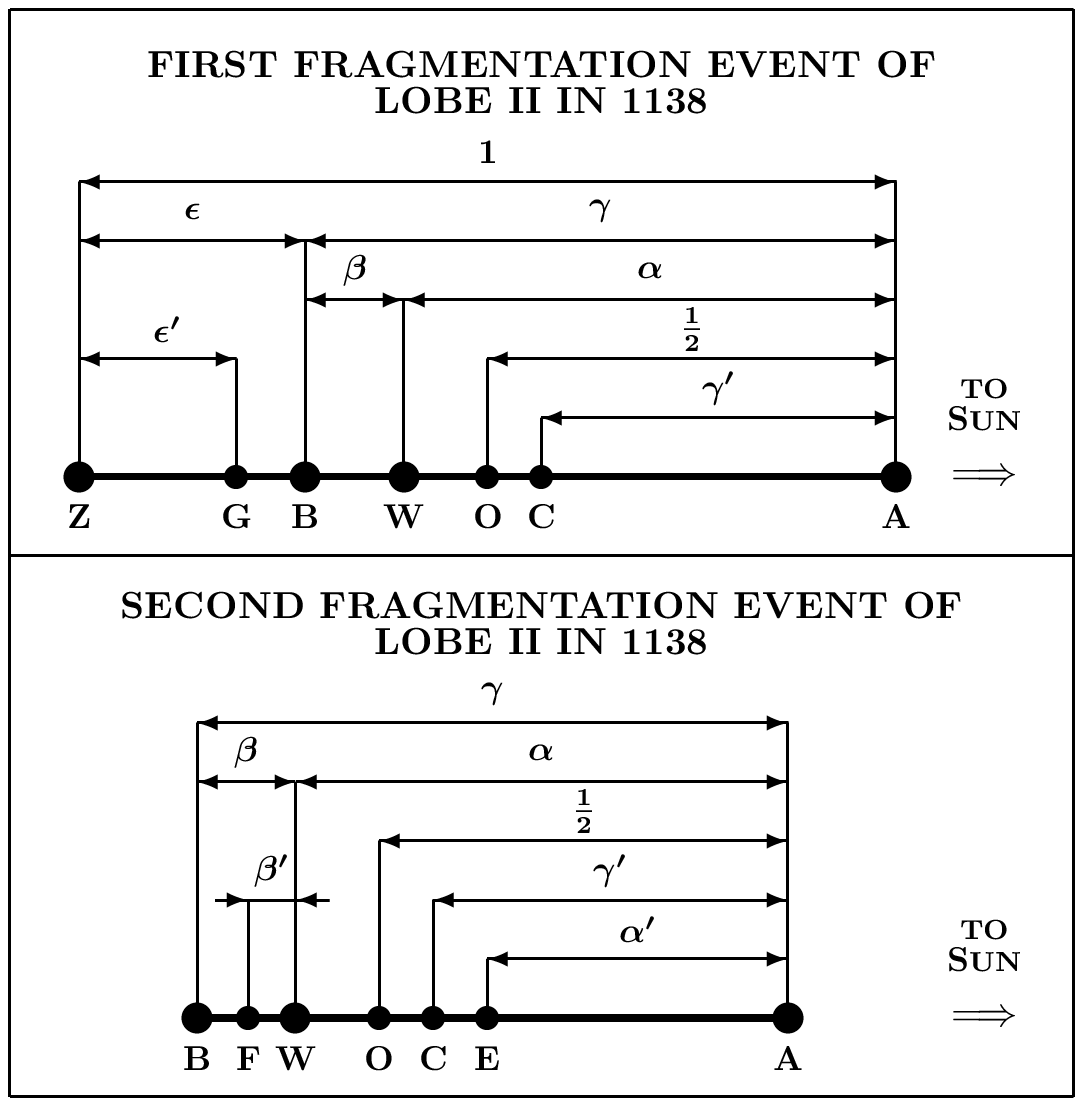}}}
\vspace{-5.65cm}
\caption{Relative sizes and center-of-mass positions of the Lobe II parent and its
fragments in the two fragmentation events in 1138.  In the first event (top), the
long axis of the prolate-spheroidal parent is unity, the parent of the 1882 sungrazer
($\alpha$) and Ikeya-Seki ($\beta$) is $\gamma$ (being truncated at {\bf B}, the size
of the minor fragment is $\epsilon$. The position of the parent's center of mass is
at {\bf O}, the center of mass of the parent of the 1882 sungrazer and Ikeya-Seki is
at {\bf C}, and the center of mass of the minor fragment is at {\bf G}.  In the second
fragmentation event, Ikeya-Seki separated from the 1882 comet (bottom); their centers
of mass are at {\bf F} and {\bf E}, respectively.{\vspace{-0.1cm}}}
\end{figure}

The major fragment, which was to split into the 1882 comet and Ikeya-Seki, continued to
orbit the Sun as a single body, until (32) was satisfied.  To determine the distance,
$\Psi$, between the centers of mass of the 1882 sungrazer and Ikeya-Seki, the following
conditions apply:
\begin{eqnarray}
\Re(\alpha^\prime) & = & {\textstyle \frac{1}{2}} \Re(\alpha), \nonumber \\
\Re(\beta^\prime \!+\! \alpha) & = & {\textstyle \frac{1}{2}} \left[ \Re(\alpha) \!+\!
 \Re(\gamma) \right], \nonumber \\
\Psi & = & \left[ (\beta^\prime \!+\! \alpha) \!-\! \alpha^\prime \right] {\cal D}.
\end{eqnarray}
Since \mbox{$\alpha = 0.60190$}, we obtain \mbox{$\Re(\alpha) = 0.65073$}, from the
first equation of (40) \mbox{$\alpha^\prime = 0.38135$}, from the second equation
\mbox{$\beta^\prime = 0.05756$}, and from the third equation \mbox{$\Psi = 23.1$ km}.
Inserted into (32), the required fragmentation distance is 
\mbox{$r_{\rm frg} = 2.40 \, q = 0.0193 \; {\rm AU} = 4.15$ {\Rsun}}.

\begin{table*}[t]
\vspace{-4.2cm}
\hspace{0.53cm}
\centerline{
\scalebox{1}{
\includegraphics{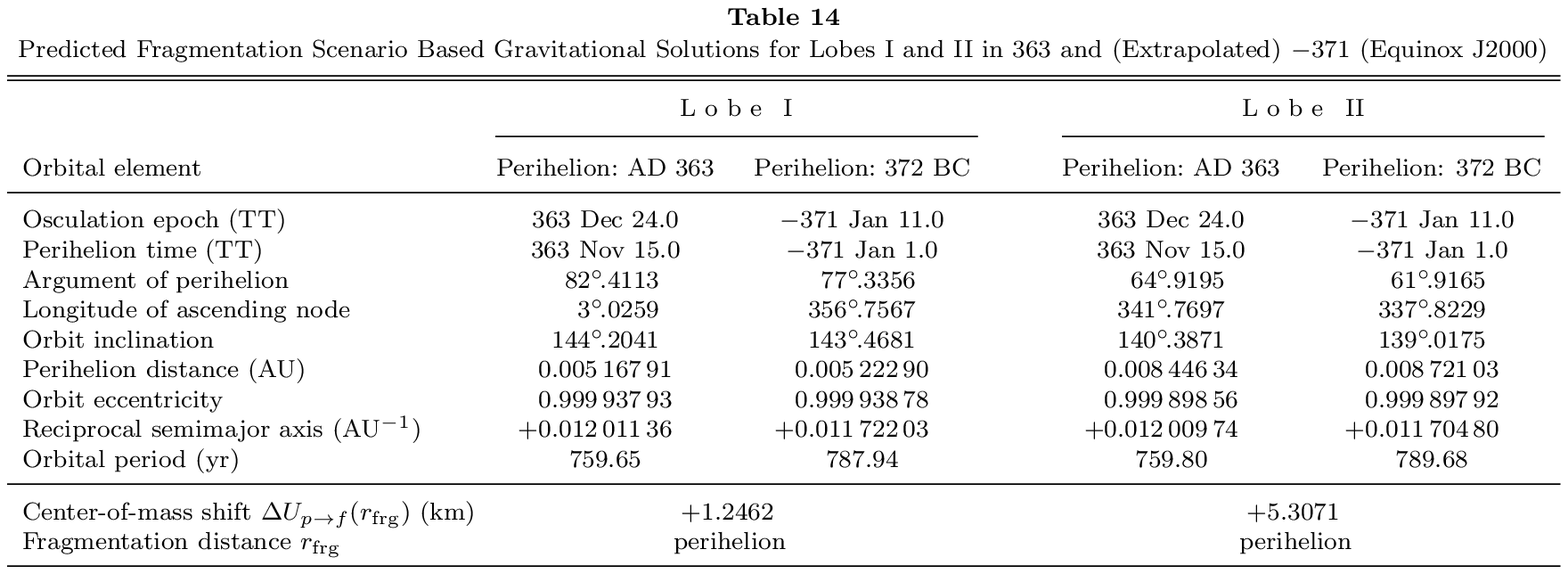}}}
\vspace{-18.15cm}
\end{table*}

The reader may notice that the procedure is approximate, one reason being that
the condition (32), which applies to the center of mass of the 1882~sungrazer, was
assigned instead to the center of mass of the 1882/1965 pair.  Since the distance
between them amounted to \mbox{$(\gamma^\prime \!-\! \alpha^\prime) {\cal D} = 4.61$
km}, the center-of-mass shift of the parent at perihelion was only \mbox{$5.25 \!-\!
4.61 = 0.64$ km}, so that \mbox{$K = 0.0107$}, \mbox{$\Re(\gamma) = 0.9712$}, and
\mbox{$\gamma = 0.89853$}.  The parent's size then equaled merely $\sim$67~km.
These improved results (and further iterations) have of course no effect on our
simulation of the motion of Lobe~II before 1138.  

%
\section{Kreutz Sungrazers at Perihelion in AD 363,\\Aphelion Breakup of
 Contact-Binary Parent,\\and Its Pre-Fragmentation Orbit} 
In the context of the contact-binary model, it was postulated that the largest surviving
mass of Lobe~I appeared most recently as the Great March Comet of 1843 and, before, as
the Great Comet of 1106.  Similarly, the largest surviving mass of Lobe~II appeared most
recently as the Great September Comet of 1882 and, before, as the Chinese comet of 1138.
The arrival of Lobe~II lagged behind Lobe~I {\vspace{-0.04cm}}by 39$\frac{1}{2}$~yr in
the 19th century and by 32$\frac{1}{2}$~yr in the 12th century.

\subsection{Kreutz Sungrazers at Perihelion in AD~363} 
The spectacle of brilliant daylight comets in AD~363, recorded for posterity by
Ammianus Marcellinus, a Roman historian,\footnote{See Paper~1 for details and useful
references on this subject.} was simulated in this feasibility study as the first
perihelion appearance of the Kreutz system after the progenitor's fragmentation at
large heliocentric distance.  It differed from the subsequent perihelion returns (in
the 12th and 19th centuries) in that {\it both\/} Lobe~I and Lobe~II contributed
virtually simultaneously to the event.  We have shown that the perihelion returns
\mbox{1843--1106--363} and \mbox{1882--1138--363} could both be linked in the
framework of either the sublimation scenario or the fragmentation scenario, even
though the high value of the critical nongravitational parameter for Lobe~II made
the veracity of the sublimation scenario suspect.

Since fragmentation events at large heliocentric distance are known to have a fairly
minor effect on the perihelion time, we used the fragmentation-scenario tools to
derive the center-of-mass shifts $\Delta U$ in an effort to determine the preliminary
sets of orbital elements for Lobe~I and Lobe~II in the period of time between
363~November~15 and the previous perihelion on -371~January~1.  The results are
presented in Table~14.  The two orbits for the progenitor in 372~BC are of course
artificial; the ultimate orbit is going to be the product of simulation of the
process of near-aphelion fragmentation (Sections 7.2 and 7.3).

Comparison of Tables 10 and 14 for Lobe I and Tables~13 and 14 for Lobe~II reveals
a major difference between the 12th century returns on the one hand and AD~363 on
the other hand:\ the center of mass of the {\it major\/} fragment was located
sunward of the parent object in 1106 and 1138 (\mbox{$\Delta U_{p \rightarrow f}
< 0$}; Scenario~I in Figure~2),{\nopagebreak} but antisunward in AD~363 (\mbox{$\Delta
U_{p \rightarrow f} > 0$}; Scenario~II in Figure~2).  This difference has an important
ramification.  Scenario~I implies that a typical {\it minor\/} fragment, or most
minor fragments, should have the center(s) of mass on the antisunward side of the
parent's center of mass, should end up in highly elongated orbits, with the periods
much longer than \mbox{700--800}~yr, and should not return to perihelion until
some time in the distant future.  By contrast, Scenario~II implies that minor
fragments have typically the centers of mass on the sunward side of the parent's
center of mass, ending up in orbits with the periods shorter, and potentially
much shorter, than \mbox{700--800}~yr.  Such fragments might have become sungrazers
in their own right during the centuries following the year 363, starting as early
as the 6th century.

To inspect this issue quantitatively, we employ a relationship analogous to (25),
expressing the orbital period of a fragment, $P_{\rm frg}$, as a function of the
orbital period of the parent, $P_{\rm par}$, rather than vice versa:
\begin{equation}
P_{\rm frg} = P_{\rm par} \left(\! 1 - \frac{2 \Delta U_{p \rightarrow f}(r_{\rm
 frg})}{r_{\rm frg}^2} P_{\rm par}^{\frac{2}{3}} \!\right)^{\!\!-\frac{3}{2}} \!\!\! .
\end{equation}

Suppose, for the sake of argument, that Lobes~I and II were in AD~363 nearly 100~km
across each and that a reasonable upper limit on the center-of-mass shift of a
fragment was about 45~km.  Table~15 presents the orbital periods that the fragments
acquire as a function of (i)~the heliocentric distance at fragmentation ($q$ being
again the perihelion distance) and (ii)~the center-of-mass shift relative to the
parent comet in AD~363 (see Table~14).  It is astonishing to see that in an extreme
case the orbital period could be as short as one-fifth the parent's orbital period
($\sim$760~yr) for Lobe~I and less than one-half of it for Lobe~II.

In a second experiment, we picked up four of the potential Kreutz sungrazers from
Hasegawa \& Nakano's (2001) list in the 7th through 11th century and computed the
center-of-mass shifts that would be required if the comets were fragments of Lobe~I or
Lobe~II released at the 363 perihelion.  It is seen that their orbits could readily
be accommodated this way, with the exception of the comet of 607, which could not
be a fragment of Lobe~II.

From (41) it is obvious that the relation between the orbital periods of the parent
and a fragment is governed by the ratio of $\Delta U/r_{\rm frg}^2$.  The effect drops
very{\vspace{-0.05cm}} rapidly with increasing distance from the Sun but also with
decreasing size of the object and its orbit.  In practice it has no meaningful
application to fragmenting comets other than sungrazers.   

\subsection{Line of Intersection of Lobes' Orbital Planes and\\Positions of Lobes at
 Fragmentation Time} 

In order to successfully simulate the separation of the two lobes from the
contact-binary progenitor near aphelion, three conditions have to be satisfied in
the barycentric system:\ (i)~the point of separation must be located on the line
of intersection of the lobes' orbital planes; (ii)~the lobes must be at the same
barycentric distance; and (iii)~they must reach the point simultaneously.  Given
that the tabulated orbits are available at an osculation epoch near perihelion,
it is essential that before the first of the three conditions is addressed, the
orbits be integrated to a new osculation epoch chosen near aphelion to make the
simulation procedure tractable.

%
%

The line of intersection is obviously close to the lines of apsides of the two
lobes, as their longitudes and latitudes of perihelion in AD~363 were very similar,
282$^\circ\!$.35 and +35$^\circ\!$.43 for Lobe~I and 283$^\circ\!$.05 and
+35$^\circ\!$.27 for Lobe~II.  To find the line of intersection is straightforward.
In the ecliptic coordinate system, the unit vector {\boldmath $R$} normal to the
orbital plane of a comet is a function of the longitude of the ascending node,
$\Omega$, and the orbital inclination, $i$:
\begin{equation}
\mbox{\boldmath $R$} = \left( \!\! \begin{array}{c}
R_x \\
R_y \\
R_z \end{array}
\!\! \right) = \left( \!\! \begin{array}{c}
\sin \Omega \sin i \\
-\cos \Omega \sin i \\
\cos i \end{array}
\!\! \right). 
\end{equation}
Referring to the vector normal to the orbit of Lobe~I as \mbox{{\boldmath
$R$}({\scriptsize \bf I}) = [$R_x$({\scriptsize I}), $R_y$({\scriptsize I}),
$R_z$({\scriptsize I})]}, and to the orbit of Lobe II as \mbox{{\boldmath
$R$}({\scriptsize \bf II}) = [$R_x$({\scriptsize II}), $R_y$({\scriptsize
II}), $R_z$({\scriptsize II})]}, a vector along the line of intersection,
\mbox{{\boldmath $L$} = $(L_x, L_y, L_z)$} is given by their vector product:
\begin{equation}
\mbox{\boldmath $L$} = \mbox{\boldmath $R({\scriptstyle \bf I})$}
 \mbox{\large \boldmath $\times$} \mbox{\boldmath $R({\scriptstyle \bf II})$},
\end{equation}
so that
\begin{displaymath}
L_x = \left| \begin{array}{cc}
R_y(\mbox{\scriptsize I})  & R_z(\mbox{\scriptsize I}) \\
R_y(\mbox{\scriptsize II}) & R_z(\mbox{\scriptsize II})
\end{array}
\right|,
\end{displaymath}

\vspace{-0.25cm}
\begin{displaymath}
L_y = \left| \begin{array}{cc}
R_z(\mbox{\scriptsize I})  & R_x(\mbox{\scriptsize I}) \\
R_z(\mbox{\scriptsize II}) & R_x(\mbox{\scriptsize II})
\end{array}
\right|,
\end{displaymath}

\vspace{-0.1cm}
\begin{equation}
L_z = \left| \begin{array}{cc}
R_x(\mbox{\scriptsize I})  & R_y(\mbox{\scriptsize I}) \\
R_x(\mbox{\scriptsize II}) & R_y(\mbox{\scriptsize II})
\end{array}
\right|. 
\end{equation}
Inserted into the equations for the position, one can use any two coordinates
\begin{table}[t]
\vspace{-4.2cm}
\hspace{5.21cm}
\centerline{
\scalebox{1}{
\includegraphics{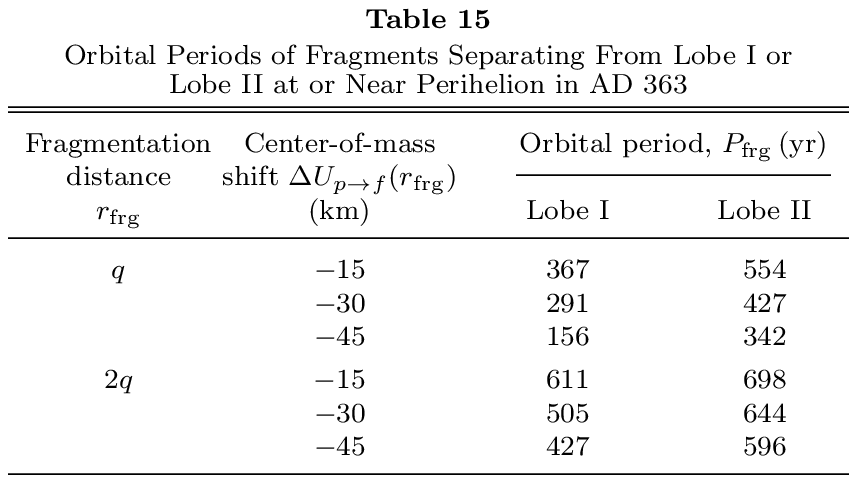}}}
\vspace{-20cm}
\end{table}
to calculate the true anomaly, $u$, of either lobe's position on the line of
intersection.  For Lobe~I, for example, \mbox{$u = u_{\rm I}$} and
\begin{equation}
\left( \!\! \begin{array}{c}
 L_x \\
 L_y \\
 L_z
\end{array}
\!\! \right) = c_0 \left( \!\! \begin{array}{cc}
P_x({\scriptstyle \rm I}) & Q_x({\scriptstyle \rm I}) \\
P_y({\scriptstyle \rm I}) & Q_y({\scriptstyle \rm I}) \\
P_z({\scriptstyle \rm I}) & Q_z({\scriptstyle \rm I})
\end{array}
\!\! \right) \!\times\! \left( \!\! \begin{array}{c}
\cos u_{\rm I} \\
\sin u_{\rm I}
\end{array}
\!\! \right) \!, 
\end{equation}
where $c_0$ is a constant and $P_x$, \ldots, $Q_z$ are the components, in the
ecliptic coordinate system, of the unit vectors:\ {\boldmath $P$} pointing
to perihelion and {\boldmath $Q$} 90$^\circ$ ahead of {\boldmath $P$} in the
orbital plane in the direction of motion.  The components are functions of
the angular orbital elements:
\begin{equation}
\left( \!\! \begin{array}{cc}
P_x & \!\!Q_x \\
P_y & \!\!Q_y \\
P_z & \!\!Q_z
\end{array}
\!\! \right) \!\!\:=\!\!\: \left( \!\! \begin{array}{ccc}
\cos \Omega & \!-\sin \Omega & \!0 \\
\sin \Omega & \!\cos \Omega  & \!0 \\
     0      & \!     0       & \!1
\end{array}
\!\! \right) \!\!\:\times \! \left( \!\! \begin{array}{cc}
\cos \omega        & \!-\sin \omega \\
\sin \omega \cos i & \!\cos \omega \cos i \\
\sin \omega \sin i & \!\cos \omega \sin i
\end{array}
\!\! \right) \!\!\: . 
\end{equation}
Combining in (45), the expressions for, say, $L_x$ and $L_y$, we get for the true
anomaly of the line of intersection in the orbit of Lobe~I
\begin{equation}
\tan u_{\rm I} =  \frac{L_x P_y({\scriptstyle \rm I}) - L_y P_x({\scriptstyle \rm
 I})}{L_y Q_x({\scriptstyle \rm I}) - L_x Q_y({\scriptstyle \rm I})} \,. 
\end{equation}
Since the position is near aphelion, one has to add 180$^\circ$ to the value obtained
from (47) when the tangent is negative (pre-aphelion location), or subtract
180$^\circ$ when it is positive (post-aphelion location).  This true anomaly gives
the barycentric distance and time of Lobe~I on the line of intersection.  The
same procedure provides the same data for Lobe~II.

\begin{table}[b]
\vspace{-3.6cm}
\hspace{5.21cm}
\centerline{
\scalebox{1}{
\includegraphics{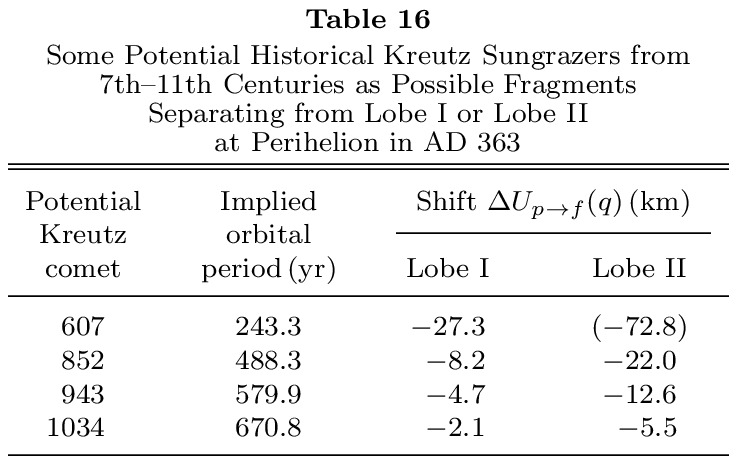}}}
\vspace{-21.1cm}
\end{table}

We next used the elements from Table~14 to determine the position of the line of
intersection of the orbital planes of the lobes.  We first integrated their motions from
perihelion in AD~363 to the previous aphelion and converted the heliocentric system
to the barycentric system.  For an osculation epoch at the end of 5~BC, the two orbits
and the gap between them are shown in Table~17.  The ecliptic components of the unit
vectors {\boldmath $R$}({\scriptsize \bf I}), {\boldmath $R$}({\scriptsize \bf II}),
{\boldmath $L$}, and the lobes' true anomalies, $u_{\rm I}$, $u_{\rm II}$,
and barycentric distances, $r_{\rm I}$, $r_{\rm II}$, are summarized in Table~18.
The angle between the orbital planes was 13$^\circ\!$.23, while the line of
intersection made angles of 0$^\circ\!$.2 and 0$^\circ\!$.4 with the lines of
apsides of Lobe~I and Lobe~II, respectively.

\begin{table*}
\vspace{-4.2cm}
\hspace{0.75cm}
\centerline{
\scalebox{1}{
\includegraphics{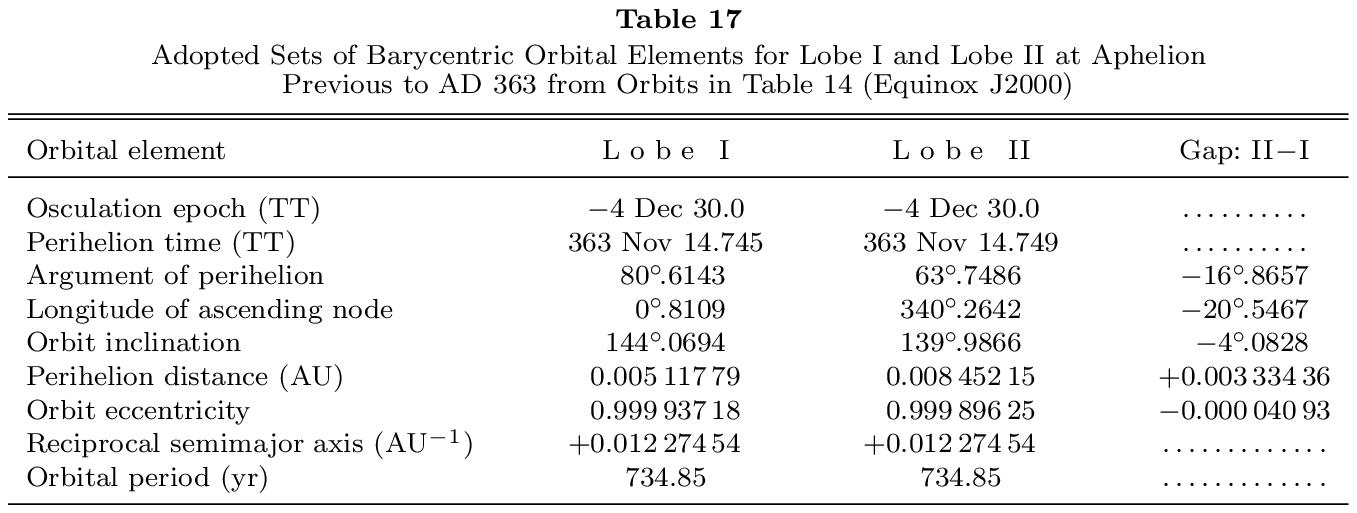}}}
\vspace{-19.5cm}
\end{table*}

Under rigorous conditions, it would next be a matter of iteration to bring first the
barycentric distances and then the times into harmony.  One possible approach would
be to keep these two quantities constant for one lobe, to select this time as the
osculation epoch, and to iterate these quantities for the other lobe:\ the barycentric
distance by varying $\Delta U$ of the pre-363 orbit, the time by shifting the perihelion
time in AD~363.

For two reasons we did not pursue this line of attack.  One reason was the tiny angles
between the positions of the line of intersection of the lobes' orbital planes and the
lines of apsides, which indicated --- given the approximations involved --- that,
angularly, the point of fragmentation was virtually indistinguishable from aphelion.
The other reason was that, in the context of the proposed contact-binary model, both
lobes were believed to have undergone, before reaching their perihelion in AD~363,
events of secondary fragmentation, which were bound to modify, however slightly, the
positions of their orbital planes and thereby introduce additional uncertainties into
the orbit computations.  Instead, we {\it assumed\/} that the breakup of the progenitor
into the two lobes occurred at aphelion and that Lobe~I began to move in an orbit
described by the set of elements listed for it in Table~17.  This provided us with an
opportunity to simulate effects by the separation velocity (Section~7.3).
 
\subsection{Orbit of Progenitor Prior to Fragmentation and\\Separation Velocities of
 the Lobes} 
Let at the adopted fragmentation time of $-$4~Dec~30, \mbox{{\boldmath $U_{\bf
I}$} = $(x_{\rm I}, y_{\rm I}, z_{\rm I})$} and \mbox{{\boldmath $V_{\bf I}$} =
$(\dot{x}_{\rm I}, \dot{y}_{\rm I}, \dot{z}_{\rm I})$} be, respectively, the
orbital-position and orbital-velocity vectors of Lobe~I; and let \mbox{{\boldmath
$U_{\bf II}$} = $(x_{\rm II}, y_{\rm II}, z_{\rm II})$} and \mbox{{\boldmath
$V_{\bf II}$} = $(\dot{x}_{\rm II}, \dot{y}_{\rm II}, \dot{z}_{\rm II})$} be the
orbital-position and orbital-velocity vectors of Lobe II.  Whereas the two position
vectors were equal (neglecting the lobes' dimensions), the velocity vectors were not.
Following Figure~5, the progenitor's orbit was fully described by the position
vector {\boldmath $U_{\rm par}$} and the velocity vector {\boldmath $V_{\rm par}$},
an average of the velocity vectors of the lobes.  In the ecliptic coordinate system,
\begin{eqnarray}
\mbox{\boldmath $U_{\bf par}$} & = & \mbox{\boldmath $U_{\bf I}$} = \mbox{\boldmath
 $U_{\scriptsize \bf II}$} = (x_{\rm par}, y_{\rm par}, z_{\rm par}) \nonumber \\
\mbox{\boldmath $V_{\bf par}$} & = & {\textstyle \frac{1}{2}} \left( \mbox{\boldmath
 $V_{\scriptsize \bf I} + V_{\bf II}$} \right) = (\dot{x}_{\rm par}, \dot{y}_{\rm par},
 \dot{z}_{\rm par}). 
\end{eqnarray}
By integrating these position and velocity vectors from the osculation epoch back in
time, the progenitor's orbit at the previous perihelion could readily be determined.

The derivation of the progenitor's velocity by averaging the lobes' velocities
implied that the fragments parted at equal rates in the exactly opposite
directions, as depicted in Figure~5.  The differences between the lobes' orbital
velocities and the velocity of the progenitor were the lobes' {\it separation
velocities\/}:
\begin{eqnarray}
\mbox{\boldmath $V_{\bf sep}({\scriptstyle \bf I})$} & = & \mbox{\boldmath
 $V_{\bf sep}$} = (\dot{x}_{\rm I} \!-\! \dot{x}_{\rm par}, \dot{y}_{\rm I}
 \!-\! \dot{y}_{\rm par}, \dot{z}_{\rm I} \!-\! \dot{z}_{\rm par}) \nonumber \\
\mbox{\boldmath $V_{\bf sep}({\scriptstyle \bf II})$} & = & \mbox{\boldmath
 $-V_{\bf sep}$} = (\dot{x}_{\rm II} \!-\! \dot{x}_{\rm par}, \dot{y}_{\rm II}
 \!-\! \dot{y}_{\rm par}, \dot{z}_{\rm II} \!-\! \dot{z}_{\rm par}). \nonumber \\[-0.1cm]
 & & 
\end{eqnarray}

Given that the investigation of the separation-velocity vector was conducted in the
coordinate system~of~RTN (radial-transverse-normal), rotating with the comet along
its orbit, it was first necessary to convert the vector's components into the ecliptic
system.  Writing $V_{\rm R}$ for the radial, $V_{\rm T}$ for the transverse, and
$V_{\rm N}$ for the normal component of the separation velocity of Lobe~I relative
to the parent and \mbox{$\dot{x}_{\rm sep} = \dot{x}_{\rm I} \!-\! \dot{x}_{\rm
par}$}, etc., for its ecliptic components, the transformation formulas were:
\begin{displaymath}
\left( \!\! \begin{array}{c}
\dot{x}_{\rm sep} \\
\dot{y}_{\rm sep} \\
\dot{z}_{\rm sep}
\end{array}
\!\! \right) = \left( \!\! \begin{array}{ccc}
P_x({\scriptstyle \rm I}) & Q_x({\scriptstyle \rm I}) & R_x({\scriptstyle \rm I}) \\
P_y({\scriptstyle \rm I}) & Q_y({\scriptstyle \rm I}) & R_y({\scriptstyle \rm I}) \\
P_z({\scriptstyle \rm I}) & Q_z({\scriptstyle \rm I}) & R_z({\scriptstyle \rm I})
\end{array}
\!\! \right) \\[-0.1cm]
\end{displaymath}
\begin{equation}
{\hspace{2.85cm}} \times \!\! \left( \!\! \begin{array}{ccc}
\cos u_{\rm I} & -\sin u_{\rm I} & \! 0 \\
\sin u_{\rm I} &  \cos u_{\rm I} & \! 0 \\
      0        &        0        & \! 1
\end{array}
\! \right) \!\! \times \!\! \left( \!\! \begin{array}{c}
V_{\rm R} \\
V_{\rm T} \\
V_{\rm N}
\end{array}
\!\! \right) \!. 
\end{equation}
From the separation velocity $\dot{x}_{\rm sep}$, etc., of Lobe~I relative to the
parent, the components of the parent's orbital velocity at the time of fragmentation
equaled \mbox{$\dot{x}_{\rm par} = \dot{x}_{\rm I} \!-\! \dot{x}_{\rm sep}$}, etc.,
and the components of the orbital velocity of Lobe~II equaled \mbox{$\dot{x}_{\rm II}
= \dot{x}_{\rm I} \!-\! 2 \dot{x}_{\rm sep}$}, etc.

\begin{table}[b]
\vspace{-3.8cm}
\hspace{5.25cm}
\centerline{
\scalebox{1}{
\includegraphics{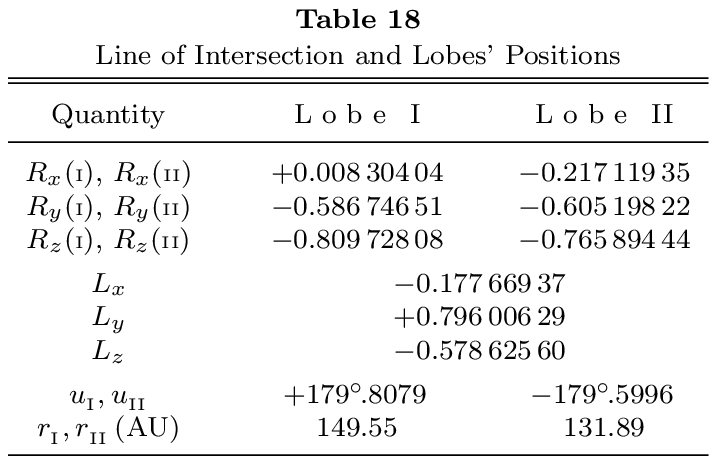}}}
\vspace{-21.1cm}
\end{table}

We were now ready to simulate the orbital conditions at the {\it primary fragmentation
event that~gave~birth to the Kreutz system\/}:\ At aphelion, on $-$4 December~30,
Aristotle's comet, modeled as a bilobed rotating object, was to split into two
fragments (lobes) that were to end up in the orbits presented in Tables 14 and
17.\footnote{A possible independent separation of the neck was not simulated in
this exercise.}  The goal was to fit, as closely as possible, the orbits of Lobes~I
and II in AD~363 and 5~BC as products of the contact-binary parent's pre-fragmentation
orbit by determining the differential momenta (expressed in terms of the separation
velocity vectors) that Lobes~I and II acquired (in the opposite directions) at the
breakup.

In the case of the parent's orbit being unknown, the problem could equivalently be
addressed by starting from the orbit of one lobe --- say, Lobe~I --- and by searching
for a separation velocity vector {\boldmath $V_{\bf sep}$} [see (49)], such that
its subtraction from the lobe's orbital-velocity vector determined the parent's orbit,
and its another subtraction approximated the orbit of the other lobe --- Lobe~II ---
at perihelion in AD~363 (Table~14).  When successful, this exercise demonstrates that
the forthright splitting of the progenitor into the two lobes is compatible with their
orbital properties.

To assure that the fragments arrived at their perihelion in AD~363 in fairly
tight formation (within days of one another, thereby seen simultaneously), we set
\mbox{$V_{\rm R} = 0$}, in line with Papers~1 and 2.  We began to address the task
by adopting for the other two separation velocity components of Lobe~I the values
from Table~6 of Paper~2, \mbox{$V_{\rm T} = -1.86$ m s$^{-1}$} and \mbox{$V_{\rm N}
= -1.80$ m s$^{-1}$}.  These velocities had of course been derived from the
19th-century osculating orbits of the 1843 and 1882 sungrazers, thereby ignoring
all time dependent effects, including the two-millenia long, continuous planetary
perturbations of the two sungrazers and their precursors.

Contrary to the arguments associated with the quest to explain the orbital
differences among Kreutz sungrazers (and especially the discrete populations) by a
cumulative effect of the planetary perturbations over longer periods of time, our
results show that the gaps between the orbital elements of the two lobes, or
equivaently, the 1843 and 1882 sungrazers, were being {\it contracted\/} with time.
Indeed, while the difference between the inclinations in the 4th century was about
4$^\circ$, it was reduced to less than 2$^\circ\!$.5 in the 19th century; similarly,
in the same period of time the difference in the longitude of the ascending node
dropped from more than 21$^\circ$ to 16$^\circ$, and the difference in the perihelion
distance from 0.7\,{\Rsun} to 0.5\,{\Rsun}.  These systematic trends manifestly
demonstrate that the planetary perturbations could not possibly account for the
existence of the Kreutz system's populations.

The diminishing differences between the orbital elements of the two giant sungrazers
intervened in our efforts to simulate the initial magnitude of the separation velocity
between the lobes in that the numbers based on the 19th century orbits turned out to
have been much too low.  It was necessary to iterate the orbital solution, with the
result showing eventually that the transverse component of the separation velocity
was about 26~percent higher and the normal component 54~percent higher than estimated
in Paper~2.

The outcome of our computer simulations of the birth of the two most massive fragments
of the Kreutz system as products of the bilobed progenitor's splitting is presented
in Tables~19 and 20.  Although we made no effort to strictly optimize the solution,
we found that at the adopted fragmentation time of $-$4~December~30.0~TT, essentially
the point of aphelion, either lobe separated with a velocity, relative to the
progenitor's center of mass, of
\begin{equation}
| \mbox{\boldmath $V$}_{\bf sep} | = 3.63 \; {\rm m \: s}^{-1}. 
\end{equation}
The transverse and normal components of the separation velocity of Lobe~I were,
respectively, \mbox{$V_{\rm T}({\scriptstyle \rm I}) = -2.35$\,m\,s$^{-1}$}
and \mbox{$V_{\rm N}({\scriptstyle \rm I}) = -2.77$\,m\,s$^{-1}$};
the radial component, as already noted, was forced to be nil.  Thanks
to this condition, Lobe~II --- whose transverse and normal
components of the separation velocity were of the same magnitude, but of the
opposite sign --- reached its perihelion in AD~363 only about 4.2~days after Lobe~I.
If the radial component were not nil --- or close to nil --- the two lobes would
have arrived at perihelion weeks, possibly months, apart and they would not have
been perceived as an essentially single event, contrary to the impression that one
gets from the Roman historian's account.

We note that even though the gaps between the orbital elements of the two lobes
(Table~17) were greater than between the osculating orbits of the 1843 and 1882
sungrazers, the separation velocity that accounts for the gaps is still low.
Interpreting it as the progenitor's rotation velocity, we could estimate the
compatible rotation period.  If either lobe is assumed to be about 50~km in
diameter, so is the distance between the centers of mass.  If the circumference,
\mbox{$2\pi\!\cdot\!50 = 314$ km}, is described in one day, this equals
3.63~m\,s$^{-1}$, derived in (51).  The rotation period of 24~hr is rather on
the long side.

The 4-day gap between the perihelion times of the two lobes is no surprise.  In
Paper~2 we anticipated a gap of 2.9~days with a lower separation velocity, showing
that the gap is an effect of the transverse component.  The standard time of
363~Nov~15.0~TT that we have used throughout this paper in search of the various
orbital solutions was essentially a default value, dictated by the narrative on the
observation of the daylight comets.  We could have used another date nearby with the
virtually identical outcome.  Indeed, a change in Lobe~I's orbital period of one
month between AD~363 and 1106 is triggered by a radial shift in the center-of-mass
position of only 1.7~meters at perihelion, a trivial effect.

Either lobe is believed to have undergone events of secondary fragmentation not long
after their birth.  According to the model in Paper~2, Lobe~I was likely to have
produced the precursors of Populations~Pre-I and Pe, respectively, whereas Lobe~II was
apparently the parent to Fragment~IIa and Precursor~IIa$^{\mbox{\boldmath $\!\ast$}}$.
In the light of these events, the agreement in Table~19 between the orbital elements
of Lobe~II at the AD~363 and 5~BC epochs derived from the back integration of the
precursor orbit of the Great September Comet of 1882 on the one hand and from the
precursor orbit of the Great March Comet of 1843 and the fragmentation parameters
of the progenitor on the other hand is astonishingly good.  The high degree of
correspondence demonstrates in the very least the feasibility ---  and in fact
fundamental veracity --- of the proposed fragmentation hypothesis.

\begin{table*}
\vspace{-4.2cm}
\hspace{0.53cm}
\centerline{
\scalebox{1}{
\includegraphics{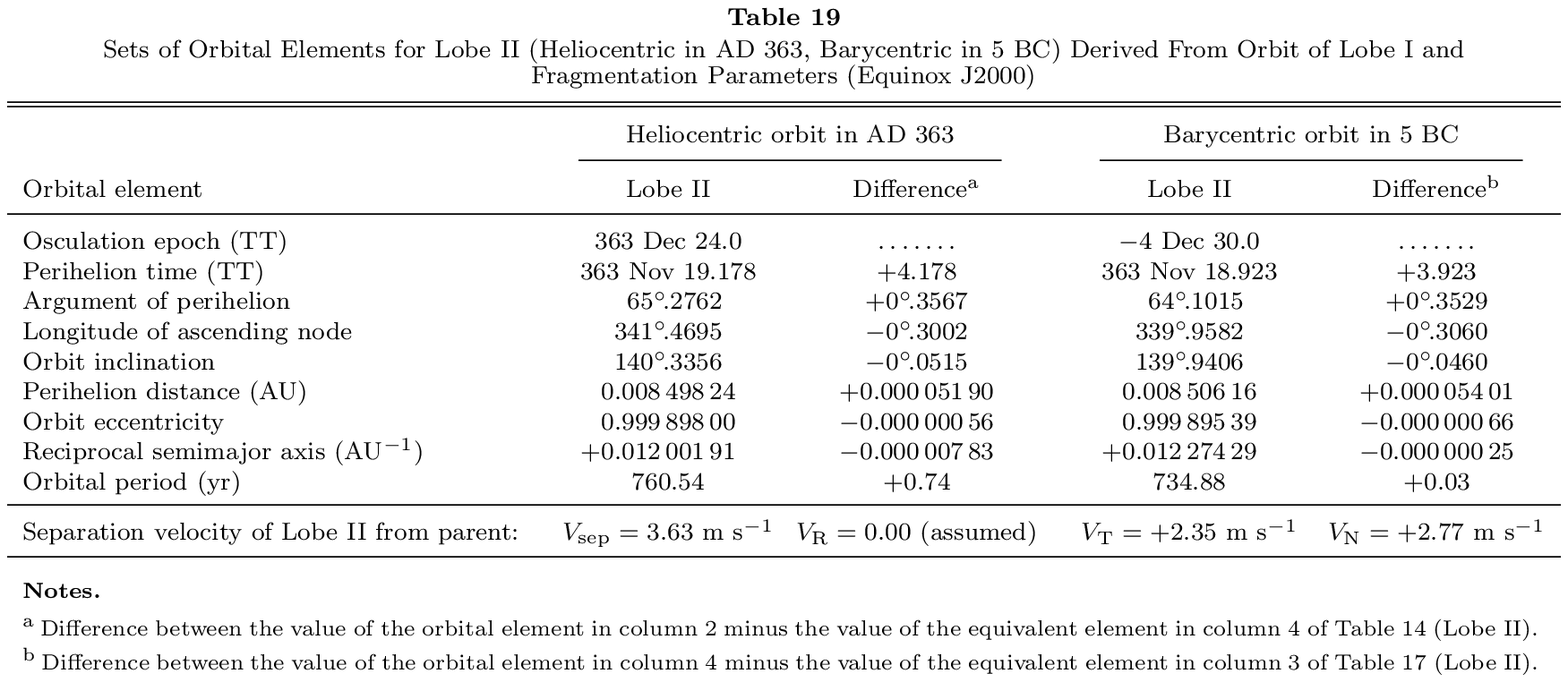}}}
\vspace{-17.25cm}
\end{table*}

Another major conclusion is that Aristotle's comet, which according to historical
accounts appeared in the winter at the end of 373~BC or the beginning of 372~BC,
fits our orbit simulation scheme as the putative contact-binary progenitor.  We no
longer see any reason to doubt this object's significance in regard to the Kreutz
system.  We integrated the orbit of the progenitor back in time over two revolutions
about the Sun, showing the result in Table~20.  It is perhaps somewhat surprising
to see the orbit of Aristotle's comet resembling the orbit of the Great September
Comet of 1882 in both the angular elements and the period, with only the perihelion
distance deviating a little toward that of the 1843 sungrazer.

Several accounts of historical comets in Hasegawa's (1980) catalog between 1141~BC
and 1201~BC (recorded in Babylonia, Greece, Troy, and Assyria) and another object
between{\vspace{-0.04cm}} 1921~BC and 1950~BC (recorded in
Chaldea)\footnote{Observations must have been made near the end of the Sumerian era;
Chaldea existed in the same region about a millennium or more later.} --- most copied
from Pingr\'e's (1783) cometography --- led us to integrate the orbit from 372~BC
back in time over two revolutions about the Sun.  While we felt that there was no
point in deriving $\Delta U$ shifts to bring the computed perihelion times into
better agreement with the recorded times, it is clear that the differences of not
more than several decades would require fairly minor values of the shift corrections.
One cannot rule out that especially the Greek comet seen either in 1176 or 1201~BC
and the one in the 20th century~BC could have been previous appearances of
Aristotle's comet.

The somewhat longer than expected period of time between Aristotle's comet and
the candidate objects in the late 12th century~BC is not surprising, if
some of the 16~comets in Hasegawa's catalog between 532 and 394~BC were siblings
of Aristotle's comet whose birth dated back to the 12th century~BC.  Although
speculative, this argument is supported by six Kreutz candidates on Hasegawa \&
Nakano's (2001) list between AD~101 and 252, potentially their next returns to
perihelion.

The agreement between the predicted perihelion time of Aristotle's comet and the
alleged appearance of the Sumerian comet, to better than 50~years is excellent
given the poor dating of those times.  It suffices to mention that the difference
between two historical scales, the so-called short and middle chronologies, is
64~years.

In any case, the Kreutz sungrazer system in our terminology refers to fragments
that were derived from either Lobe~I or Lobe~II (or the neck, if it ever
established its own hierarchy of products) at or after the breakup of Aristotle's
comet into the lobes at large heliocentric distance about two millennia ago.  Any
sungrazers that were given birth in the course of previous fragmentation events
involving Aristotle's comet or its predecessor, during which the object's
contact-binary nature remained intact, are obviously related to the Kreutz
sungrazers by virtue of being derived from the common parent, but they stay
outside our area of interest.

\begin{table*}
\vspace{-4.2cm}
\hspace{0.52cm}
\centerline{
\scalebox{1}{
\includegraphics{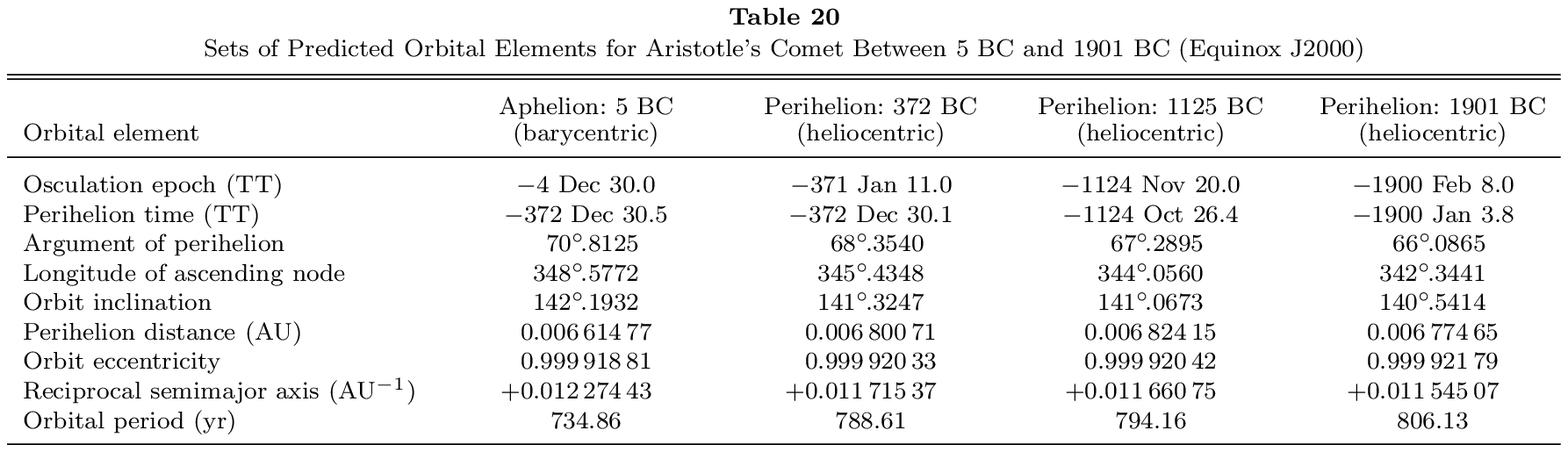}}}
\vspace{-19.5cm}
\end{table*}

\section{Comments on Potential Orbital History of\\Sungrazing Comet Pereyra}
%
Next to the fragments of the Great September Comet of 1882 and comet Ikeya-Seki,
Pereyra (C/1963~R1) is the only other major Kreutz sungrazer whose orbital period is
known with relatively high accuracy.\footnote{Comet Lovejoy (C/2011 W3), whose orbital
period has also been determined rather accurately (Sekanina \& Chodas 2012), disintegrated
shortly after perihelion and we do not rank it as a major Kreutz sungrazer.}  Pereyra's
osculating orbit (Marsden 1967, Marsden et al.\ 1978) was classified by Marsden as
Population (or Subgroup) I, even though it does not resemble the orbit of the 1843
sungrazer as closely as does, for example, the Great Southern Comet of 1880 (C/1880~C1).
The notable difference is in the perihelion distance, which is about 10~percent smaller
and treacherously close to the Sun's radius.  Also, Pereyra's orbital period of 900~yr is
longer than for any other bright Kreutz sungrazer with a known period.  Although the comet
was observed exclusively after perihelion,\footnote{The comet was discovered three weeks
after perihelion.} the derived orbital period is deemed relevant to the comet's preperihelion
motion because its nucleus apparently did not fragment.\footnote{Roemer (1963, 1965)
reported a possible secondary nucleus 0$^\prime\!$.1 from the primary on 1963~November~9,
nearly 80~days after perihelion, which was never confirmed.  Its existence is highly
unlikely, because a genuine companion (i)~should have been detected much earlier and
(ii)~this long after perihelion the separation distance from the primary should have been
substantially greater.}

The odd orbital properties make Pereyra's comet a maverick that is difficult to
compartmentalize among the bright Kreutz sungrazers.  The long orbital period is particularly
perplexing because it suggests the previous perihelion about 50~years prior to the 1106
comet.  Marsden wrestled with the problem, noting an agreement between the orbits of the
1843 and 1880 sungrazers at their previous return to perihelion when he assigned an orbital
period near 370~yr to the former--- about one half the period we adopt --- and slightly over
400~yr to the latter.  He combined the evolution of this pair with that of comet Pereyra,
whose orbital period was more than twice as long.  The solutions that Marsden deemed the
most promising required that the parent --- which he called the ``Combo'' --- should have
passed perihelion either in mid-February 1072 when Pereyra separated and in mid-March 1463
when the 1880 comet separated from the 1843 comet; or in early December 1114 when Pereyra
separated and in late October 1487 for the second fragmentation event.  The second solution
--- the more probable one according to Marsden --- implied for Pereyra a barycentric orbital
period of 849~yr, which was by nearly 60~yr shorter than the most probable value.  As
Marsden himself noted, no sungrazer was observed in either 1463 or 1487.  One may add
that none was seen in 1072 or 1114 either (see England 2002).  Not to mention
that a subsequent revision of the orbit of the 1843 sungrazer (Sekanina \& Chodas 2008)
ruled out the possibility of its orbital period having been anywhere near 400~yr.

We have now revisited the issue of Pereyra's sungrazer by first computing the time of its
previous perihelion, based on Marsden's osculating elements.  Our orbit integration showed
that the perihelion time should have occurred on 1057~April~7, so that its barycentric
orbital period was 906.4~yr.  Marsden determined the orbital period with a mean error of
$\pm$17~yr.  Accordingly, a perihelion passage between the years 1040 and 1074 should
satisfy the predicted perihelion time within $\pm$1$\sigma$.

Hasegawa \& Nakano (2001) have on their list of potential Kreutz sungrazers an exceedingly
bright Chinese-Korean comet,\,allegedly of absolute magnitude\,$-$1,\,which was to pass
perihelion nominally on 1041~August~4.  The historical records noted that the comet
was first seen on 1041~September~1.  We propose that Pereyra's comet separated from this
parent at its 1041 return to perihelion and examine the ramifications of this hypothesis.

\begin{table*}
\vspace{-4.2cm}
\hspace{0.6cm}
\centerline{
\scalebox{1}{
\includegraphics{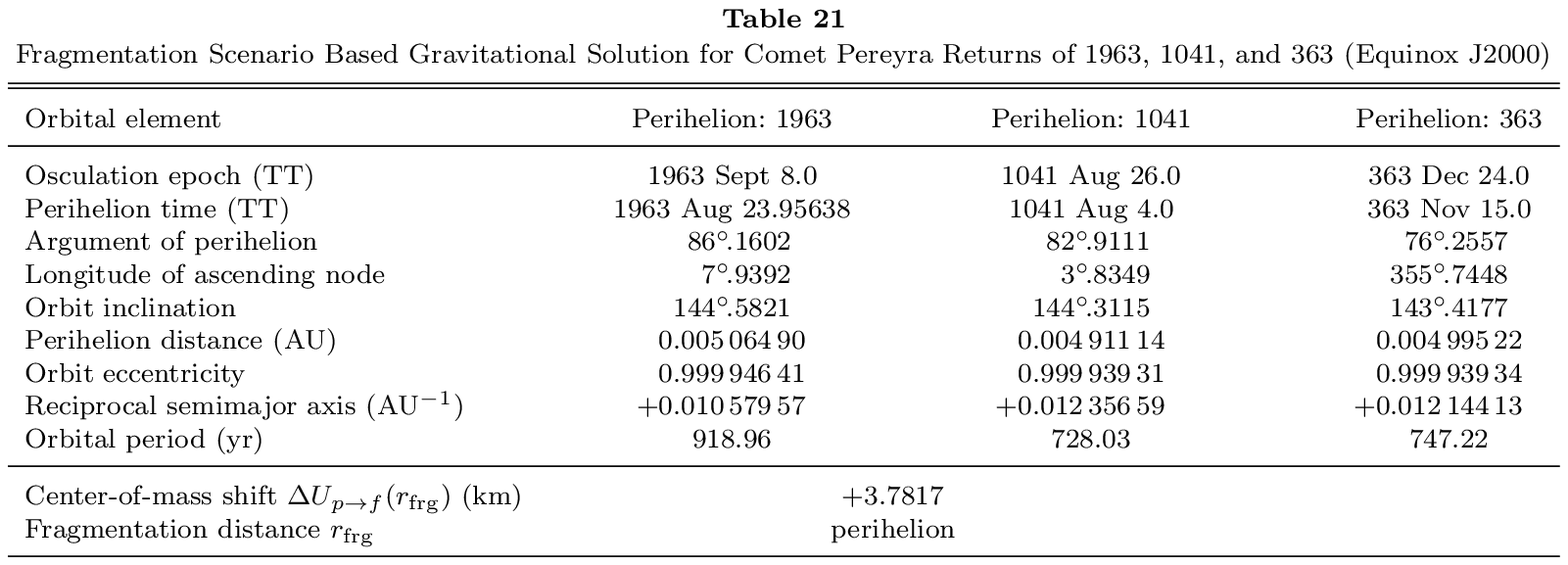}}}
\vspace{-18.9cm}
\end{table*}

We integrated the gravitational orbit of Pereyra's comet, adjusted to fit the perihelion
of the 1041~comet, and found that its previous perihelion would have occurred on
158~November~11.  While Hasegawa \& Nakano tabulate a possible Kreutz comet in January
133 and another one in October 191, we prefer a linkage to Lobe~I in November 363, so
that Pereyra would be its indirect, second-generation product.

Table 21 shows that the proposed history of Pereyra's comet is entirely feasible. A notable
feature is a rapid rotation of the orbital plane:\ between 1041 and 363 the nodal line
advanced at an incredible rate of more than 8$^\circ$ per revolution.  The sudden jump in
the orbital period of more than 200~yr was triggered by a radial shift of the center of
mass by less than 4~km at the 1041 perihelion, when the comet was barely 40,000~km above
the photosphere.  The orbital elements at perihelion in AD~363 show that Pereyra's
precursor obviously separated from Lobe~I, as predicted in Papers~1 and 2, where it was
referred to as Fragment~Pe.  The differences in the elements relative to Lobe~I are,
however, much smaller than the differences between Lobes~I and II.  For example, the
gap in the nodal longitude is 7$^\circ$ compared to 21$^\circ$, in the inclination
0$^\circ\!$.8 compared to 3$^\circ\!$.8, and the perihelion distance 0.04\,{\Rsun}
compared to 0.7\,{\Rsun}.

Even though the gaps between the 363 orbital elements of Lobe~I and Pereyra's precursor
were much smaller than between those of Lobes~I an II, they were large enough that
obviously the precursor arrived at perihelion in AD~363 as a separate
fragment and that instead of Lobe~I's path from the apparition of 363 to 1106 it
followed an alternative path to 1041.  In other words, unlike for other
Population~I fragments (such as C/1880~C1), the orbital evolution of Pereyra became
divorced from that population's parent already before the first passage through
perihelion.  This explains why comet Pereyra was among Population~I members a
maverick:\ it was their sibling, but a more distant sibling.

The reader may notice a remarkable similarity in the time of the year between the
arrivals of the 1041 comet and the 1138 comet.  The adopted perihelion dates differ
by three days, the date of the first observation by two days.  The length of the
period of observation of the 1041 comet is disputed:\ the quoted Chinese source
says more than 90~days, while the Korean source says ``20 and some odd~days,'' a
little less than for the 1138~comet.  A large uncertainty was obviously involved.  The
observations of the 1138 comet were terminated, in part, because of interference from
moonlight, when the predicted apparent magnitude was 4; after the observing conditions
improved, there evidently was nobody to pick up the comet again.  It seems that this
happened with the 1041~comet in Korea as well, given that the Moon phases were moved
forward one week in 1041 compared to 1138.  Unlike the Koreans, the Chinese must
have observed the {\it tail\/} of the 1041~comet until its head was of apparent
magnitude~$\sim$7, as prominent sungrazers sometimes have been (see Paper~2).  In any
case, the head of the 1041~comet could not possibly have been of an absolute magnitude
of $-$1.  Adopting for this object the same post-perihelion rate of fading as for the
1138 comet, apparent magnitude~7 at the end of November 1041 gives with Hasegawa \&
Nakano's ephemeris the post-perihelion absolute magnitude \mbox{$H_0^+ \!= 1.7$} and,
with the comet's perihelion distance of 0.00491~AU, its{\vspace{-0.01cm}} preperihelion
absolute magnitude becomes \mbox{$H_0^- \!= 4.0$}, or about 1.2~mag fainter than we
have assigned to the 1138~comet.

\section{Aristotle's Comet:\ Why A Contact Binary and Why Aphelion Breakup?} 
Besides the arguments favoring the idea of a contact-binary model for the
progenitor that rely on the intrinsic properties of the Kreutz system and have
been presented in Paper~1, this paradigm is likewise supported by independent
circumstantial evidence.  Among the six comets imaged at closeup by cameras on
board spacecraft (1P/Halley, 19P/Borrelly, 81P/Wild, 9P/Tempel, 103P/Hartley,
67P/Churyumov-Gerasimenko) and one by radar (8P/Tuttle), the nuclei of at least
two (8P, 67P) were found to be bilobed, while the nuclei of 103P (described as
peanut-shaped) and perhaps even 1P and 19P had similar outlines (e.g., Harmon
et al.\ 2010, A'Hearn et al.\ 2011, Jutzi \& Asphaug 2015, Jorda et al.\ 2016).

Another line of evidence that generally corroborates the contact-binary model is
the existence of {\it persisting\/} twin objects both among short-period comets
(such as 3D/Biela, observed double at two consecutive returns to perihelion) and
among comets of longer period (such as the pair of C/2002~A1 and C/2002~A2, both
under observation over more than two years) as well as among the Kuiper Belt
objects (e.g., Goldreich et al.\ 2002), including the Pluto--Charon pair.

Discovered several years ago, the Kuiper Belt object \mbox{(486958) Arrokoth =
2014~MU$_{69}$} (a.k.a.\ Ultima Thule) was imaged by the New Horizons Mission
(Stern et al.\ 2019), offering arguably the most robust piece of supporting
circumstantial evidence for at least two reasons:\ (i)~the overall length of this
contact-binary object equaled, according to Stern et al., 36~km and the larger,
highly-flattened lobe's maximum diameter was 22~km, comparable to the crudely
estimated dimensions of the Great September Comet of 1882 (Sekanina 2002), a
major surviving mass of the presumed Kreutz progenitor; and (ii)~the likelihood
that contact binaries like Arrokoth are common among transneptunian objects in
general and Kuiper Belt objects in particular, a notion expressed independently
by several researchers (e.g., Rickman et al.\ 2015, Stern et al.\ 2019, Benecchi
et al.\ 2019).  Contact-binary objects are believed to have initially been
separate objects orbiting about one another at very low velocities (a few meters
per second), eventually collapsing gravitationally into a single body (e.g.,
Stern et al.\ 2019, Grundy et al.\ 2020, McKinnon et al.\ 2020). 

Extending this rationale to its seemingly logical conclusion, one would argue
that the Kreutz progenitor was a Kuiper Belt object.  Unfortunately, since
Kuiper Belt objects are not perturbed into retrograde orbits, this hypothesis
has to be abandoned.  Instead, a plausible region of origin could be the inner
Oort Cloud, the source of Halley-type comets (Levison et al.\ 2001). Inner Oort
Cloud objects may have an affinity for forming contact binaries like Kuiper Belt
objects, although the lower spatial density offers a less favorable environment.
Once the progenitor enters --- presumably thanks to mostly indirect perturbations
by the outer planets, Jupiter in particular --- an orbit that is inclined
90$^\circ$ to 100$^\circ$ to the ecliptic and has an orbital period of
$\sim$800~yr and perihelion distance of less than 2~AU, a process of migration
governed by continuing long-term indirect planetary perturbations was shown by
Bailey et al.\ (1992) to have been progressively reducing the perihelion distance
over a period on the order of a million years, injecting eventually the object
into the observed retrograde sungrazing orbit.

Fern\'andez et al.\ (2021) have recently considered three different models for
the Kreutz system progenitor, concluding that the most promising one was an
Oort-Cloud comet, which was first injected into an Earth-crossing orbit and
then perturbed into a sungrazing orbit by the Lidov-Kozai mechanism.  While
this model has some attractive features, it wrestles with the problem of a
rate at which such massive objects could be delivered into these orbits.  The
study does not address the issue of discrete populations of sungrazer fragments
and deals only with near-perihelion tidally-triggered fragmentation.   
  
There indeed is a universal consensus that the Sun's tidal forces, possibly in
concert with other forces, instigate perihelion fragmentation of the Kreutz comets.
One such other potential trigger is the Sun's extreme heat that foments stress
because of the propagation of a thermal wave through the nucleus, unless massive
amounts of ejected dust in the atmosphere make it optically thick, thereby
protecting the nucleus from direct exposure.  These effects are of course
superposed on presumably less robust effects of the nucleus' rotation and/or
precession.   

This conclusion is based on a plausible premise that a nucleus breaks
up, if stress in its interior, induced by the Sun tidally, and potentially
exacerbated thermally, exceeds the strength of the nucleus.  It is fitting
to ask what happens if the {\it magnitude of the combined tidal, thermal,
and rotational stresses does not exceed the nucleus strength\/}.  One expects
that the object then survives perihelion passage, though {\it not necessarily
intact\/}:\ some, however minor, damage is being sustained at each close
approach to the Sun and the extent of this damage adds up, revolution after
revolution.  Indeed, the egress of a sungrazer out of the solar corona is
always anticipated with some trepidation over its survival.  The observations
offer only one of three answers:\ (i)~the object emerges with a single nuclear
condensation, in which case it is said to have survived; or (ii)~it emerges
with two or more condensations, indicating that its nucleus split; or (iii)~it
emerges with no condensation, the sign of disintegration.  What remains
hidden in the case of survival is the {\it extent of damage\/}:\ the
observations do not discriminate between a structurally intact, nearly intact,
moderately damaged, or heavily damaged (but still unbroken) nucleus.

Although hidden to the observers, the genuine physical condition of the nucleus
at the end of its extremely close approach to the Sun's photosphere makes all
the difference in the comet's future life.  An intact or nearly intact object
is obviously much more likely to successfully complete the hundreds-of-years
lasting revolution about the Sun than an object that barely held together at
the start of the long journey.  The gradual weakening of structural cohesion
of a piece of material under terrestrial conditions (on vastly shorter time
scales) is known as {\it material fatigue\/}; there is no reason why the
process could not apply, on another time scale, to objects in space as well.

When the celestial object is a contact binary, the neck is likely to be more
susceptible to damage than the lobes.  In general, the neck is subjected to
either compressive or tensile stress, depending on the size and mass (and thus
the density) of the lobes as well as the rate of rotation.  The magnitude of
the force also depends on the cross-sectional area of the neck.  For Arrokoth,
\mbox{McKinnon} et al.\ (2020) determined that the neck was being subjected to
compression if the object's density was greater than 0.25~g~cm$^{-3}$, which
was likely to be the case. Judging from the estimated giant size of the nucleus
of C/1882~R1, the neck of the proposed Kreutz progenitor should have been
subjected to an even higher level of compressive stress, unless the rotation
was much faster than Arrokoth's.

Except in the Sun's proximity, the force affecting the Kreutz progenitor's neck
should not have changed materially over the period of time from the instance
of gentle merger of the two lobes until after the process of the object's slow
migration into the sungrazing orbit was long underway.  Bailey et al.'s (1992)
computations suggest that the perihelion distance had dropped to $\sim$0.1~AU
approximately 50~revolutions and to $\sim$0.01~AU some 25~revolutions about
the Sun before the minimum distance was reached.  While the object was
repeatedly --- every 700--800~years or so --- approaching the Sun in an orbit
whose perihelion distance followed this downward spiral, the near-perihelion
environment was subjecting the neck to three {\it cyclic\/} effects that were
gradually accumulating from orbit to orbit and, as a result, weakening the
bond between the pair of lobes:\ two such effects, as already noted, were,
respectively, tensile stress, prompted by the Sun's tidal force, and thermal
stress, triggered by an enormous heat pulse, both strongly perihelion-distance
dependent; a third was an imprint of similarly varying rates of vigorous
sublimation of volatile substances from the neck, thereby reducing its
cross-sectional area with time.

The neck of the bilobed Kreutz progenitor is thus presumed to have been
subjected to an essentially constant load, with dozens of overlapping {\it
cyclic\/} stress peaks centered, once every 700--800~years, on the time of
perihelion, the magnitude of which was growing progressively from orbit to
next orbit as the perihelion distance was steadily decreasing toward the
presumed present value slightly exceeding 1\,{\Rsun}.  It is this cyclic
component of the load that represents what one can refer to as an extreme
case of the already noted material fatigue, even though the duration of the
cycle is vastly longer and the number of cycles orders of magnitude lower
than in applications to fabricated materials, metals in particular, under
usual terrestrial conditions.

In summary, fatigue of material is a process of progressive, localized
deformation and/or weakening in an object subjected to a cyclic load below
the static material strength, which prompts the development, growth, and
propagation of cracks from cycle to cycle, culminating in the object's
fracture, a.k.a. fatigue failure (see, e.g., Suresh 2014).  Since fatigue
is an irreversible stochastic process with a degree of randomness, the
lifetime, defined by the number of cycles to failure, is rather poorly
determined, especially in the presence of an appreciable background stress.
Usually associated with tensile stresses, fatigue has several phases, but
most time is consumed in the crack growth phase.  Once a crack has begun,
each loading cycle enlarges the crack by a small amount until it reaches
a critical point when the stress intensity factor of the crack exceeds the
fracture toughness of the material producing its rapid propagation that
nearly instantly completes the fracture.

Because of the limited number of orbits with perihelia very close to the
Sun during the orbital evolution, the neck of the Kreutz progenitor was
exposed to {\it low-cycle fatigue\/} (when the number of cycles is
$\ll$10$^5$).  And since the location of the point of failure is
essentially random in time, the location of aphelion close to 160~AU from
the Sun implies that there is a $\sim$50~precent chance of fracture at
heliocentric distances $>$140~AU, a $\sim$60~percent chance at $>$130~AU,
a $\sim$70~percent chance at $>$110~AU, and an $\sim$80~percent chance at
$>$90~AU from the Sun.  Accordingly, the probability of breakup at large
heliocentric distance is high.

It is noted that fragmentation events at arbitrary (and generally large)
distances from the Sun coexist with events of perihelion fragmentation
because of strongly heterogeneous structure and morphology of the nuclei:\
the former incidents are their response to an {\it accumulated effect of
cyclic episodes of deformation, the magnitude of which is smaller than
a fracture limit\/}, whereas the latter occurrences are their response
to {\it discrete instances of damage whose magnitudes reach or exceed a
relevant fracture limit\/}.  Accordingly, while new damage (cracks
initiation) is always inflicted over a short period of time around
perihelion, the ultimate response (completed cracks propagation) may
or may not be constrained to this orbital arc, depending on the magnitude
of the injury.  Hence, a Kreutz sungrazer may fragment at any point of
the orbit, including aphelion.

\section{Conclusions} 
In broad terms, the purpose of this paper was to conduct a feasibility study in support
of the contact-binary model for the orbital history of the Kreutz sungrazer system.
While the problem potentially has a large number of solutions, each of which satisfies
a set of boundary conditions, our task was to demonstrate that the particular solution
that we chose to pursue was one of them.  Our approach was based on the assumption that
the largest surviving masses of the two lobes of the contact-binary progenitor were
the Great March Comet of 1843 (Lobe~I) and the Great September Comet of 1882 (Lobe~II),
associated with Populations~I and II, respectively.  We accepted Marsden's (1967) result
on the undisputed genetic connection of the 1882 sungrazer and Ikeya-Seki's comet, which
separated from their common parent at the previous perihelion passage.

We settled the notorious, protracted dispute over the position of the Great Comet of
1106 in the hierarchy of the Kreutz system.  We developed a new technique for deriving
the time of the previous return of Ikeya-Seki'comet to perihelion, based on linking the
1965 preperihelion astrometric observations with the early post-perihelion ones.  The
evidence pointed unequivocally to the conclusion that the comet was at perihelion
around the year 1139, more than three decades after the appearance of the 1106 comet.
A close relationship between the 1106 comet on the one hand and Ikeya-Seki (and the
1882~sungrazer) on the other hand was thus ruled out.  By contrast, we see no obstacle
for accepting the 1106~comet as the previous appearance of the Great March Comet of
1843, supported by the orbital computations by \mbox{Sekanina} \& Chodas (2008).

We also settled the other tenacious problem, previously brought up by Marsden (1989):\
Should there have been a second parent comet in the early 12th century to account for
two populations (or subgroups) of the Kreutz sungrazers?  In the case we deal with here,
this presents a dilemma:\ Where was the parent comet of the 1882/1965 pair, if it was not
the 1106~comet?  Our --- first ever --- reference to the Chinese comet of 1138 as the
pair's parent did:\ (i)~answer the above question; (ii)~show the time of its appearance
consistent with the prediction from our analysis of comet Ikeya-Seki; (iii)~turn
out to be in excellent agreement with Marsden's (1967) prediction of the previous
appearance of the 1882~sungrazer; (iv)~substantially improve the orbital agreement
between the 1882~sungrazer and Ikeya-Seki in comparison with Marsden's (1967) results
for their assumed common origin in 1115; and (v)~demonstrate that an intrinsically
bright Kreutz sungrazer that was at perihelion in early August and unobserved in
broad daylight, was not necessarily missed, as was often claimed in the literature.

We are rather confident that the Chinese comet of 1138 was the 12th-century ``missing''
major sungrazer,~even though we admit that our arguments are unfortunately based
on a minimum amount of data.  Comparison with the naked-eye's limiting magnitude
for stellar objects showed that this comet's {\it post-perihelion intrinsic\/}
brightness was apparently greater than that of the 1843~sungrazer and could have
been comparable to the brightness of the 1882~sungrazer.  Its {\it apparent\/}
brightness suffered from extremely unfavorable geometric conditions during the
object's departure from perihelion on the far side of the Sun.  The description
of this comet (Ho 1962) is similar to the account, in a Korean source, of a
suspected Kreutz sungrazer of 1041 (Hasegawa \& Nakano 2001), which appeared at
the same time of the year. 

Two modes of orbit integration that we applied in an effort to link consecutive
returns of the Kreutz sungrazers to perihelion were referred to as, respectively,
a {\it sublimation scenario\/} and a {\it fragmentation scenario\/}.  In the first
mode the orbital period was being affected continuously by an outgassing-driven
nongravitational acceleration, requiring no fragmentation.  The other mode was
applicable only in the presence of near-perihelion tidally-driven fragmentation,
requiring no nongravitational effects; the orbital period was altered suddenly.
A parent's breakup essentially along a plane perpendicular to the radius vector
resulted in the generation of two or more fragments whose centers of mass were located
at slightly different heliocentric distances relative to the parent's center of mass,
yet at the time of breakup they had the same orbital velocity.  The fragments, with
no momentum exchange involved, were thus bound to end up in different orbits; the
period was the longer the farther from the Sun was the center of mass.  The enormity
of this effect at perihelion in {\it sungrazing\/} orbits is illustrated by noting
that a shift of one kilometer (a few percent of the size of a major sungrazer's
nucleus) in the radial direction could easily change a 700~yr period by $\sim$40~yr.

The Kreutz sungrazers are of course subjected to both modes of orbit transformation, and
comet Ikeya-Seki was an example on which the interaction between them was illustrated.
We showed that at the time of tidal splitting fragment~B was at the sunward end of the
parent nucleus and its orbital period would have been shorter than that of the main
fragment~A, if it were not subjected to a significant nongravitational acceleration.
Accordingly, the orbital period of the {\it parent\/} was shorter than that of
fragment~A.  This explains why the back integration of the orbit of this fragment indicates
the previous perihelion around the year 1115, while the comet actually passed perihelion
(as part of its own parent comet) more than 20~years later.

We did not mix the two modes of orbit integration, the sublimation and
fragmentation scenarios, in our orbit simulation computations.  We demonstrated
that either of them successfully linked the three returns of the largest
surviving masses of Lobe~I, \mbox{1843--1106--363}, and Lobe~II,
\mbox{1882--1138--363}, even though the sublimation scenario required an
improbably high nongravitational acceleration to accomplish the linkage for
Lobe~II.  The fragmentation scenario did not need center-of-mass shifts greater
than a few kilometers at perihelion, deemed very modest and quite acceptable.
For Lobe~II we described the sequence of two fragmentation events in the
year 1138:\ in the primary event at perihelion the arriving parent, presumed
to have been more than 60~km across, shed a minor fragment into an orbit
of long period, while its major fragment split tidally again, about 2.5~hr
later, into the 1882~sungrazer and Ikeya-Seki's comet.

In connection with the sublimation scenario we point out that the current
computer techniques, aiming to account for variations in the nongravitational
acceleration whose effects are asymmetric relative to perihelion, are
inadequate and need a conceptual upgrade.  Refined methods should unquestionably
have a favorable effect on simulating the orbital evolution of the Kreutz
sungrazers in the future.

While the sublimation scenario failed to fit the sungrazers' orbital changes
prior to AD~363, the fragmentation scenario --- even without the benefit of
incorporating any nongravitational motion --- allowed us to set up a method
of successfully simulating the orbits of Lobes~I and II toward their point
of birth, essentially at aphelion previous to the perihelion passage in
AD~363.  We adopted a default perihelion time of 363~Nov~15.0 (which was
a convenient standard to which the arriving sungrazers' orbits were referred
to rather than the actual time of perihelion passage) and computed for either
lobe a fragmentation scenario based gravitational orbit that linked its
perihelion time in AD~363 with the perihelion time expected for Aristotle's
comet --- the proposed contact-binary progenitor --- in 372~BC.  It turned
out that the line of intersection of the orbital planes of the lobes, on
which they had to be located at the time of their separation from the common
parent, made such extremely small angles with the lobes' lines of apsides
that the assumption of a breakup at aphelion was bound to be a very good
approximation.

With this aphelion breakup of Aristotle's comet set at the very end of 5~BC
at a heliocentric distance of 163~AU, we further assumed that the two lobes
were of equal dimensions, separating with equal velocities in the opposite
directions, and that the progenitor was spinning along an axis pointing at
the Sun.  The radial component of the separation velocity was then nil, while
the transverse component determined the perihelion distances as well as
perihelion times of the lobes and the normal component governed the spatial
positions of their orbital planes.  A separation velocity of 3.63~m~s$^{-1}$
fitted the angular elements to about 0$^\circ\!$.3, demonstrating that the
existence of the two major populations of the Kreutz system as the product of
fragmentation of a progenitor comet at large heliocentric distance is feasible
and fundamentally correct.  The procedure also resulted in a predicted set of
orbital elements for Aristotle's comet.  

In the light of the well-known difficulties that efforts to compartmentalize
the orbital characteristics of the Kreutz comet discovered by Pereyra in
1963 had provoked in the past, we offer a novel explanation to the oddities
of this maverick sungrazer.  With its derived period exceeding 900~yr, a
plausible previous appearance was the comet of 1041 that passed perihelion in
early August according to Hasegawa \& Nakano (2001), a case of interest in
its own right.  Integration of Pereyra's motion suggests that its precursor
had separated from Lobe~I already before the perihelion passage in AD~363.
This circumstance allowed Pereyra's comet to bypass the normal evolutionary
path for members of Population~I via the Great Comet of 1106 and instead to
follow an alternative path via the comet of 1041.  This example illustrates
the diversity of orbital tracks that the Kreutz sungrazers could heed, thus
explaining the large number of discrete populations presented in Paper~1.

The Kreutz sungrazers are still believed to fragment mostly at close proximity
of perihelion due to the Sun's tidal forces.  This may very well be true.
However, the existence of discrete populations of fragments with large gaps
between their angular elements and perihelion distances cannot be a product of
perihelion fragmentation, alone or enhanced with effects of the indirect planetary
perturbations over periods of time that are not excessively long.  We argue that
the sungrazers are in fact subject to fragmentation along the entire orbit around
the Sun, including aphelion, because of material fatigue of their nuclei.
The perihelion region is where a Kreutz sungrazer is exposed to a cyclic
load, whose magnitude is below the static material strength, yet it prompts the
growth and propagation of cracks, resulting eventually in fatigue failure.  This
condition is reached randomly, at an arbitrary time.  Since sungrazers spend most
time at large heliocentric distance, fatigue fracture far from the Sun is likely.
The probability for a rotating contact-binary Kreutz sungrazer may further be
augmented by continuous rotation stress and the potentially brittle nature of
the connecting neck.
 
\vspace{0.3cm} 
%
The first author was in charge of this study's strategy, the second author was
responsible for performing the orbital computations.  This research was carried out
in part at the Jet Propulsion Laboratory, California Institute of Technology, under
contract with the National Aeronautics and Space Administration.\\
\begin{center}
{\large \bf Appendix} \\[0.2cm]
KREUTZ'S ORBIT FOR PRIMARY FRAGMENT OF\\THE GREAT SEPTEMBER COMET OF 1882 \\
\end{center}

Because the {\it Catalogue of Cometary Orbits\/} by Marsden \& Williams (2008)
lists for the 1882 sungrazer's~primary nucleus (fragment B or No.\,2) the
relativistic orbit by Hufnagel (1919), Kreutz's (1891) definitive nonrelativistic
orbit for this fragment, converted to equinox J2000, is not widely accessible.
The purpose here is to alleviate this problem.  The following set of orbital
elements is that published by Kreutz on page 35 of his Paper~II, but updated
using current conventions: \\[-0.3cm]

(1)~Equinox 1882.0, employed by Kreutz, has~been~replaced by J2000. \\[-0.3cm]

(2)~Berlin Mean Time has been converted to Universal Time by adding 12~hr and
subtracting the longitude of Berlin Observatory{\vspace{-0.05cm}} (IAU code 548; active
1835--1913), 0$^{\rm h}$53$^{\rm m}$34$^{\rm s}\!$.93 east of Greenwich (Gould~1853).
Universal Time has then been converted to Terrestrial Time (TT) by adding a correction
$\Delta T$ of $-$5$^{\rm s\!}$.20 from {\it Astronomical Almanac\/}'s online tabular
data,\footnote{See a website:\ {\tt https:/$\!$/webspace.science.uu.nl/$\sim$gent0113/}
{\tt deltat/deltat.htm}.} based on the work by Stephenson \& Morrison (1984).\\[-0.3cm]

(3)~The orbit was integrated from Kreutz's osculation epoch of 1882 Sept 20.5 Berlin
Mean Time to a standard osculation epoch of 1882~Oct~2.0~TT = JD\,2408720.5. \\[-0.3cm]

(4)~The {\it probable\/} errors of the elements used by Kreutz have been converted
to the {\it mean\/} errors by applying a multiplication factor of 1.4826. \\[-0.3cm]

(5)~A typographical error that we detected in Kreutz's paper has been corrected.
It affected the published~value of eccentricity, which was inconsistent with the
listed logarithm of eccentricity as well as with the ratio of the perihelion distance
to semimajor axis.\\[-0.3cm]

The orbital elements:\\[-0.5cm]
\begin{center}
{\hspace{0.96cm}}Osculation epoch 1882 October 2.0 TT \\[0.1cm]
{\hspace{0.96cm}}$T\!$ = $\!$1882 Sept 17.724040$\:\!\pm\,$0.000047 TT\\[-0.22cm]
\begin{displaymath}
\begin{array}{rcrcl}
\omega & \!\!=\!\! & \;\:69^\circ\!.58477 & \!\!\!\pm\!\!\! & 0^\circ\!.00225 \\
\Omega & \!\!=\!\! & 347^\circ\!.65640 & \!\!\!\pm\!\!\! & 0^\circ\!.00279 \\
i & \!\!=\!\! & 142^\circ\!.01104 & \!\!\!\pm\!\!\! & 0^\circ\!.00068 \\
q & \!\!=\!\! & 0.00775020 & \!\!\!\pm\!\!\! & 0.00000091 \; {\rm AU}\\
e & \!\!=\!\! & 0.99990790 & \!\!\!\pm\!\!\! & 0.00000021 \\[-0.065cm]
1/a & \!\!=\!\! & 0.01188356 & \!\!\!\pm\!\!\! & 0.00002894 \; {\rm AU}^{-1}\\
P & \!\!=\!\! & 771.93 \!\!\:\pm\!\!\: 2.\rlap{82 {\rm yr}}
\end{array}
\end{displaymath}
\end{center}

The elements are based on 18 normal places in right ascension and declination
(between 1882 Sept~8 and 1883 Mar~3), on an observation of the comet's ingress
on the Sun's disk on 1882 Sept~17, and on 13 normal places in the coordinate
perpendicular to the nuclear line (between 1882 Oct~1 and 1883 May~26).  The
32~normal places, of which only five were preperihelion ones, were made up
of a total of slightly more than 600 individual astrometric observations,
extending over 260~days, from 1882 Sept 8 through 1883 May 26.  The mean
residual for a normal place of unit weight was $\pm$1$^{\prime\prime}\!$.51.

It is interesting that in his first paper Kreutz (1888) considered the No.\,2
fragment to have been a ``center-of-gravity'' of the disintegrated nucleus,
because he was able to successfully link the preperihelion observations of
the single nucleus with the post-perihelion observations of this fragment.
At the time he anticipated that no other fragment would provide a similarly
satisfactory linkage, but after convincing himself, in his second~paper
(Kreutz 1891), that the fragments Nos.\,3 and 4 did~as well, he changed
his mind and concluded that the center of gravity could have been located
anywhere along the nuclear train between the fragments Nos.\,2 and 4.
This led him to maintain that the comet's pre-fragmentation orbital
period was between 770 and about 1000~years.  Yet, it appears that he
attached more weight to the lower limit, which follows from the fact that
in~his~third and final paper (Kreutz 1901) he returned to the problem
of the comet of 1106.  Even though he did mention suggestions by some
astronomers that this comet might have been an earlier appearance of the
1843 sungrazer, Kreutz's preference in favor of the identity between
the 1882 and 1106 comets based on the latter's daytime observations gets
across as substantial. \\[0cm]

\begin{center}
{\footnotesize REFERENCES}
\end{center}
\begin{description}
{\footnotesize
\item[\hspace{-0.4cm}]
A'Hearn, M. F., Belton, M. J. S., Delamere, W. A., et al.\ 2011,{\linebreak}
 {\hspace*{-0.6cm}}Science, 332, 1396
\\[-0.57cm]
\item[\hspace{-0.3cm}]
Andrews, A. D. 1965, IAU Circ., 1937
\\[-0.57cm]
\item[\hspace{-0.3cm}]
Bailey, M. E., Chambers, J. E., \& Hahn, G. 1992, A\&A, 257, 315
\\[-0.57cm]
\item[\hspace{-0.3cm}]
Benecchi, S. D., Porter, S. B., Buie, M. W., et al.\ 2019, Icarus,{\linebreak}
 {\hspace*{-0.6cm}}334, 11
\\[-0.57cm]
\item[\hspace{-0.3cm}]
Bessel, F. W. 1836, Astron. Nachr., 13, 345
\\[-0.57cm]
\item[\hspace{-0.3cm}]
England, K. J. 2002, J. Brit. Astron. Assoc., 112, 13
\\[-0.57cm]
\item[\hspace{-0.3cm}]
Everhart, E., \& Raghavan, N. 1970, AJ, 75, 258
\\[-0.57cm]
\item[\hspace{-0.3cm}]
Fern\'andez, J. A., Lemos, P., \& Gallardo, T. 2021, Mon. Not. Roy.{\linebreak}
 {\hspace*{-0.6cm}}Astron. Soc., 508, 789
\\[-0.57cm]
\item[\hspace{-0.3cm}]
Festou, M., Rickman, H., \& Kam\'el, L. 1990, in Asteroids, Comets,{\linebreak}
 {\hspace*{-0.6cm}}Meteors III, ed. C.-I. Lagerkvist, H. Rickman, B. A. Lindblad,{\linebreak}
 {\hspace*{-0.6cm}}\& M. Lindgren (Uppsala: Uppsala University), 313
\\[-0.57cm]
\item[\hspace{-0.3cm}]
Gingerich, O. 1965, IAU Circ., 1943 and 1946
\\[-0.57cm]
\item[\hspace{-0.3cm}]
Goldreich, P., Lithwick, Y., \& Sari, R. 2002, Nature, 420, 643
\\[-0.57cm]
\item[\hspace{-0.3cm}]
Gould, B. A. 1853, AJ, 3, 17
\\[-0.57cm]
\item[\hspace{-0.3cm}]
Grundy, W. M., Bird, M. K., Britt, D. T., et al.\ 2020, Science,{\linebreak}
 {\hspace*{-0.6cm}}367, 3705
\\[-0.57cm]
\item[\hspace{-0.3cm}]
Hall, M. 1883, Observatory, 6, 233
\\[-0.57cm]
\item[\hspace{-0.3cm}]
Harmon, J. K., Nolan, M. C., Giorgini, J. D., et al.\ 2010, Icarus,{\linebreak}
 {\hspace*{-0.6cm}}207, 499
\\[-0.57cm]
\item[\hspace{-0.3cm}]
Hasegawa, I. 1980, Vistas Astron., 24, 59
\\[-0.57cm]
\item[\hspace{-0.3cm}]
Hasegawa, I., \& Nakano, S. 2001, Publ.\,Astron.\,Soc.\,Japan, 53, 931
\\[-0.57cm]
\item[\hspace{-0.3cm}]
Hirayama,\,T., \& Moriyama,\,F.\ 1965, Publ.\,Astron.\,Soc.\,Japan, 17,{\linebreak}
 {\hspace*{-0.6cm}}433
\\[-0.57cm]
\item[\hspace{-0.3cm}]
Ho, P.-Y. 1962, Vistas Astron., 5, 127
\\[-0.57cm]
\item[\hspace{-0.3cm}]
Hubbard, J. S. 1852, AJ, 2, 153
\\[-0.57cm]
\item[\hspace{-0.3cm}]
Hufnagel, L. 1919, Astron. Nachr., 209, 17
\\[-0.57cm]
\item[\hspace{-0.3cm}]
Jorda, L., Gaskell, R., Capanna, C., et al.\ 2016, Icarus, 277, 257
\\[-0.57cm]
\item[\hspace{-0.3cm}]
Jutzi, M., \& Asphaug, E. 2015, Science, 348, 1355
\\[-0.57cm]
\item[\hspace{-0.3cm}]
Kanda, S. 1935, Nihon Temmon Shiryo (Astronomical Materials{\linebreak}
 {\hspace*{-0.6cm}}in Japanese History), Tokyo. (In Japanese)
\\[-0.57cm]
\item[\hspace{-0.3cm}]
Kreutz, H. 1888, Publ. Sternw. Kiel, 3
\\[-0.57cm]
\item[\hspace{-0.3cm}]
Kreutz, H. 1891, Publ. Sternw. Kiel, 6
\\[-0.57cm]
\item[\hspace{-0.3cm}]
Kreutz, H. 1901, Astron. Abhandl., 1, 1
\\[-0.57cm]
\item[\hspace{-0.3cm}]
Kronk, G. W. 1999, Cometography:\ Volume 1 (Ancient--1799).{\linebreak}
 {\hspace*{-0.6cm}}Cambridge, UK:\ Cambridge University Press, 580\,pp
\\[-0.57cm]
%
%
\item[\hspace{-0.3cm}]
Levison, H. F., Dones, L., \& Duncan, M. J. 2001, AJ, 121, 2253
\\[-0.57cm]
\item[\hspace{-0.3cm}]
Lourens, J.\,v.\,B.\ 1966, Mon.\,Not.\,Astron.\,Soc.\,South Africa,~25,~52
\\[-0.57cm]
\item[\hspace{-0.3cm}]
Marsden, B. G. 1965, IAU Circ., 1947
\\[-0.57cm]
\item[\hspace{-0.3cm}]
Marsden, B. G. 1967, AJ, 72, 1170
\\[-0.57cm]
\item[\hspace{-0.3cm}]
Marsden, B. G. 1989, AJ, 98, 2306
\\[-0.57cm]
\item[\hspace{-0.3cm}]
Marsden, B. G., \& Williams, G. V. 2008, Catalogue of Cometary{\linebreak}
 {\hspace*{-0.6cm}}Orbits 2008, 17th ed.  Cambridge, MA:\ Minor Planet Center/{\linebreak}
 {\hspace*{-0.6cm}}Central Bureau for Astronomical Telegrams, 195pp
\\[-0.57cm]
\item[\hspace{-0.3cm}]
Marsden, B. G., Sekanina, Z., \& Everhart, E. 1978, AJ, 83, 64
\\[-0.57cm]
\item[\hspace{-0.3cm}]
Marsden, B. G., Sekanina, Z., \& Yeomans, D. K. 1973,~AJ,~78,~211
\\[-0.57cm]
\item[\hspace{-0.3cm}]
McKinnon, W. B., Richardson, D. C., Marohnic, J. C., et al.\ 2020,{\linebreak}
 {\hspace*{-0.6cm}}Science, 367, 6620
\\[-0.57cm]
\item[\hspace{-0.3cm}]
Nicolai, B. 1843, Astron. Nachr., 20, 349
\\[-0.57cm]
\item[\hspace{-0.3cm}]
Pereyra, Z. M. 1971, AJ, 76, 495
\\[-0.57cm]
\item[\hspace{-0.3cm}]
Pingr\'e,\,A.\,G.\,1783,\,Com\'etographie\,ou\,Trait\'e\,historique\,et\,th\'eorique{\linebreak}
 {\hspace*{-0.6cm}}des com\`etes.\ Tome Premier.\ Paris:\ Imprimerie Royale
\\[-0.57cm]
\item[\hspace{-0.3cm}]
Pohn, H. 1965. IAU Circ., 1937
\\[-0.57cm]
\item[\hspace{-0.3cm}]
Porter, J. G. 1967, Quart. J. Roy. Astron. Soc., 8, 274
\\[-0.57cm]
\item[\hspace{-0.3cm}]
Rickman, H. 1986, in The Comet Nucleus Sample Return Mission,{\linebreak}
 {\hspace*{-0.64cm}}ESA\,SP-249,\,ed.\,O.\,Melita\,(Noordwijk,\,Netherlands:\,ESTEC),\,195
\\[-0.57cm]
\item[\hspace{-0.3cm}]
Rickman, H., Froeschl\'e, C., Kam\'el, L., \& Festou, M. C. 1992, AJ,{\linebreak}
 {\hspace*{-0.6cm}}102, 1446
\\[-0.64cm]
\item[\hspace{-0.3cm}]
Rickman, H., Kam\'el, L., Festou, M. C., \& Froeschl\'e,{\vspace{-0.07cm}} C. 1987,
in{\linebreak}
 {\hspace*{-0.6cm}}Diversity and Similarity{\vspace{-0.07cm}} of Comets, ESA SP-278,
 ed.\,E.\,J.\,Rolfe \&{\linebreak}
 {\hspace*{-0.6cm}}B. Battrick (Noordwijk, Netherlands:\ ESTEC), 471}
\vspace{-0.38cm}
\end{description}
\pagebreak
\begin{description}
{\footnotesize
\item[\hspace{-0.3cm}]
Rickman, H., Marchi, S., A'Hearn, M. F., et al.\ 2015, A\&A, 583,{\linebreak}
 {\hspace*{-0.6cm}}A44
\\[-0.57cm]
\item[\hspace{-0.3cm}]
Roemer, E. 1963, Publ. Astron. Soc. Pacific, 75, 535
\\[-0.57cm]
\item[\hspace{-0.3cm}]
Roemer, E. 1965, AJ, 70, 397
\\[-0.57cm]
\item[\hspace{-0.3cm}]
Roemer, E., \& Lloyd, R. E. 1966, AJ, 71, 443
\\[-0.57cm]
\item[\hspace{-0.3cm}]
Schaefer, B. E. 1993, Vistas Astron., 36, 311
\\[-0.57cm]
\item[\hspace{-0.3cm}]
Schaefer, B. E. 1998, Sky Tel., 95, 57; code{\vspace{-0.02cm}} version
 by L. Bogan{\linebreak}
 {\hspace*{-0.6cm}}at {\tt https://www.bogan.ca/astro/optics/vislimit.html}
\\[-0.57cm]
%
%
\item[\hspace{-0.3cm}]
Sekanina, Z. 1982, in Comets, ed. L. L. Wilkening{\vspace{-0.02cm}} (Tucson:\
 Univer-{\linebreak}
 {\hspace*{-0.6cm}}sity of Arizona Press), 251
\\[-0.57cm]
\item[\hspace{-0.3cm}]
Sekanina, Z. 1988, AJ, 86, 1455
\\[-0.57cm]
\item[\hspace{-0.3cm}]
Sekanina, Z. 1992, in Observations and Physical{\vspace{-0.01cm}} Properties
 of{\linebreak}
 {\hspace*{-0.6cm}}Small Solar System Bodies, Proc.\,Li\`ege
 Internat.\,Astrophys.\,Coll.{\linebreak}
 {\hspace*{-0.6cm}}No.\,30, ed.\,A.\,Brahic, J.-C.\,G\'erard, \& A.\,Surdej
 {\vspace{-0.02cm}}(Li\`ege:\,Universit\'e{\linebreak}
 {\hspace*{-0.6cm}}de Li\`ege, Institut d'Astrophysique), 133
\\[-0.57cm]
\item[\hspace{-0.3cm}]
Sekanina, Z. 1993, AJ, 105, 702
\\[-0.64cm]
\item[\hspace{-0.3cm}]
Sekanina, Z. 2002, ApJ, 566, 577}
\vspace{0.6cm}
\end{description}

\newpage
\begin{description}
{\footnotesize
%
\item[\hspace{-0.3cm}]
Sekanina, Z. 2021, eprint arXiv:2109.01297 (Paper 1)
\\[-0.59cm]
\item[\hspace{-0.3cm}]
Sekanina, Z. 2022, eprint arXiv:2202.01164 (Paper 2)
\\[-0.59cm]
\item[\hspace{-0.3cm}]
Sekanina, Z., \& Chodas, P. W. 2002, ApJ, 581, 760
\\[-0.59cm]
\item[\hspace{-0.3cm}]
Sekanina, Z., \& Chodas, P. W. 2004, ApJ, 607, 620
\\[-0.59cm]
\item[\hspace{-0.3cm}]
Sekanina, Z., \& Chodas, P. W. 2007, ApJ, 663, 657
\\[-0.59cm]
\item[\hspace{-0.3cm}]
Sekanina, Z., \& Chodas, P. W. 2008, ApJ, 687, 1415
\\[-0.59cm]
\item[\hspace{-0.3cm}]
Sekanina, Z., \& Chodas, P. W. 2012, ApJ, 757, 127 (33\,pp)
\\[-0.59cm]
\item[\hspace{-0.3cm}]
Stephenson, F. R., \& Morrison, L.\,V. 1984, in Rotation in the Solar{\linebreak}
 {\hspace*{-0.6cm}}System, ed. R. Hide (London: Royal Society), 165
\\[-0.59cm]
\item[\hspace{-0.3cm}]
Stern, S. A., Weaver, H. A., Spencer, J. R., et al.\ 2019, Science,{\linebreak}
 {\hspace*{-0.6cm}}364, 9771
\\[-0.59cm] 
\item[\hspace{-0.3cm}]
Strom, R. 2002, A\&A, 387, L17
\\[-0.59cm]
\item[\hspace{-0.3cm}]
Suresh, S. 2004, Fatigue of Materials. (2nd ed.)  Cambridge, UK:{\linebreak}
 {\hspace*{-0.6cm}}Cambridge University
 Press, 703pp
\\[-0.59cm]
\item[\hspace{-0.3cm}]
Tammann, G. A. 1966, IAU Circ., 1952
\\[-0.68cm]
\item[\hspace{-0.3cm}]
Yeomans, D. K., \& Chodas, P. W. 1989, AJ, 98, 1083}
\vspace{0.25cm}
\end{description}
\end{document}